\def\Fig#1{Fig.~\ref{#1}}
\def\Figs#1#2{Figs.~\ref{#1} and \ref{#2}}
\def\BE{\begin{equation}}
\def\EE#1{\label{#1}\end{equation}}
\def\be{\begin{align}}
\def\LA{\langle}
\def\ra{\rangle}
\def\i{{\rm i}}
\def\e{{\rm e}}
\def\d{{\rm d}}
\def\Nabla{\bm\nabla}
\def\rot{\Nabla_{\bf x}\times}
\def\ROT{\Nabla_{\bf X}\times}
\def\rf#1{(\ref{#1})}
\def\R{{\mathbb R}}
\def\T{{\mathbb T}}
\def\bz{\bm\zeta}
\def\A{\boldsymbol{\mathfrak{A}}}
\def\D{\mathfrak{D}}
\def\Db{\boldsymbol{\mathfrak{D}}}
\def\L{\boldsymbol{\mathfrak{L}}}
\def\P{\boldsymbol{\mathfrak{P}}}
\def\F{\boldsymbol{\mathfrak{F}}}
\def\kk{{\bf k}}
\documentclass[preprint2,10pt,useAMS,usenatbib]{mAAStex}
\usepackage{amsmath,amssymb,graphicx,bm,url,textcomp}
\begin{document}

\shorttitle{Negative magnetic eddy diffusivities}
\shortauthors{A.~Andrievsky, A.~Brandenburg, A.~Noullez, V.~Zheligovsky}
\title{Negative magnetic eddy diffusivities from test-field method and multiscale stability theory}
\author{Alexander Andrievsky$^1$, Axel Brandenburg$^{2,3}$,
Alain Noullez$^{4}$, and Vladislav Zheligovsky$^{1,4}$}
\affil{
$^1$Institute of earthquake prediction theory and mathematical geophysics
Russian Ac.~Sci.,\\
84/32 Profsoyuznaya St., 117997 Moscow, Russia\\
$^2$Nordita, KTH Royal Institute of Technology and Stockholm University,\\
Roslagstullsbacken 23, SE-10691 Stockholm, Sweden\\
$^3$Department of Astronomy, Stockholm University, AlbaNova University Center,
SE-10691 Stockholm, Sweden\\
$^4$Laboratoire Lagrange, Universit\'e C\^ote d'Azur, Observatoire de la C\^ote
d'Azur,\\CNRS, Blvd de l'Observatoire, CS 34229, 06304 Nice cedex 4, France
}

\begin{abstract}
The generation of a large-scale magnetic field in the kinematic regime
in the absence of an $\alpha$-effect is investigated by following
two different approaches: the test-field method and the
multiscale stability theory relying on the homogenisation technique.
Our computations of the magnetic eddy diffusivity tensor
of the parity-invariant flow~IV of G.O.~Roberts and the modified
Taylor--Green flow confirm the findings
of previous studies, and also explain some of their apparent contradictions.
The two flows have large symmetry
groups; this is used to considerably simplify the eddy diffusivity
tensor. Finally, a new analytic result is presented: upon
expressing the eddy diffusivity tensor in terms of solutions
to auxiliary problems for the adjoint operator, we derive relations
between the magnetic eddy diffusivity tensors that arise for mutually reverse
small-scale flows $\bf v(x)$ and $-\bf v(x)$.
\end{abstract}


\maketitle

\section{Introduction}

It is well-known that at sufficiently high Reynolds number turbulence
is characterised by a hierarchy of fluctuations
interacting on a wide range of space and time scales. When this happens
in a flow of conducting fluid, magnetic field generation commences
if the magnetic Reynolds number is sufficiently high \citep{Moff}. As predicted
by the magnetic induction equation governing the process of generation, small
scales also develop in the generated magnetic field. The interaction of fine
structures of flow and magnetic field usually influences
the evolution of their large-scale parts. In particular, by Parker's
hypothesis, such an interaction may give rise to a mean electromotive force
(e.m.f.), parallel to the large-scale magnetic field.

In astrophysics, when the generation of the geomagnetic or solar magnetic
field is under investigation, fine structures are generally of lesser
interest than global ones. With present-day computers, it is impossible
to resolve structures over the whole range of interacting scales;
by choosing the domain of integration of the equations of magnetohydrodynamics,
we can only focus on the large or small scales.
However, in simulations of the global picture it is desirable to take
into account the integral influence of physical processes at small scales.

Since the 1960s, German scientists (\citealt{SKR66}; see also \citealt{KR80})
were developing the {\it theory
of mean-field electrodynamics} (MFE), a first attempt supposed to advise
how to do this. Perhaps, the best introduction to the ideas on which
this theory is built is by one of its founders Karl-Heinz \cite{Ra07}.
The three-dimensional magnetic and flow velocity fields,
$\bf b$ and $\bf v$, are decomposed into ``mean'', $\overline{\bf b}$ and
$\overline{\bf v\vphantom{t}}$, and ``fluctuating'', $\bf b'$ and $\bf v'$, fields:
$${\bf b}=\overline{\bf b}+{\bf b'},\qquad{\bf v}=\overline{\bf v\vphantom{t}}+{\bf v'}.$$
Any averaging procedure is deemed acceptable provided it satisfies
the Reynolds rules (see \citealt{Ra07}), e.g.,
planar averaging over any pair of Cartesian variables,
one-dimensional averaging along any given direction, or ensemble
averaging for turbulent flows.
The equations for mean magnetic field and fluctuations take the form
\be{\partial\overline{\bf b}\over\partial t}=&\,\eta\nabla^2\overline{\bf b}
+\Nabla\times(\overline{\bf v\vphantom{t}}\times\overline{\bf b}+\overline{\bf v'\times b'}),\label{bmean}\\
{\partial{\bf b'}\over\partial t}=&\,\eta\nabla^2{\bf b'}
+\Nabla\times\big(\overline{\bf v\vphantom{t}}\times{\bf b'}+{\bf v'}\times\overline{\bf b}
+({\bf v'\times b'})'\big).\label{bprim}\end{align}
Here ${\bf f'}\equiv{\bf f}-\overline{\bf f}$ denotes the fluctuating part
of a vector field~$\bf f$. The problem then reduces to the use of~\rf{bprim}
for expressing the mean e.m.f.~$\overline{\bf v'\times b'}$ in terms
of $\overline{\bf b}$ and $\overline{\bf v\vphantom{t}}$.
For simplicity, we henceforth assume that $\overline{\bf v\vphantom{t}}=0$
and $\bf v'$ is steady.
In MFE, for homogeneous stationary turbulence,
the mean e.m.f.~is usually expressed in terms of the mean magnetic field as
\be\overline{\bf v'\times b'}=\int\!\!\!\int\big(&{\cal K}_\alpha
({\bf x}-\bm\xi,t-\tau)\overline{\bf b}(\bm\xi,\tau)\label{coint}\\
-&{\cal K}_\eta({\bf x}-\bm\xi,t-\tau)\Nabla\times\overline{\bf b}(\bm\xi,\tau)\big)
\,\d\bm\xi\,\d\tau\nonumber\end{align}
when averaging is planar (${\cal K}_\alpha$ and ${\cal K}_\eta$ do not
depend on the spatial variables over which the e.m.f.~is averaged in the
l.h.s.) --- in general, $\bm\eta$ should be defined as a rank 3 tensor
acting on $\Nabla\overline{\bf b}$.
Our task is to determine the kernels. In Fourier space, \rf{coint} implies
\BE{\cal F}_{\kk,\omega}(\overline{\bf v'\times b'})=
{\bm\alpha}(\kk,\omega){\cal F}_{\kk,\omega}\overline{\bf b}
-{\bm\eta}(\kk,\omega){\cal F}_{\kk,\omega}(\Nabla\times\,\overline{\bf b}).\EE{xemf}
Here, following \cite{BRS08}, we have denoted
\BE{\cal F}_{\kk,\omega}{\bf f}\equiv\int\!\!\!\int
\e^{-\i(\kk\cdot{\bf x}-\omega t)}\,{\bf f}({\bf x},t)\,\d{\bf x}\,\d t,\EE{Fou}
$${\bm\alpha}(\kk,\omega)={\cal F}_{\kk,\omega}
{\cal K}_\alpha({\bf x},t),\quad{\bm\eta}(\kk,\omega)=
{\cal F}_{\kk,\omega}{\cal K}_\eta({\bf x},t).$$
In the limit $\kk\to0$ and $\omega\to0$,
$\bm\alpha$ and $\bm\eta$ describe the (magnetic\footnote{This
paper is devoted to the study of magnetic
$\alpha$-effect and magnetic eddy diffusivity exclusively --- as opposed
to the hydrodynamic $\alpha$-effect known as the AKA-effect
(see \citealt{FSS87,DF91}),
or combined $\alpha$-effect and eddy diffusivity emerging in large-scale
perturbations of magnetohydrodynamic regimes (see Chaps.~6--9 in \citealt{Zh}).
Note that the expression ``magnetic $\alpha$-effect'' is sometimes used with
a different meaning, designating a term proportional to current helicity that
quenches against the kinetic $\alpha$-effect. With this disclaimer in mind,
we omit the attribute ``magnetic'' from now on when referring
to the $\alpha$-effect and eddy diffusivity.}) $\alpha$-effect and
eddy diffusivity correction\footnote{We use here the terminology of the
multiscale stability theory. In fact, the ``corrections'' can be much larger
than the molecular diffusivity which they ``correct'' ---
the turbulent diffusivity can be by orders of magnitude
larger than the molecular diffusivity.} tensors.

The {\it test-field method\/}\footnote{Not to be confused with Kraichnan's
``test-field model'' of turbulence \citep{kra}, used by \cite{SLF} as a method
for closure of the hierarchy of moment equations.} (TFM) for computing
$\bm\alpha$ and $\bm\eta$ was developed within the MFE paradigm.
To the best of our knowledge, it was first proposed
by \cite{Sch05,Sch07}. Perhaps, the most detailed description of the TFM
procedure applied by \cite{BrMi} is found in \cite{Br08}. The recipe is to solve
equation \rf{bprim} for zero-mean magnetic perturbation $\bf b'$, where
$\overline{\bf b}$ is a test field.
The initial condition for $\bf b'$ can be any solenoidal small-scale zero-mean field
(for instance,~0). For space-periodic magnetic fields, the test fields
\BE\overline{\bf b}=\cos(\kk\cdot{\bf x})\,{\bf e}_n\qquad\mbox{and}\qquad
\overline{\bf b}=\sin(\kk\cdot{\bf x})\,{\bf e}_n,\EE{ritf}
are chosen.
By using sufficiently many independent test fields, we obtain a linear
system of equations that relates $\overline{{\bf v'}\times{\bf b'}}$
through the unknown coefficients of $\bm\alpha$ and $\bm\eta$
to $\overline{\bf b}$.
This system can be solved to obtain $\bm\alpha$ and~$\bm\eta$.
Similarly, the temporal dependence of the kernels in \rf{coint}
can be ``probed'' in Fourier space by considering the test fields
\BE\overline{\bf b}=\cos(\kk\cdot{\bf x})\,\e^{-\i\omega t}\,{\bf e}_n\quad
\mbox{and}\quad
\overline{\bf b}=\sin(\kk\cdot{\bf x})\,\e^{-\i\omega t}\,{\bf e}_n.\EE{mem}

In kinematic dynamo problems, where the evolution of a weak magnetic field
is studied (so that its influence on the flow via the Lorentz force can be
neglected), the flow velocity, $\bf v$, is known a priori. It can be a stationary
field, often supposed to have a vanishing average ($\overline{\bf v\vphantom{t}}=0$),
as have the flows that we consider in this paper. Alternatively, it can be
a time-dependent flow, for instance, supplied by an independent hydrodynamic
simulation. The kinematic dynamo problem is an instance of the full
magnetohydrodynamic (MHD) stability problem that focuses on the stability of
non-magnetic states; the flow and magnetic field perturbations then
decouple since the Lorentz force is quadratic in the magnetic field.
In a general setup, one considers the stability
of an MHD regime featuring a non-vanishing magnetic field that affects the
flow, and therefore perturbations involve both the flow and magnetic field
that cannot be disentangled.

MHD perturbations involving much larger spatial and temporal scales
than those of the perturbed MHD regimes (which, e.g., can be periodic
or quasi-periodic in space, and steady or periodic in time) can also be
explored by an approach known as the {\it multiscale stability theory} (MST).
It originates from the studies
of hydrodynamic stability \citep{DF91} and kinematic dynamo \citep{La} and
relies on mathematically precise asymptotic methods for {\it homogenisation
of elliptic operators}. An introduction to MST can be found in \cite{Zh};
the linear MHD stability problem for large-scale perturbations was
considered by \cite{Zh03} (see also Chap.~6 of \citealt{Zh}).
Here we will only consider the kinematic dynamo problem, and focus
on the generation of a magnetic field involving large scales by a small-scale
fluid flow. For a steady flow, the dynamo problem
can be reduced to the eigenvalue problem for the magnetic induction operator:
\BE\eta\nabla^2{\bf b}+\Nabla\times({\bf v}\times{\bf b})=\lambda{\bf b}\EE{eig}
(here $\eta$ denotes the magnetic molecular diffusivity
and $\lambda$ is the eigenvalue).

We assume that a {\it large-scale magnetic mode} $\bf b(X,x)$ depends
on {\it fast}, $\bf x$, and {\it slow}, ${\bf X}=\varepsilon\bf x$,
spatial variables, the flow depends only on
$\bf x$, and the scale ratio $\varepsilon$ is small. We proceed
by expanding a mode $\bf b(X,x)$ and the associated eigenvalue $\lambda$ (its
real part is the growth rate of the mode) in power series in $\varepsilon$,
\BE{\bf b}=\sum_{n=0}^\infty{\bf b}_n({\bf X},{\bf x})\,\varepsilon^n,
\qquad\lambda=\sum_{n=0}^\infty\lambda_n\varepsilon^n,\EE{bexp}
and deriving a hierarchy of equations that the eigenvalue equation yields
in successive orders $\varepsilon^n$. As it turns out, we can find each term
of the expansions by solving successively equations from this hierarchy.
For parity-invariant flows, that we will mostly consider, the series
for the eigenvalue involves only even powers of $\varepsilon$
(see Section~3.5 of \citealt{Zh}).

The first equation in the hierarchy shows that the leading terms ${\bf b}_0$
and $\lambda_0$ in the expansion~\rf{bexp} are, respectively, a small-scale
eigenfunction and the associated eigenvalue of the operator of magnetic
induction. The asymptotic expansion can be developed for any eigenvalue
$\lambda_0$. For small scale ratios $\varepsilon$, the growth rate
may exceed Re($\lambda_0$) due to the interaction of the fluctuating components
of the magnetic field and of the small-scale flow, but the corrections are
at best linear in the small parameter $\varepsilon$ and hence small. We are
mostly interested in the case where no small-scale magnetic field
is generated and $\lambda_0=0$, since then the presence of large spatial scales
can, in principle, result in the onset of magnetic field generation, i.e., in
a qualitative change in the behaviour of the MHD system. (The case of
an oscillatory small-scale kinematic dynamo occurring for imaginary $\lambda_0$
was considered in Section 3.8 of \citealt{Zh}; it is, actually, algebraically
much simpler.) For $\lambda_0=0$, the first
term ${\bf b}_0$ is a linear combination of neutral small-scale magnetic modes
with coefficients depending on the slow variable. These coefficients,
called {\it amplitudes}, are determined from the solvability conditions
for the higher-order small-scale equations from the hierarchy. When the problem
is considered in a three-dimensional periodic domain, the kernel
of the magnetic induction operator comprises three neutral magnetic modes whose
averages are the unit Cartesian coordinate vectors (generically, the kernel is
three-dimensional). The amplitudes of these modes can clearly be interpreted as
the Cartesian components of the mean magnetic field. Furthermore, by the theorem
on the Fredholm alternative \citep[see, e.g.,][]{SG},
the solvability condition consists
of the orthogonality of the inhomogeneous term to the kernel of the operator adjoint
to the operator of magnetic induction. Generically, this amounts to vanishing
of the integral of the inhomogeneous term over the periodicity box.
As a result, when $\lambda_0=0$, equations for the amplitudes can be interpreted
as mean-field equations, where the respective terms describe
the $\alpha$-effect or the eddy diffusivity effect.

The MST analysis reveals the non-universal character of \rf{coint}. This asymptotic
equality can be rigorously derived for a multiscale kinematic dynamo and
volume averaging in the generic case, when the kernel of the magnetic induction
operator comprises three magnetic modes with non-vanishing linearly independent
averages. However, \rf{coint} does not necessarily hold for other types
of averaging, or when the dimension of the kernel is higher --- in the latter
case, amplitudes of all neutral modes are involved in \rf{coint},
as this happens, e.g., for translation-invariant convective dynamos
(see, e.g., \citealt{CZ}). For MHD turbulence, \rf{coint} is likely
to stem, for various averaging procedures, from the ergodic properties
of the respective MHD dynamical system, but, to the best of our knowledge,
this equality was never fully demonstrated in the context of MFE
at the mathematical level of rigour; it remains a phenomenological property
of turbulence (such as, for instance, the Kolmogorov law).

The standard $\alpha$-effect and eddy diffusivity, arising in the limits
$\kk\to0$ and $\omega\to0$, are an idealisation in which nonlinear terms,
higher spatial derivatives and temporal derivatives of the magnetic field
are omitted in the expression \rf{coint} for the e.m.f.
This is justified if the mean fields vary sufficiently slowly
in space and time, i.e., on scales much larger and longer than
those of the fluctuations.
While this simplification may be permissible in some cases,
e.g., for forced turbulence with sufficient scale separation,
for certain flows, such as the Roberts and Otani flows, it is not \citep{HB09}.
A particularly striking example are flows II and III of
\cite{GOR}; for describing the nature of the dynamo in those flows,
it is crucial to retain the convolution in time
in the integral operators in \rf{coint} \citep{RDRB14}. Then
the electromotive force at a given time depends on the
magnetic field also at earlier times, so the system possesses ``memory''.
It is important to realise that the memory effect does occur even
for steady flows such as those considered here.
Excluding the memory effect from consideration more often
results in quantitative distortions, such as too high
an estimate for the critical dynamo number \citep{RB12}, rather than
in qualitative changes.

We note in passing that,
instead of implementing an integral transform in both space and time,
which is cumbersome, it is convenient to solve an
evolution equation for the e.m.f.~$\overline{\bf v'\times b'}$.
Such an equation was first derived by \cite{BF02} using the
$\tau$ approximation, which captures temporal nonlocality,
i.e., the memory effect \citep{HB09}.
This was then extended by \cite{RB12} to capture also spatial nonlocality.
Usually this also yields a satisfactory (at least qualitatively) description
of the unusual phenomena related to the memory effect, such as the ones
encountered in flows II and III of G.O.~Roberts \citep{RDRB14}.
These ideas will turn out to be important in Section~\ref{DNSandTFM}, when we
compare the magnetic field for the modified Taylor--Green flow (mTG) obtained
from direct numerical simulations (DNS) with that found by TFM.

We have thus two independent theories: MFE, physical in spirit,
especially when making simplifying assumptions regarding the kernel in the
integral equation~\rf{coint} for MHD turbulence, and MST, which
yields a mathematically rigorous derivation of equations for similar
quantities from first principles. MST has a narrower scope, being
applicable to treat only linear and weakly nonlinear MHD stability problems.
While MST applies specifically to the limit $\kk\to0$ and $\omega\to0$,
TFM can be applied to non-infinitesimal $|\kk|$ and $\omega$.
It can therefore be used to assemble the kernels
${\cal K}_\alpha$ and ${\cal K}_\eta$.
In other words, MST strives to describe an influence of the flow,
characterised by certain temporal and spatial scales, on magnetic fields
involving much larger scales; TFM is more ambitious in trying to assess
the influence of both larger and smaller hydrodynamic scales on magnetic
field of a given scale.
Although the limits $\kk\to0$ and $\omega\to0$ can be numerically expensive
for TFM, a comparison with MST is possible.

Recently \cite{BrMi} applied TFM to compute the magnetic eddy diffusivity in flows
previously employed in the studies of \cite{La} and \cite{GOR} with the use
of MST and a similar approach. \cite{Du07} observed in simulations
the beginning of magnetic field generation by mTG when
increasing the magnetic Reynolds number starting from small values,
which the authors cautiously attributed to the onset of the action of negative
magnetic eddy diffusivity investigated by \cite{La}. \cite{BrMi}
found, in agreement with \cite{GOR}, that the so-called flow~IV of G.O.~Roberts
(further referred to as R-IV)
does yield negative magnetic eddy diffusivity, but they failed to reproduce
the results of \cite{La} on the presence of negative magnetic eddy diffusivity
in mTG. We resolve this controversy in the present paper and show that
in a suitable parameter range eddy diffusivity is negative,
however, the relevant
TFM averaging is not over the horizontal plane (which is applicable
for R-IV), but one along the vertical direction,
or a planar one over any of the other two Cartesian coordinate planes
such that the average still depends on one of the two horizontal directions.

The need for the cross-examination stems from the fact that some applications
of the MFE ideas can fail to conform with the mathematical structure of problems
under consideration. For instance,
the mean e.m.f.\ computed as ``an average over the lower half-volume, the upper
half-volume, or, better still, one half of the difference of these two'' was
used in the~studies of \cite{CH06,CH08} of the $\alpha$-effect in convective
dynamo in a layer. How could these procedures possibly help to track
the evolution of the mean magnetic field? Such an averaging does not obey
the Reynolds rules, namely, because averaging and taking the spatial gradient do not commute,
and turns the midplane into an artificial boundary.
In each half-cell the mean field depends only on the horizontal variables.
The opposite $\alpha$-effect values in two adjacent half-cells force us
to assume opposite mean fields over and below the~midplane, in order to avoid
singularities in the $\alpha$-effect operator at the midplane. This inevitably
implies the existence of a boundary layer at the midplane. However, nothing
resembling a boundary-layer kind of behaviour of magnetic field
in the numerical solutions was reported {\it ibid.}, clearly showing that
averaging over a half-cell is unnatural and incompatible with the physics
of the problem,
and is also inappropriate from the mean-field electrodynamics perspective.
The $\alpha$-effect operator must be calculated by averaging over the entire
periodicity cell; the observed ``antisymmetry of $\alpha$ about the midplane''
\citep{CH06} simply implies that in these dynamos the relevant $\alpha$-effect
is zero (i.e., the $\alpha$-effect operator is not involved in the equations
for the evolution of the mean field), and the essential eddy effect is eddy
diffusivity. Furthermore, the convective dynamos considered {\it ibid.} are
translation-invariant, and hence some amplitudes, essential in the description
of the large-scale modulation of the generated instability modes, cannot be
interpreted as mean fields\footnote{We will see in Section \ref{math}
that a large-scale magnetic mode has the structure
${\bf b}=\sum_n^NB^n({\bf X})\widetilde{\bf S}_n({\bf x})+{\rm O}(\varepsilon)$, where
$N=\dim\ker\L$ is the number of independent small-scale neutral magnetic modes
$\widetilde{\bf S}_n({\bf x})$.
By normalizing the small-scale modes, we can impose the conditions
$$\overline{\widetilde{\bf S}}_n=\left\{
\begin{array}{ll}{\bf e}_n&\mbox{~for }n\le K,\\
0&\mbox{~for }K+1\le n\le N.
\end{array}\right.$$
We then find $\overline{\bf b}=\sum_n^KB^n({\bf X}){\bf e}_n$; thus, for $n\le K$
the amplitudes $B^n({\bf X})$ have the sense of the mean components of the mean field
$\overline{\bf b}$; for $n\ge K+1$ no any such or similar interpretation is possible.}
(see Chaps.~8 and 9 in \citealt{Zh}); neglecting these modes is also likely
to affect the results of \cite{CH06,CH08}. As a result, no sound
conclusions concerning the $\alpha$-effect, intended for astrophysical
or general MHD applications, can be drawn from the findings of those two papers.

Our paper is organised as follows. In Section~\ref{math} we remind the reader of
the MST formalism for the large-scale kinematic dynamo. In Section~\ref{flowIV}
we calculate, in the MST framework, the operator of magnetic eddy diffusivity
for R-IV using its many symmetries, and state results of the computation
of its two coefficients. In Section~\ref{TGf} we discuss how the symmetries
of mTG reduce the number of auxiliary problems involved in MST computations
of eddy diffusivity, and present numerical results. Despite using
algorithms that differ drastically from those used by \cite{La}, we reproduce
the results of this paper with 4 significant digits. In Section~\ref{TGtest}
we explain why no large-scale dynamo was found for mTG
by \cite{BrMi}, and show that eddy diffusivities obtained by TFM with
an alternative planar averaging qualitatively agree with the MST values.
In Section~\ref{TGnum} we show that the growth rates of large-scale dynamo
modes have the symmetry properties implied by the structure of the eddy
diffusivity operator. In Section \ref{work} we demonstrate that the TFM
procedure with the spatial averaging reproduces the MST $\alpha$-effect and
eddy diffusivity tensors, and consider analytically and numerically
the difference of the two approaches for a planar averaging using mTG as
an example. Concluding remarks end the paper.

\section{The mathematical theory of generation\\of large-scale magnetic field}
\label{math}

We review here the results of application of MST for the investigation
of large-scale magnetic field generation by small-scale steady flow
of electrically conducting incompressible fluid \citep{La,zpf,Zh}. We consider
the kinematic dynamo problem as a problem of determination of the spectrum
of the magnetic induction operator, which enables us to find growing large-scale
modes even when in addition a small-scale
dynamo operates. For the sake of simplicity, both the large-scale magnetic mode
$\bf b(X,x)$ and the flow $\bf v(x)$ are assumed to be $2\pi$-periodic in each
fast spatial variable~$x_i$. The mode is solenoidal and satisfies the eigenvalue
equation~\rf{eig} for the magnetic induction operator.

{\it1. Magnetic $\alpha$-effect.}
Generically, the average of the leading term in the expansion~\rf{bexp} of a
magnetic mode, ${\bf B(X)}=\LA{\bf b}_0({\bf X},{\bf x})\ra$, and the leading
term in the expansion of the associated eigenvalue, $\Lambda=\lambda_1$, are
a solution to the eigenvalue problem for the $\alpha$-effect {\it operator},
\BE\ROT\A{\bf B}=\Lambda{\bf B},\EE{alpei}
in the subspace of solenoidal fields, $\Nabla_{\bf X}\cdot{\bf B}=0$. Here
the {\it tensor of magnetic $\alpha$-effect}, $\A$, is the $3\times3$ matrix
whose $n$-th column is $\LA{\bf v}\times{\bf S}_n\ra$, $\LA\cdot\ra$ denotes
the average over the periodicity cell $\T^3=[0,2\pi]^3$ of the fast variables,
$$\LA{\bf f}\ra({\bf X})=(2\pi)^{-3}\int_{\T^3}{\bf f}({\bf X},{\bf x})\,\d{\bf x},$$
vector fields ${\bf S}_n({\bf x})$ are zero-mean solutions to {\it auxiliary
problems of type I}:
\be&\L{\bf S}_n=-{\partial{\bf v}\over\partial x_n}\label{Seq0}\\
\Leftrightarrow\quad&\L({\bf S}_n+{\bf e}_n)=0,\label{Seq}\end{align}
$$\L{\bf b}\equiv\eta\nabla_{\bf x}^2{\bf b}+\rot({\bf v}\times{\bf b})$$
is the small-scale magnetic induction operator, and ${\bf e}_n$ are unit vectors
of the Cartesian coordinate system. ${\bf S}_n({\bf x})$ are solenoidal.

Let $\bf B(X)$ be a solenoidal space-periodic solution to the eigenvalue
problem~\rf{alpei} whose associated eigenvalue is $\Lambda$. Then
${\bf B}(\mu{\bf X})$ is also a solenoidal solution to~\rf{alpei} whose associated
eigenvalue is $\mu\Lambda$; for any integer $\mu$, positive or negative, this mode
possesses the spatial periodicity of the original mode $\bf B(X)$. Thus, a mean
field, that is initially an infinite sum of modes defined by~\rf{alpei}, grows
in general superexponentially; consequently, the large-scale magnetic field
grows and destabilises the MHD system on time scales that are intermediate
between the fast time $t$ and the slow time $T=\varepsilon t$ (unless all modes
defined by~\rf{alpei} are associated with imaginary eigenvalues $\Lambda$).

{\it2. Magnetic eddy diffusivity.}
A field $\bf f$ is {\it parity-invariant}, if
\BE{\bf f}(-{\bf x})=-{\bf f}({\bf x}),\EE{par}
and {\it parity-antiinvariant}, if
$${\bf f}(-{\bf x})={\bf f}({\bf x}).$$
For parity-invariant flows $\bf v$, parity-invariant and
parity-antiinvariant vector fields constitute invariant subspaces of the
magnetic induction operator $\L$. Hence, vector fields ${\bf S}_n({\bf x})$
are parity-antiinvariant, and the $\alpha$-effect is absent: $\A=0$.
The magnetic field~\rf{bexp} is then
\be{\bf b(X,x)}=\left.\sum_{n=1}^3\right(&B^n({\bf X})({\bf S}_n({\bf x})
+{\bf e}_n)\label{order1}\\
&+\left.\varepsilon\sum_{m=1}^3{\partial B^n\over\partial X_m}({\bf X})
\,{\bf G}_{mn}({\bf x})\!\right)\!+{\rm O}(\varepsilon^2),\nonumber
\end{align}
where vector fields ${\bf G}_{mn}({\bf x})$ are zero-mean solutions
to {\it auxiliary problems of type II}:
\BE\L{\bf G}_{mn}=-2\eta{\partial{\bf S}_n\over\partial x_m}
-{\bf e}_m\times({\bf v}\times({\bf S}_n+{\bf e}_n)).\EE{auxII}
${\bf G}_{mn}({\bf x})$ are parity-invariant.

The solenoidal mean part of the leading term in the expansion~\rf{bexp}
of the mode, and the leading term in the expansion of the associated
eigenvalue, $\Lambda=\lambda_2$, are a solution to the eigenvalue
problem for the operator of magnetic eddy diffusivity:
\BE\eta\nabla^2_{\bf X}{\bf B}+\ROT\sum_{n=1}^3\sum_{m=1}^3\Db_{mn}
{\partial B^n\over\partial X_m}=\Lambda{\bf B}.\EE{eddei}
Here, $\Db$ is the tensor of eddy diffusivity correction,
\BE\Db_{mn}=\LA{\bf v}\times{\bf G}_{mn}\ra.\EE{Ddef}

We assume that the mean fields reside and are bounded in the entire space $\R^3$.
Hence, solutions to the eigenvalue problem~\rf{eddei} are Fourier
harmonics\footnote{The vector $\varepsilon\bf q$ is analogous to the wave
vector $\kk$ referred to in the exposition of TFM in the Introduction.}
\BE{\bf B(X)}=\widetilde{\bf B}\,\e^{\i\bf q\cdot X},\qquad
\widetilde{\bf B}\cdot{\bf q}=0.\EE{Fouha}
Here, $\widetilde{\bf B}=(\widetilde B^1,\widetilde B^2,\widetilde B^3)$ and
${\bf q}=(q_1,q_2,q_3)$ are constant vectors satisfying the conditions
$|{\bf q}|=1$,
$\widetilde{\bf B}\cdot{\bf q}=0$ (solenoidality of the mean magnetic mode) and
\BE-\eta\widetilde{\bf B}-{\bf q}\times\sum_{n=1}^3\sum_{m=1}^3\Db_{mn}
\widetilde B^nq_m=\Lambda\widetilde{\bf B}.\EE{eiH}
Solenoidality of the modes implies
\BE\widetilde{\bf B}=\beta_t{\bf T}+\beta_p{\bf P},\EE{Btopo}
where
\BE{\bf T}=(-q_2,q_1,0),\quad{\bf P}=(q_1q_3,q_2q_3,-(q_1^2+q_2^2))\EE{topo}
(this is equivalent to decomposing the mode into the toroidal and poloidal
(this is equivalent to decomposing the mode into toroidal and poloidal
components). Substituting~\rf{Btopo} into~\rf{eiH} and scalar multiplying
by $\bf T$ and $\bf P$, we recast~\rf{eiH} into an equivalent eigenvalue problem
in the coefficients $\beta_t$ and $\beta_p$:
\be-\!\sum_{m,l,n}\!\D^l_{mn}P^l(\beta_tT^n+\beta_pP^n)q_m&=(q_1^2+q_2^2)(\eta
+\Lambda)\beta_t,\label{Frst}\\
\sum_{m,l,n}\!\D^l_{mn}T^l(\beta_tT^n+\beta_pP^n)q_m&=(q_1^2+q_2^2)(\eta
+\Lambda)\beta_p.\label{Scnd}\end{align}
Taking into account the symmetries of the generating flow can considerably
simplify the eigenvalue problem~\rf{Frst}--\rf{Scnd} (see Sections~\ref{flowIV}
and~\ref{TGf}).

Eigenvalues $\Lambda$ depend on the wave vector $\bf q$ of the large-scale
amplitude modulation: $\Lambda=\Lambda(\bf q)$. If the real part of
$\Lambda(\widetilde{\bf q})$ is the maximum of Re($\Lambda(\bf q)$) over unit
wave vectors~$\bf q$, then $\eta_{\rm eddy}=-\Lambda(\widetilde{\bf q})$
is called the {\it minimum magnetic eddy diffusivity}. When
Re$(\eta_{\rm eddy})>0$,
generation of large-scale magnetic field by the mechanism of negative
eddy diffusivity is possible. From a physicist's point of view, this mechanism
is important only if the flow $\bf v$ does not generate small-scale magnetic
fields (i.e., fields of the same spatial periodicity, as that of
the flow), because otherwise small-scale magnetic fields grow and destabilise
the MHD system on time scales of the order of unity, which is faster than
the growth of the large-scale field in the slow time $T=\varepsilon^2t$.
This can also be interpreted as follows: when only the small-scale dynamo
is acting, the magnetic field can involve Fourier harmonics of arbitrarily large
wave lengths (compatible with the boundary conditions, i.e., not exceeding
the size of the periodicity box when periodicity conditions in space are
considered), but they decay and are unimportant for generation. By contrast,
when the small-scale dynamo is inactive, the presence of large scales
in the field becomes a key ingredient, without which the mechanism of negative
eddy diffusivity cannot make a dynamo work. It can also happen that the small-
and large-scale mechanisms coexist and are acting simultaneously.

{\it3. Computation of the eddy diffusivity tensor.}
The load of computation of the tensor of eddy diffusivity correction
is halved, if instead of computing the fields ${\bf G}_{mn}$
one solves {\it auxiliary problems for the adjoint operator} \citep{Zh}:
\BE\L^*{\bf Z}_l={\bf v}\times{\bf e}_l,\EE{adj0}
for zero-mean fields ${\bf Z}_l$, $1\le l\le3$, the adjoint operator being
$$\L^*\,{\bf z}\equiv\eta\nabla_{\bf x}^2{\bf z}-{\bf v}\times(\rot{\bf z}),$$
since, as it is easy to see from~\rf{Ddef},~\rf{auxII} and~\rf{adj0},
\BE\D^l_{mn}=\!\left\LA\!{\bf Z}_l\cdot\!\left(\!2\eta{\partial{\bf S}_n\over\partial x_m}
+{\bf e}_m\times({\bf v}\times({\bf S}_n\!+{\bf e}_n))\!\right)\!\right\ra.\EE{Dlmk}

{\it4. Relations between tensors of magnetic eddy diffusivity
correction for mutually opposite
flows.} The average~\rf{Dlmk} can be expressed in terms of solutions to the
auxiliary problems for the adjoint operator. We decorate by the superscript
``minus'' the quantities pertinent to the reverse flow $-\bf v$:
$$\L^-{\bf b}\equiv\eta\nabla_{\bf x}^2{\bf b}-\rot({\bf v}\times{\bf b}),$$
$$\L^-({\bf S}^-_n+{\bf e}_n)=0,\qquad(\L^-)^*({\bf Z}^-_l+{\bf e}_l)=0.$$
Clearly,~\rf{adj0} implies
\BE\L^-(\Nabla_{\bf x}\times{\bf Z}_l+{\bf e}_l)=0,\EE{adj}
and hence for all $l$,
\BE\Nabla_{\bf x}\times{\bf Z}_l={\bf S}^-_l\quad\Rightarrow\quad
{\bf Z}_l=\eta^{-1}\nabla_{\bf x}^{-2}({\bf v}\times({\bf S}^-_l+{\bf e}_l)),\EE{ZlviaSl}
where $\nabla_{\bf x}^{-2}$ denotes the inverse Laplacian in the fast variables.
Using the analogues of these relations for the flow $\bf v$ to eliminate
${\bf S}_n$ in~\rf{Dlmk}, we obtain
\BE{\D^l_{mn}\!=\!\eta\left\LA\!{\bf Z}_l\cdot\!\left(\!2\,\Nabla_{\bf x}\times
{\partial{\bf Z}^-_n\over\partial x_m}-{\bf e}_m\times\nabla_{\bf x}^2{\bf Z}^-_n\!\right)\!\right\ra}.\EE{ZlZk}
Applying standard vector analysis transformations, we can express this average
as an integral of the scalar product of ${\bf Z}^-_n$ and a field
resulting from the action of a differential operator on ${\bf Z}_l$.
By self-adjointness of the Laplacian and the curl, and antisymmetry
of the triple product with respect to permutation of its factors, we find
\BE\D^l_{mn}=-(\Db^-)^n_{ml}.\EE{opf}

When small-scale magnetic fields are not generated (i.e., all eigenvalues
of the small-scale magnetic induction operator have non-positive real parts),
the auxiliary problems can be solved numerically by computing
${\bf S}_n+{\bf e}_n$ and $\Nabla_{\bf x}\times{\bf Z}_l+{\bf e}_l$ as
small-scale dominant eigenmodes of the magnetic induction operators $\L$ and
$\L^-$, respectively, (see~\rf{Seq} and~\rf{adj})
in the subspace of solenoidal vector fields whose average can be non-zero.
The same small-scale eigenvalue code is applied to solve all these six
eigenproblems, the flow being reversed, $\bf v\to-v$, when computing
$\Nabla_{\bf x}\times{\bf Z}_l$.

\section{Generation of large-scale magnetic field\\by R-IV}\label{flowIV}

\cite{GOR} studied how simple flows depending on two
spatial variables $x_1$ and $x_2$ (deemed horizontal), such as~\rf{iceIX}
(see below), generate magnetic fields, whose dependence on~$x_3$
enters via the factor $\e^{\i\varepsilon x_3}$. Here, $\varepsilon$ is a small
parameter; thus this work is clearly in the multiscale spirit, although he
did not present the complete multiscale formalism, nor derived the operator
of eddy diffusivity. His flow~IV (labelled here R-IV)
lacks the $\alpha$-effect; it is
the first known example of a dynamo exploiting the mechanism of negative
eddy diffusivity, as was suggested previously on general grounds \citep{zpf}.
To the best~of our knowledge, \cite{BrMi} were the first to identify and study
in detail this mechanism for R-IV. It should be emphasised that flows
II and III are also non-helical dynamos, thus indicative of a negative eddy
diffusivity effect; however, later those flows turned out to have
positive eddy diffusivity, and their dynamo action was identified as being
due to turbulent pumping with a time delay \citep{RDRB14}.

We follow \cite{BrMi} in investigating large-scale generation
by R-IV. In the spatial variables introduced
by \cite{Tilg} (rotated by $45^\circ$ about the vertical axis with respect
to the variables used by \citealt{GOR}), its Cartesian components are
\be v_1&=\sqrt{2}\sin x_1\cos x_2,\nonumber\\
v_2&=-\sqrt{2}\cos x_1\sin x_2,\label{iceIX}\\
v_3&=\sin x_1.\nonumber\end{align}
It is clearly incompressible and parity-invariant (see \rf{par}), thus lacking
an $\alpha$-effect.

\subsection{The effect of symmetries}\label{symIV}

The symmetries of the flow control the structure of the tensor of eddy
diffusivity correction $\Db$.

{\it1. Translation antiinvariance with respect to the shift by half a period
in $x_1$} of R-IV:
$${\bf v}(x_1,x_2,x_3)=-{\bf v}(x_1+\pi,x_2,x_3).$$
(Note that the nonlinearity in the Navier--Stokes equation is not invariant
for the antisymmetry of this type, making this choice of flow somewhat
academic.) Hence, applying the operation of shift by half a period
in the direction $x_1$, which we denote by~$\widehat{\ }$:
$$\widehat{\bf f}(x_1,x_2,x_3)\equiv{\bf f}(x_1+\pi,x_2,x_3),$$
to the eigenvalue problem~\rf{adj}, we find
\BE{\bf Z}^-_n=\widehat{\bf Z}_n.\EE{Zl}
Substituting this into~\rf{ZlZk}, using the self-adjointness of the Laplacian,
the curl and operator $\widehat{\ }$, and integrating by parts
in $x_m$ the first term in~\rf{ZlZk}, we obtain
\BE\D^l_{mn}=-\D^n_{ml},\EE{antiDIV}
and $\D^n_{mn}=0$ for any flow possessing translation antiinvariance with respect
to the shift by half a period in one of the spatial variables.

{\it2. Symmetry in $x_2$} of R-IV:
\be v^1(x_1,-x_2,x_3)&=v^1(x_1,x_2,x_3),\nonumber\\
v^2(x_1,-x_2,x_3)&=-v^2(x_1,x_2,x_3),\label{sig2}\\
v^3(x_1,-x_2,x_3)&=v^3(x_1,x_2,x_3);\nonumber\end{align}
antisymmetry in $x_2$, is defined by changing here the signs in the r.h.s.~to
the opposite ones. Clearly, the curl or vector multiplication by R-IV maps
fields, symmetric in $x_2$, to fields, antisymmetric in $x_2$, and vice versa.
Consequently, fields symmetric and antisymmetric in $x_2$ constitute invariant
subspaces of the operators of magnetic induction $\L$ and $\L^-$. It follows
from~\rf{Seq0} that ${\bf S}_n$ are symmetric in $x_2$ for odd $n$ and antisymmetric
in $x_2$ for $n=2$; \rf{adj0}~implies that ${\bf Z}_l$ are antisymmetric
in $x_2$ for odd $l$ and symmetric in $x_2$ for $l=2$.

Vector multiplication by ${\bf e}_m$ also maps symmetric in $x_2$ fields
to antisymmetric ones and vice versa for odd $m$, and
does not change the symmetry and antisymmetry of a field in $x_2$
for $m=2$. Therefore,~\rf{Dlmk} implies
\BE\D^l_{mn}=0,\quad\mbox{if~~}l+m+n\mbox{~~is odd.}\EE{zeroDlmk}

{\it3. Wave vector parity.} We call ``even'' a three-dimensional vector field
depending on two spatial variables $x_1$ and $x_2$, when it is a linear
combination of harmonics $\widetilde{\bf B}_{\bf q}\,\e^{\i\bf q\cdot x}$ such
that $\widetilde B^3=0$ if $q_1+q_2$ is even and
$\widetilde B^1=\widetilde B^2=0$ if $q_1+q_2$ is odd; we call
a field ``odd'', when it is a linear combination of harmonics
$\widetilde{\bf B}_{\bf q}\,\e^{\i\bf q\cdot x}$ such that $\widetilde B^3=0$
if $q_1+q_2$ is odd and $\widetilde B^1=\widetilde B^2=0$ if $q_1+q_2$ is even.
Clearly, in this terminology R-IV~\rf{iceIX} is even.

Taking the curl or calculating the vector product with R-IV transforms
an even field
into an odd one, and vice versa. Thus, even and odd fields constitute invariant
subspaces of the magnetic induction operators $\L$ and $\L^-$. By virtue
of~\rf{Seq0} and~\rf{adj0}, ${\bf S}_n$ are even for $n=1,2$ and odd for $n=3$,
while ${\bf Z}_l$ are odd for $l=1,2$ and even for $l=3$. Vector multiplication
by ${\bf e}_m$ maps odd fields into even ones and vice versa for $m=1,2$, and
it does not change this type of ``parity'' for $m=3$. Using this,
it is easy to show that
\BE\D^2_{11}=\D^3_{32}=0.\EE{D211}

{\it4. Swapping of the horizontal coordinates $x_1\leftrightarrow x_2$.}
Since the flow and solutions ${\bf S}_1$ and ${\bf S}_2$ to auxiliary problems
of type I are independent of the vertical coordinate, equations for horizontal
components of ${\bf S}_1$ and ${\bf S}_2$ involve the vertical components
neither of the flow, nor of the respective~${\bf S}_n$. We establish
by inspection that the field $(S_2^2(x_2,x_1+\pi),S_2^1(x_2,x_1+\pi))$
satisfies the same equation as $(S^1_1({\bf x}),S^2_1({\bf x}))$, and hence
\be S_1^1(x_1,x_2)&=S_2^2(x_2,x_1+\pi),\nonumber\\
S_1^2(x_1,x_2)&=S_2^1(x_2,x_1+\pi).
\label{S1S2}\end{align}
We use the second of these relations to show that
\BE\D^3_{21}=\D^2_{13}.\EE{D321}

Denote $\psi=\sqrt{2}\sin x_1\sin x_2$; clearly,
$$(v^1,v^2,0)=\rot(\psi{\bf e}_3).$$
Since the flow is independent of~$x_3$, for $n=3$ the source term
in the r.h.s.~of~\rf{Seq0} vanishes, and hence ${\bf S}^-_3=0$. Therefore,
by virtue of~\rf{ZlviaSl} for $l=3$, ${\bf Z}_3=(2\eta)^{-1}\Nabla_{\bf x}\psi$.
Since gradients are orthogonal to solenoidal fields in the Lebesgue space,
in expression~\rf{Dlmk} for $\D^3_{21}$ the term involving the derivative
$\partial{\bf S}_1/\partial x_2$ is zero. Hence, on the one hand,
\begin{align*}
\D^3_{21}=&(2\eta)^{-1}\left\LA\rot(\psi{\bf e}_2)\cdot(
{\bf v}\times({\bf S}_1+{\bf e}_1))\right\ra\\
=&(2\eta)^{-1}\left\LA\psi{\bf e}_2\cdot\rot({\bf v}\times({\bf S}_1+{\bf e}_1))\right\ra\\
=&-(2\eta)^{-1}\left\LA\psi{\bf e}_2\cdot\eta\nabla_{\bf x}^2{\bf S}_1\right\ra\\
=&\left\LA\psi S_1^2\right\ra.\end{align*}
We have used here the self-adjointness of the curl, \rf{Seq} for \hbox{$n=1$} and
the self-adjointness of the Laplacian. On the other, by the self-adjointness
of the curl and by virtue of~\rf{Dlmk}, \rf{Zl}~for $l=2$, \rf{Seq}~for $n=2$
and the relation ${\bf S}^-_3=0$,
\begin{align*}
\D^2_{13}=&\left\LA{\bf Z}_2\cdot({\bf e}_1\times({\bf v}\times{\bf e}_3))\right\ra\\
=&-\eta^{-1}\left\LA\nabla_{\bf x}^{-2}({\bf v}\times(\widehat{\bf S}_2+{\bf e}_2))
\cdot({\bf e}_1\times\Nabla_{\bf x}\psi)\right\ra\\
=&\eta^{-1}\left\LA\nabla_{\bf x}^{-2}(\eta\nabla_{\bf x}^2\widehat{\bf S}_2)
\cdot\psi{\bf e}_1\right\ra\\
=&\LA\psi\widehat S_2^1\ra.\end{align*}
Thus,~\rf{D321} follows from~\rf{S1S2}.

{\it5. Eddy diffusivity.} We calculate now eigenvalues of the eddy
diffusivity operator~\rf{eddei}. By~\rf{antiDIV}, the sums involving $\beta_p$
and $\beta_t$ in the l.h.s.~of~\rf{Frst} and~\rf{Scnd}, respectively, vanish,
and therefore these equations yield the same eigenvalue
$$\Lambda=-\eta+(q_1^2+q_2^2)^{-1}\sum_{m,l,n}\D^l_{mn}T^lP^nq_m.$$
By virtue of~\rf{topo},~\rf{antiDIV},~\rf{zeroDlmk},~\rf{D211} and~\rf{D321},
$$\Lambda=-\eta+\D^3_{12}\,(q_1^2+q_2^2)+\D^2_{31}\,q_3^2.$$
Actually, we have calculated the symbol of the eddy diffusivity
operator acting on mean fields (defined by the l.h.s.~of~\rf{eddei}); hence
this operator for R-IV~\rf{iceIX} is
\BE(\eta-\D^3_{12})\left({\partial^2\over\partial X_1^2}
+{\partial^2\over\partial X_2^2}\right)+(\eta-\D^2_{31})
{\partial^2\over\partial X_3^2}.\EE{lamIX}
The minimum eddy diffusivity is
$$\eta_{\rm eddy}=\eta+\min(-\D^3_{12},-\D^2_{31}).$$

\subsection{Numerical results}\label{numGOR}

The coefficients $\eta-\D^3_{12}$ and $\eta-\D^2_{31}$ of the eddy
diffusivity operator~\rf{lamIX} have been computed using~\rf{Dlmk}. Solutions
${\bf S}_n$ to auxiliary problems of type I, and solutions ${\bf Z}_l$ to auxiliary
problems for the adjoint operator have been computed by optimised iterations
\citep{Zh93} as the dominant (associated with the zero eigenvalue)
eigenfunctions of the operators of magnetic induction $\L$ \rf{Seq} and $\L^-$
\rf{adj}.
Iterations were terminated when the estimate of
the dominant eigenvalue was below $10^{-10}$ in absolute value and the norm of
the discrepancy for the normalised associated eigenvector was below
$5\cdot10^{-11}$. A resolution of $64^2$ Fourier harmonics was used before
dealiasing, that was performed by discarding harmonics with wave numbers
over~28. With this resolution, energy spectra of solutions to auxiliary
problems decay by 30 orders of magnitude for $\eta=0.2$ and still by 4 orders
for $\eta=0.01$. Plots of $\eta-\D^3_{12}$ and $\eta-\D^2_{31}$ are shown in
\Fig{fig2} for $0.01\le\eta\le0.2$; in this range of molecular
diffusivities no generation of small-scale magnetic fields takes place.

Figure \ref{fig2} implies that a large-scale magnetic field is not generated for
horizontal wave vectors $\bf q$ of the harmonic large-scale modulation, but it
is generated for the vertical wave vector. We did not check if generation of
small-scale fields starts on further decreasing the molecular diffusivity; the
behaviour of plots in \Fig{fig2} suggests that it may take place, and then
$\eta-\D^3_{12}\to-\infty$ and $\eta-\D^2_{31}\to\infty$ when $\eta$ approaches
the critical value for the onset of small-scale magnetic field generation. If
this happens, the type of the generated large-scale field changes: for smaller
$\eta$, generation of large-scale magnetic field for the vertical wave vector
$\bf q$ replaces the one for horizontal wave vectors.

\setlength{\unitlength}{.01\columnwidth}
\begin{figure}[t]
\begin{picture}(100,70)(0,0)
\put(0,0){\includegraphics[width=\columnwidth]{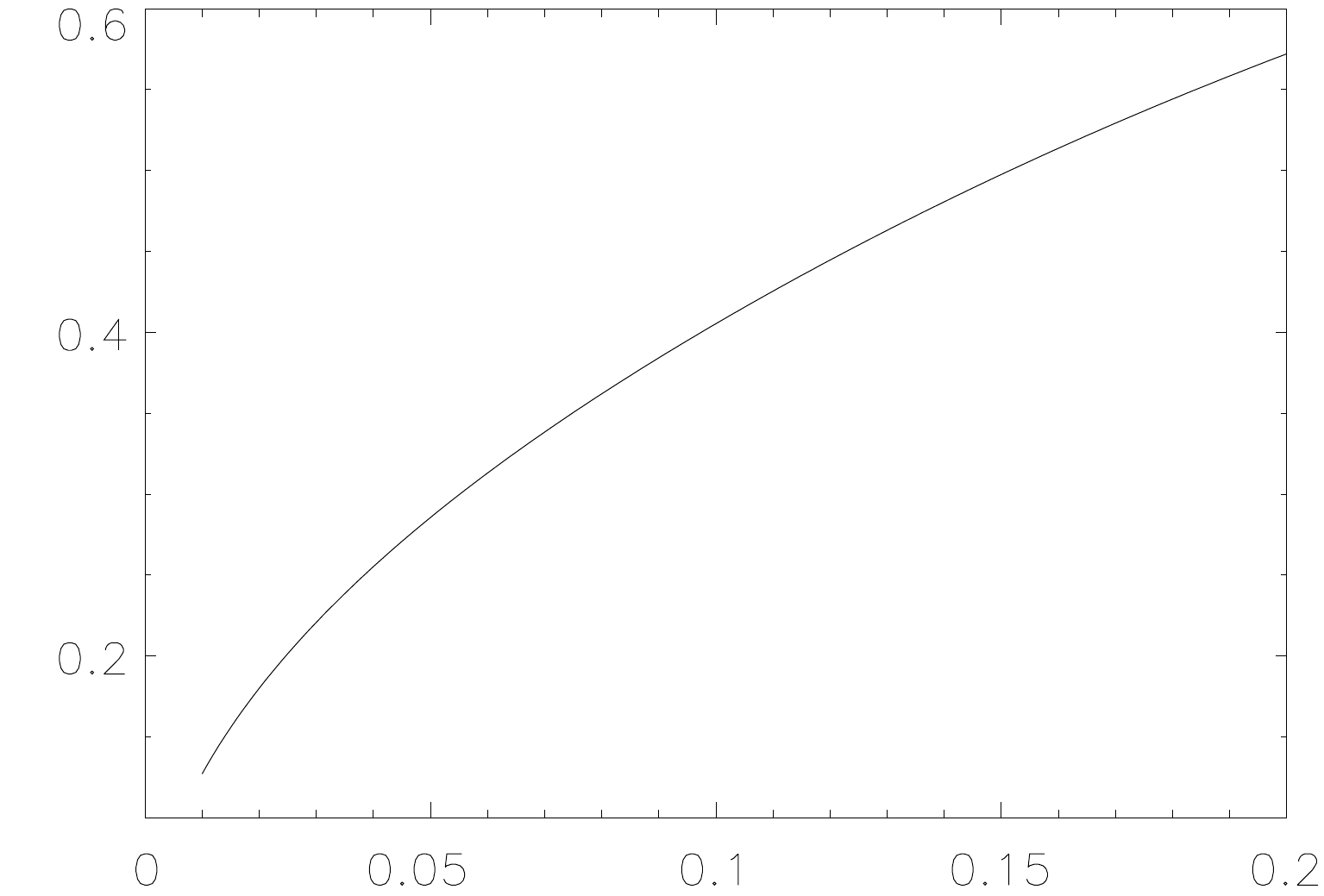}}
\put(53,9){$\eta$}
\end{picture}

\bigskip
\begin{picture}(100,70)(0,0)
\put(0,0){\includegraphics[width=\columnwidth]{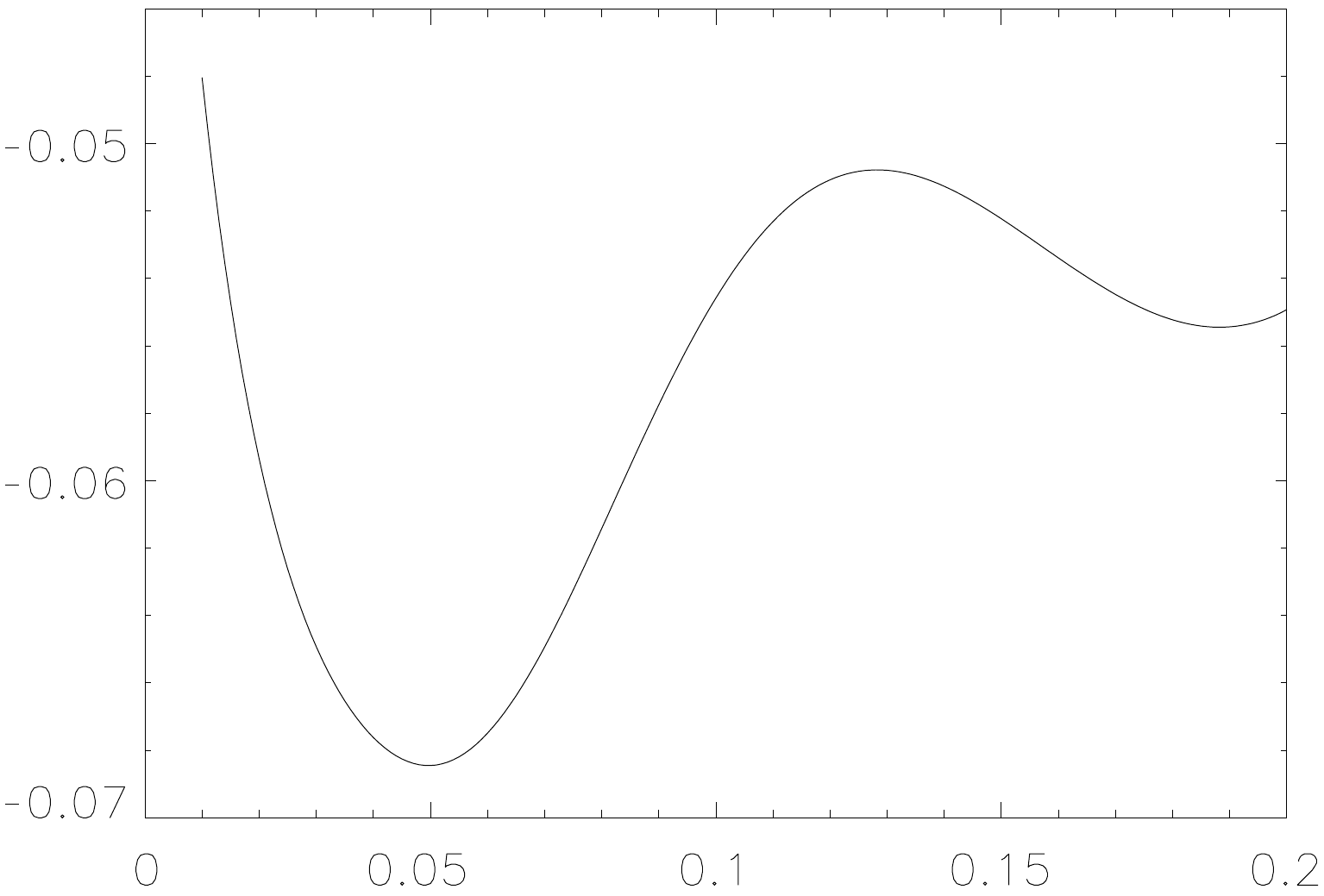}}
\put(53,9){$\eta$}
\end{picture}
\caption{Entries $\eta-\D^3_{12}$
(upper panel) and $\eta-\D^2_{31}$ (lower panel) of the eddy
diffusivity operator~\rf{lamIX} for R-IV~\rf{iceIX}.}
\label{fig2}\end{figure}

Plots of the two entries, $\eta-\D^3_{12}$ and $\eta-\D^2_{31}$, of the
eddy diffusivity tensor are shown in \Fig{fig2} for a range of
molecular diffusivities over the critical value for the onset
of generation of the small-scale magnetic field. The form~\rf{lamIX}
of the operator of eddy diffusivity corroborates the conclusions
of \cite{BrMi} that the eddy diffusivity tensor
for R-IV is diagonal and has a double eigenvalue, i.e., its action
on fields depending on the vertical slow variable (which was the object
of the studies of \citealt{GOR} and \citealt{BrMi}) is homogeneous. However,
since the two coefficients in~\rf{lamIX} are distinct (see \Fig{fig2}),
eddy diffusivity is
anisotropic, differing in the vertical and horizontal directions.
Comparison of the lower panel of Fig.~3 in \cite{BrMi} with the plot
of $\eta-\D^2_{31}$ in the lower panel of \Fig{fig2} reveals a reasonable
qualitative consistency between the large-scale magnetic field growth rates,
obtained by \cite{BrMi} in DNS, and the MST minimum
eddy diffusivity values, shown in the lower panel of \Fig{fig2},
for roughly $\eta<0.1$. However, while here we study
eddy diffusivity in the limit $\varepsilon\to0$, \cite{BrMi}
computed turbulent magnetic diffusivity for finite scale separations;
in particular, the plot in Fig.~3 {\it ibid.} shows the diagonal entry
of ${\bm\eta}(\varepsilon)$ for $\varepsilon=1$, whose behaviour is
clearly different from that of $\eta-\D^2_{31}$ presented in \Fig{fig2}.

\section{Generation of large-scale magnetic field\\by the modified Taylor--Green flow}\label{TGf}

As \cite{La} and \cite{BrMi}, we now consider large-scale generation
by the modified Taylor--Green flow (mTG), whose components are
\be v_1=&\sin x_1\cos x_2\cos x_3+a\sin 2x_1\cos 2x_3\nonumber\\
&+\,b\cos x_3(\sin x_1\cos3x_2+c\sin3x_1\cos x_2),\nonumber\\
v_2=&-\cos x_1\sin x_2\cos x_3+a\sin2x_2\cos2x_3\label{TG}\\
&-b\cos x_3(\cos3x_1\sin x_2+c\cos x_1\sin3x_2),\nonumber\\
v_3=&-a\sin2x_3(\cos2x_1+\cos2x_2)\nonumber\\
&+\,d\sin x_3(\cos x_1\cos3x_2-\cos3x_1\cos x_2).\nonumber\end{align}
The flow is incompressible for $d=b(3c-1)$, which will be henceforth assumed.
We now consider its symmetries relevant for simplification of the eigenvalue
problem~\rf{Frst}--\rf{Scnd} and calculate the eigenvalues.

\subsection{The effect of symmetries}\label{symTG}

{\it1. Symmetries in $x_i$.} A field ${\bf f}=(f^1,f^2,f^3)$ is called
{\it symmetric in $x_i$}, if for all $i$ and $j$ such that $1\le i,j\le3$
$$f^j((-1)^{\delta^1_i}x_1,(-1)^{\delta^2_i}x_2,(-1)^{\delta^3_i}x_3)
=(-1)^{\delta^j_i}\,f^j({\bf x})$$
(cf.~\rf{sig2}), and {\it antisymmetric in $x_i$}, if for all such $i$ and~$j$
$$f^j((-1)^{\delta^1_i}x_1,(-1)^{\delta^2_i}x_2,(-1)^{\delta^3_i}x_3)
=(-1)^{1-\delta^j_i}\,f^j({\bf x}),$$
where $\delta^j_i$ is the Kronecker symbol. Since mTG is symmetric in all
$x_i$, it is parity-invariant and lacks an \hbox{$\alpha$-effect}.

When a flow is symmetric in $x_i$, vector
fields possessing the symmetry or antisymmetry in $x_i$ constitute
invariant subspaces of the operators of magnetic induction $\L$ and $\L^-$.
Since all the three symmetries in $x_i$ are independent, there are eight such
invariant subspaces. We label them by 3-character strings;
{\tt A} and {\tt S} in the $i$-th entry of the label indicate that vector fields
in the invariant subspace are symmetric or antisymmetric in $x_i$,
respectively. For instance, {\tt SAA} labels the invariant subspace, in which
vector fields are symmetric in~$x_1$ and antisymmetric in $x_2$ and $x_3$.

By virtue of~\rf{Seq0} and~\rf{adj0}, invariance of the fields, symmetric
or antisymmetric in $x_i$ implies that ${\bf S}_n$
for $n\ne i$ and ${\bf Z}_i$ are symmetric in $x_i$, while ${\bf S}_i$ and
${\bf Z}_l$ for $l\ne i$ are antisymmetric in $x_i$. Consequently, $\D^l_{mn}=0$
when none of the indices $l,n$ and $m$ are equal to $i$. It follows
\BE\D^l_{mn}=\!0~~~\mbox{if~}m=n,\mbox{~or~}l=m,\mbox{~or~}l=n.\EE{eqInd}

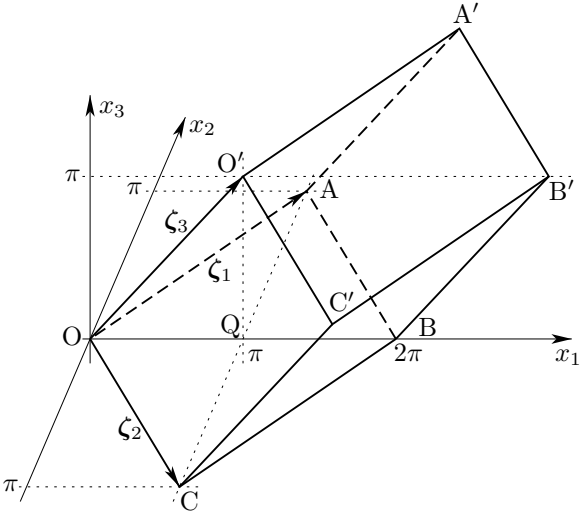
\begin{figure}[t]
\begin{picture}(100,85)(0,0)
\put(5,1){\includegraphics[width=.9\columnwidth]{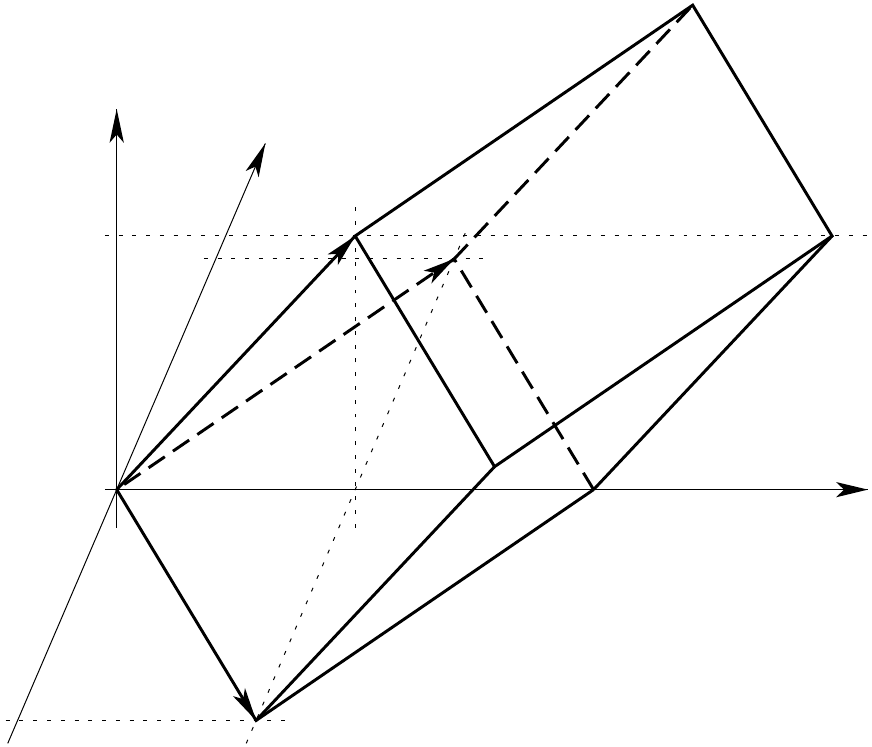}}
\put(13,53){$\pi$}
\put(23,51){$\pi$}
\put(3,3){$\pi$}
\put(12.5,26.5){O}
\put(75.5,78.5){A$'$}
\put(91,50){B$'$}
\put(37.5,54.5){O$'$}
\put(54,50.25){A}
\put(55.5,32){C$'$}
\put(70,28){B}
\put(31.5,0){C}
\put(38,28.5){Q}
\put(66,24){$2\pi$}
\put(42.5,24){$\pi$}
\put(92,24){$x_1$}
\put(18.5,64){$x_3$}
\put(33,61){$x_2$}
\put(29,45.5){$\bz_3$}
\put(21.5,12){$\bz_2$}
\put(36,37.5){$\bz_1$}
\end{picture}

\caption{An elementary periodicity cell of mTG~\rf{TG}:
a prism whose edges are periodicity vectors $\bz_i$ \rf{shf}.
The vertex O$'$ of the upper square base O$'$A$'$B$'$C$'$ projects down
along the vertical into the centre Q of the lower square base OABC
of the prism.}
\label{prsm}\end{figure}

{\it2. Swapping of the horizontal coordinates $x_1\leftrightarrow x_2$.}
The mTG has also a symmetry, which we denote
by~$\gamma$: a field $\bf f$ is $\gamma$-symmetric,~if
\begin{align*}
f^1(x_1,x_2,x_3)=f^2(x_2,x_1,x_3+\pi),\\
f^2(x_1,x_2,x_3)=f^1(x_2,x_1,x_3+\pi),\\
f^3(x_1,x_2,x_3)=f^3(x_2,x_1,x_3+\pi),\end{align*}
and $\gamma$-antisymmetric, if
\begin{align*}
f^1(x_1,x_2,x_3)=-f^2(x_2,x_1,x_3+\pi),\\
f^2(x_1,x_2,x_3)=-f^1(x_2,x_1,x_3+\pi),\\
f^3(x_1,x_2,x_3)=-f^3(x_2,x_1,x_3+\pi).\end{align*}
$\gamma$-symmetric and $\gamma$-antisymmetric fields constitute
invariant subspaces of the operators of magnetic induction $\L$ and $\L^-$.
This implies that ${\bf S}_3$ is $\gamma$-symmetric, ${\bf Z}_3$ is
$\gamma$-antisymmetric, for $n=1,2$ the field ${\bf S}_n$ is mapped
by the $\gamma$-symmetry to ${\bf S}_{3-n}$, and ${\bf Z}_n$ is mapped
by the $\gamma$-antisymmetry to~${\bf Z}_{3-n}$. (We thus need to compute
just 4 solutions to the auxiliary problems, say, ${\bf S}_1$, ${\bf S}_3$,
${\bf Z}_1$ and ${\bf Z}_3$; ${\bf S}_2$ and ${\bf Z}_2$ can then be obtained
by applying the $\gamma$-symmetry and $\gamma$-antisymmetry
to ${\bf S}_1$ and ${\bf Z}_1$, respectively.) Consequently, the remaining
non-zero entries of the eddy diffusivity correction tensor satisfy the relations
\BE\D^3_{12}=-\D^3_{21},\quad
\D^1_{23}=-\D^2_{13},\quad
\D^2_{31}=-\D^1_{32}.\EE{antiD}

When the $\gamma$-symmetry acts on a vector field, the symmetry or antisymmetry
in $x_1$ becomes a symmetry or antisymmetry, respectively, in $x_2$, and vice
versa. Thus, the $\gamma$-symmetry maps {\tt ASA} and {\tt SAA} mutually one
into another, as well as {\tt ASS} and {\tt SAS}. Since it also maps
an eigenfunction of the operator of magnetic induction, $\L$,
to an eigenfunction, restrictions of $\L$ on the two invariant subspaces,
constituting any of the two pairs, have the same spectra. The subspaces
{\tt AAA}, {\tt AAS}, {\tt SSA} and {\tt SSS} are invariant under the action
of the symmetry $\gamma$; each of them splits into invariant subspaces of~$\L$,
that consist of $\gamma$-symmetric or $\gamma$-antisymmetric fields.

{\it3. Wave number parity.} Inspection of~\rf{TG} reveals, that mTG
is comprised of Fourier harmonics $\e^{\i\bf k\cdot x}$
in which all the three wave numbers $k_i$ have the same parity, e.g., the sum
of any two wave numbers is even. Consequently, the obvious periodicity cell
$\T^3$ of the flow, which is a cube of size $2\pi$ whose edges are parallel
to the Cartesian coordinate axes, is not the smallest one. It is easily seen
that a flow possessing the parity property of this kind is invariant
under shifts along any of the periodicity vectors
\BE\bz_1=(\pi,\pi,0),\quad\bz_2=(\pi,-\pi,0),\quad\bz_3=(\pi,0,\pi).\EE{shf}
(Clearly, this translation invariance implies $2\pi$-periodi\-city in any Cartesian
variable $x_i$.) Therefore, elementary periodicity cells of the flow are prisms
whose edges are these vectors (see \Fig{prsm}).
Alternatively, one can regard the parallelepiped
$$0\le x_1\le2\pi,\qquad0\le x_2\le\pi,\qquad0\le x_3\le\pi$$
as an elementary periodicity cell of the flow, assuming the ``brick wall''
tiling of space by these cells, in which the parallelepipeds are arranged
in infinite ``bars'' parallel to the $x_1$-axis, and any two adjacent
bars are shifted along the $x_1$-axis by half a period relative each other.
The volume of the elementary periodicity cells of both types is $2\pi^3$, e.g.,
a quarter of that of $\T^3$. Nevertheless, by a small-scale dynamo we understand
the generation of magnetic fields which are $2\pi$-periodic in each variable
$x_i$.

Each invariant subspace of $\L$ considered above further splits into subspaces
of the so-called even and odd fields that are linear combinations of Fourier
harmonics such that the sums of the wave numbers $k_i+k_j$ are even or odd.
We therefore extend the labels of invariant subspaces by two additional
characters denoting the parity of the sums $k_1+k_2$ of wave numbers in the
horizontal directions (the fourth character), and the sums $k_1+k_3$ of wave
numbers in directions $x_1$ and $x_3$ (the fifth character); {\tt E} and {\tt O}
indicate even or odd such sums, respectively. For instance, the invariant
subspace {\tt SAAOE}
consists of vector fields that are symmetric in $x_1$, antisymmetric in $x_2$
and $x_3$, and are comprised of Fourier harmonics such that the sum of wave
numbers in the horizontal directions is odd and the sum $k_1+k_3$ is even;
the spectrum of $\L$ is the same in this subspace and in {\tt ASAOO}.

{\it4. Eddy diffusivity.}
For an eddy diffusivity correction tensor with the properties~\rf{eqInd}
and~\rf{antiD} stemming from the symmetries of the flow, in $x_i$ and $\gamma$,
it is straightforward, using~\rf{topo}, to reduce~\rf{Frst}--\rf{Scnd} to
\begin{align*}
(-\eta+\D^3_{12}(q_1^2+q_2^2)+\D^2_{31}q_3^2)\beta_t&=\Lambda\beta_t,\\
(-\eta+\D^1_{23}(q_1^2+q_2^2)+\D^2_{31}q_3^2)\beta_p&=\Lambda\beta_p,
\end{align*}
respectively. Therefore, the two eigenvalues are
\be\Lambda_1&=-\eta+\D^3_{12}(q_1^2+q_2^2)+\D^2_{31}q_3^2,\label{TGe1}\\
\Lambda_2&=-\eta+\D^1_{23}(q_1^2+q_2^2)+\D^2_{31}q_3^2\label{TGe2}
\end{align}
(this explains Fig.~3 in \citealt{La}). The minimum of $-\Lambda_1(\bf q)$ and
$-\Lambda_2(\bf q)$ over unit wave vectors $\bf q$ occurs for the vertical
unit vector $\widetilde{\bf q}={\bf e}_3$ or
at any horizontal unit vector $\widetilde{\bf q}=(q_1,q_2,0)$, and
$$\eta_{\rm eddy}=\eta+\min(-\D^1_{23},-\D^2_{31},-\D^3_{12}).$$

\subsection{Numerical results: eddy diffusivity}\label{TGtest}

\begin{figure}[t!]
\begin{picture}(100,70)(0,0)
\put(0,0){\includegraphics[width=\columnwidth]{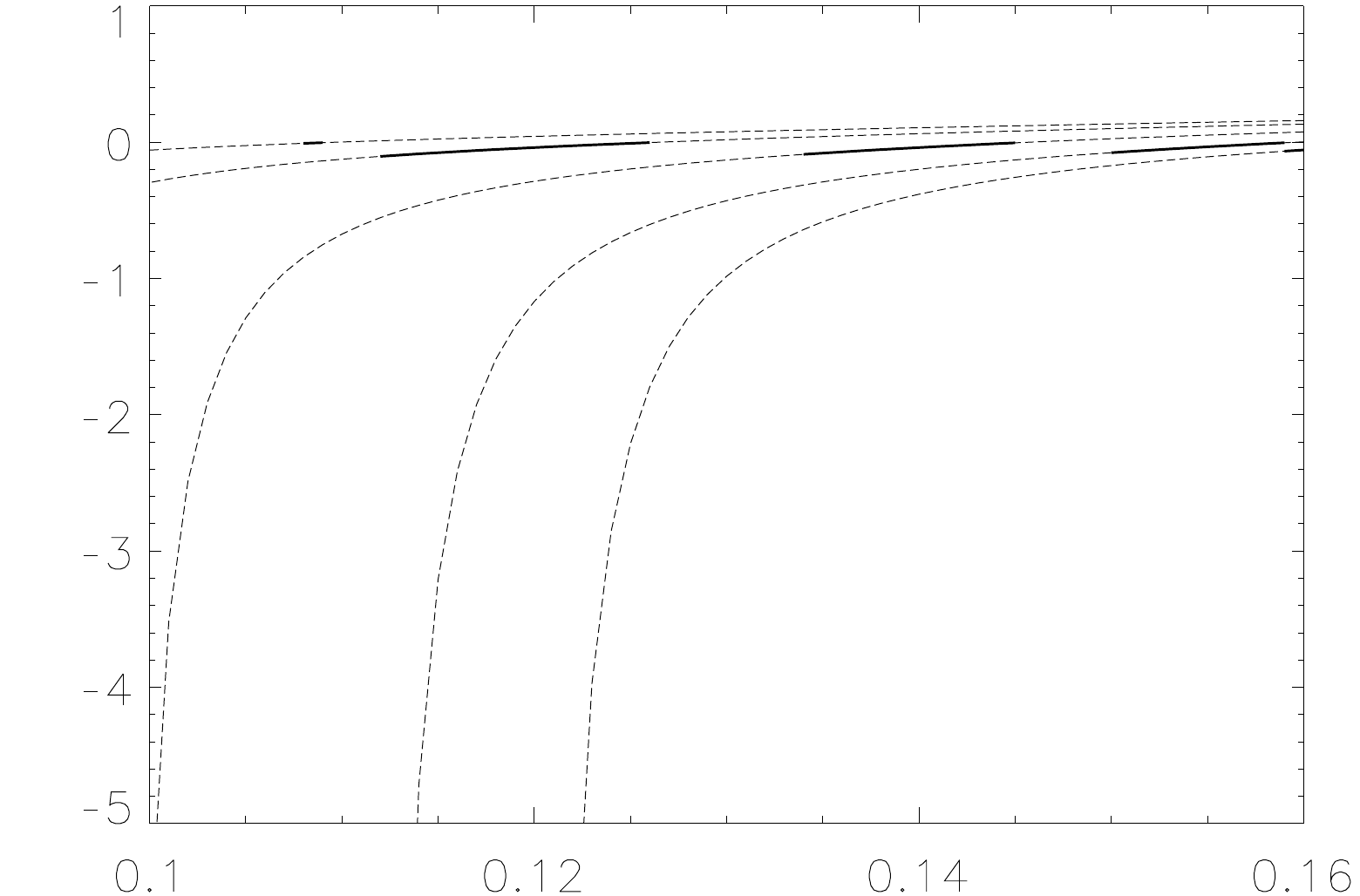}}
\put(53,8){$\eta$}
\end{picture}
\caption{Minimum eddy diffusivity $\eta_{\rm eddy}=\eta-\D^1_{23}$ (vertical
axis) in mTG~\rf{TG} for $a=b=1$. Bold solid lines:
the segments of plots for molecular diffusivities $\eta$, for which
a large-scale, but no small-scale magnetic field is generated. Lower to upper
curves: $c=0.25,\,0.3,\,0.35,\,5/13,\,0.4$\,.}
\label{eddy5}

~

\begin{picture}(100,70)(0,0)
\put(0,0){\includegraphics[width=\columnwidth]{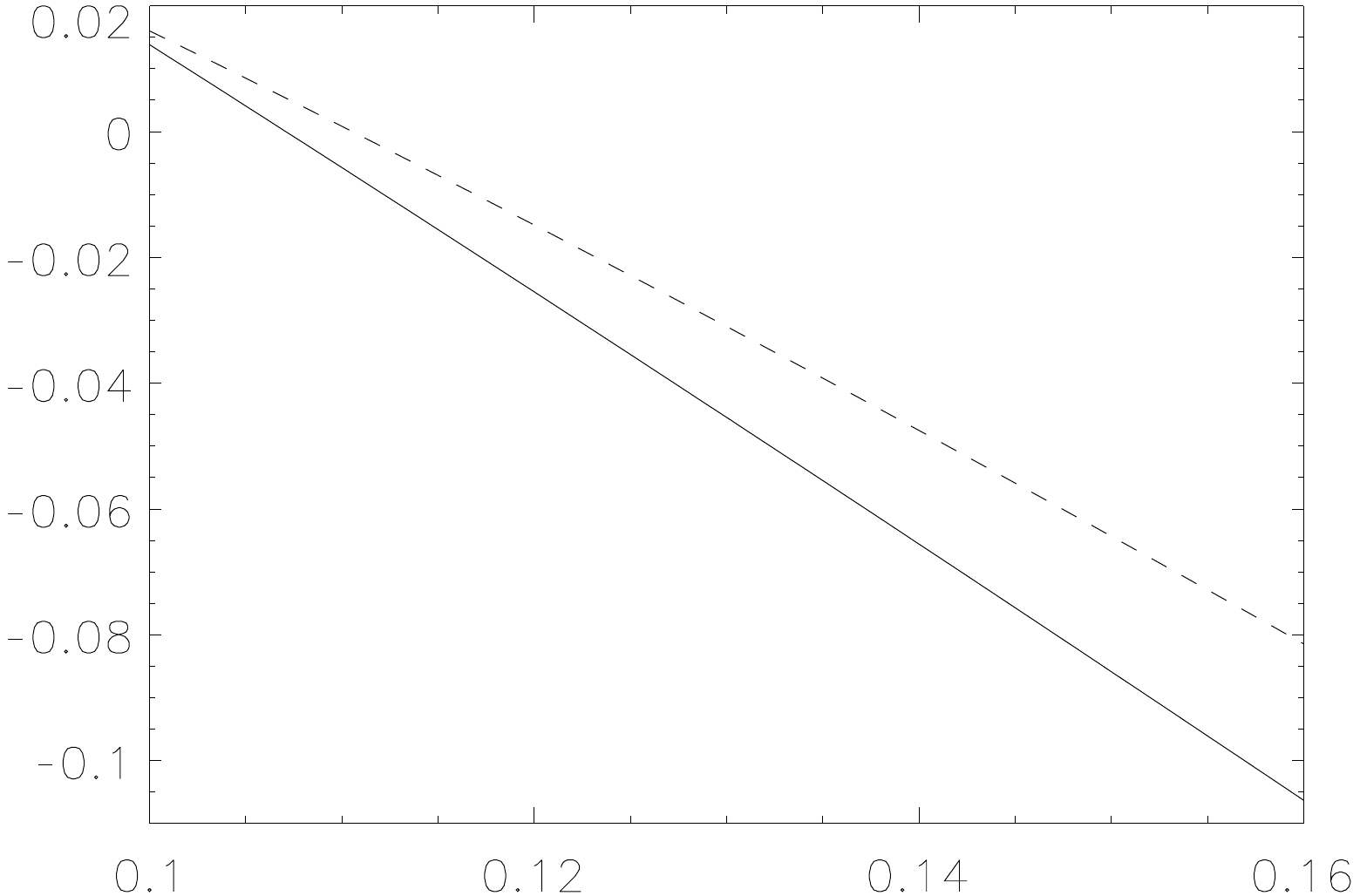}}
\put(53,8){$\eta$}
\end{picture}
\caption{Growth rates (vertical axis) of dominant small-scale magnetic
eigenmodes (solid line: the {\tt AAAEO} subspace; dashed line: the
{\tt SAAOE}/{\tt ASAOO} subspaces) generated by mTG~\rf{TG}, \rf{notte}.
}\label{short}\end{figure}

Using the same algorithms as employed for R-IV, we have computed
the eddy diffusivity tensor (see \Fig{eddy5}) for mTG
for $a=b=1$, as in \cite{La}, the coefficient $c$ and
the molecular diffusivity $\eta$ ranging in the intervals [0.25,\,0.4] step 0.05
and [0.1,\,0.16] step 0.001, respectively. Advective terms were computed
by pseudospectral methods with the resolution of $48^3$ Fourier harmonics.
Dealiasing was performed by keeping in the solution only harmonics
with wave numbers not exceeding 21. Energy spectra decaying by 7--10 orders
of magnitude, this resolution is sufficient.
As in the case of R-IV,
iterations were terminated, when an estimate
of the dominant eigenvalue was below $10^{-10}$ in absolute value and the norm
of the discrepancy for the normalised associated eigenvector was below
$5\cdot10^{-11}$. Computation of one eddy diffusivity correction tensor $\Db$
requires 10--20 minutes of a 3.9~MHz Intel Core i7 processor
(the code is sequential). We have also carried out computations for the parameter
values
\BE a=b=1,\quad c=5/13,\quad d=2/13\EE{notte}
used by \cite{La}. Although our algorithms and codes are independent
from those applied by \cite{La}, our values of $\eta_{\rm eddy}$ coincide
with those found {\it ibid.} in four significant digits.

\begin{table}[b!]
\caption{Critical molecular diffusivities $\eta$ for the onset of generation
of small-scale magnetic fields in three invariant symmetry subspaces.}
\center\begin{tabular}{|c|c|c|c|}\hline
$c$ &{\tt AAAEO}&{\tt SAAOE}/{\tt ASAOO}&{\tt SSAEE}\\\hline
0.25&0.1143&0.1580&0.1203\\
0.3 &0.1091&0.1491&0.1118\\
0.35&0.1065&0.1333&0.0985\\
5/13&0.1071&0.1105&0.0849\\
0.4 &0.1080&0.0851&0.0770\\\hline
\end{tabular}\label{tabl}\end{table}

In all runs shown in \Fig{eddy5}, we have found that $\D^1_{23}>0$ and
$\D^2_{31}<0$, $\D^3_{12}<0$. Thus, negative eddy diffusivity gives
rise to growing large-scale magnetic modes with horizontal wave vectors
of the large-scale harmonic modulation. Physically the most interesting case
occurs when generation of large-scale fields is not obstructed by generation
of small-scale fields. The segments of the plots of the minimum
eddy diffusivities corresponding to this case are shown by bold solid lines. Each
segment is bounded on the left by the critical point for the onset of generation
of small-scale magnetic field, and on the right by the point where
eddy diffusivity becomes positive. The critical values of molecular diffusivity
for the onset of generation of small-scale magnetic fields in some invariant
subspaces (see the previous section) are shown in Table~\ref{tabl}.
The dominant magnetic eigenmodes have been computed applying the algorithms
of \cite{Zh93} with a resolution of $64^3$ harmonics,
the dealiasing was performed by keeping harmonics with wave numbers
not exceeding~29. Energy spectra of the obtained eigenmodes decay by at least
11 orders of magnitude. For the considered $\eta$, the dominant eigenmodes
belong to the {\tt AAAEO} subspace (see \Fig{short};
we did not aim at computing the dominant magnetic eigenmodes in all symmetry
subspaces). The dominant eigenmodes in the {\tt AAAEO} and {\tt SSAEE} subspaces
turn out to be $\gamma$-sym\-metric. The plots of $\eta_{\rm eddy}$
have vertical asymptotes located at the critical values for the onset
of generation of the small-scale magnetic field in the {\tt SSAEE} subspace
(see \citealt{Zh} for explanations).

\subsection{Numerical results: finite scale separation}\label{TGnum}

We now consider the case of a finite (i.e., non-infinitesimal) scale
separation~$\varepsilon$. By comparing numerical solutions with
the multiscale predictions, we can roughly estimate the range of the scale
ratios $\varepsilon$, for which the asymptotic formalism qualitatively correctly
describes the large-scale dynamo driven by an array of mTG flow cells.
As established in the previous section, for a high
scale separation (i.e., in the limit of small~$\varepsilon$), a large-scale
magnetic mode generated by mTG grows the fastest, when
the unit wave vector $\bf q$ is horizontal and $\widetilde{\bf B}={\bf e}_3$
in the large-scale modulation~\rf{Fouha}. Such a mode is asymptotically close to
\be{\bf b}={\rm Re}\Big(\e^{\i\varepsilon\bf q\cdot x}
\Big(&{\bf S}_3({\bf x})+{\bf e}_3\label{eimo}\\
&+\i\varepsilon\sum_{m=1}^2q_m{\bf G}_{m3}({\bf x})
+{\rm O}(\varepsilon^2)\Big)\!\Big).\nonumber\end{align}

To study directly magnetic field generation for an arbitrary finite scale
separation $\varepsilon$, we can employ the procedure used by \cite{zpf}.
Namely, we consider the problem~\rf{eig} for a field of the form
\BE{\bf b(x)}=\e^{\i\varepsilon\bf q\cdot x}\,{\bf b'(x)},\EE{lsmm}
where $\bf q$ is
a constant unit wave vector. A small-scale (i.e., having the spatial periodicity
cell $\T^3$) vector field $\bf b'(x)$ satisfies the eigenvalue
equation\footnote{We have preserved the factor $|{\bf q}|^2$ in \rf{eee}
for this equation
to remain valid for any vector $\bf q$, and not just for a unit one.}
\be\L_{\varepsilon\bf q}{\bf b'}\equiv\,&\eta(\nabla_{\bf x}^2{\bf b'}
-\varepsilon^2|{\bf q}|^2{\bf b'})+\rot({\bf v}\times{\bf b'})\quad\label{eee}\\
&+\,\i\varepsilon({\bf q}\times({\bf v}\times{\bf b'})+2\eta\,
({\bf q}\cdot\Nabla_{\bf x}){\bf b'})=\lambda{\bf b'}\nonumber
\end{align}
and the corollary of the solenoidality condition
\BE\Nabla_{\bf x}\cdot{\bf b'}+\i\varepsilon{\bf q}\cdot{\bf b'}=0.\EE{eed}
This approach is advantageous in that it does not require performing
the asymptotic analysis of Section~\ref{math} and is applicable for all
scale ratios~$\varepsilon$, and not only very small ones. However,
it is less general in that, on the one hand, a solution to the eigenvalue
problem~\rf{eee}--\rf{eed} provides information for only one instance
of the amplitude modulation vector $\varepsilon\bf q$. On the other,
it is only applicable when tackling a linear stability problem such as
the kinematic dynamo problem studied here, but does not deliver a simplified
statement of a weakly nonlinear stability problem.

For $\varepsilon>0$, even and odd vector fields (that are linear combinations
of Fourier harmonics such that the sum of the wave numbers in the horizontal
directions is even or odd, respectively) constitute invariant subspaces
of $\L_{\varepsilon\bf q}$~\rf{eee}. If ${\bf q}={\bf e}_m$ for $i\ne m$, vector
fields, symmetric or antisymmetric in~$x_m$, also constitute invariant
subspaces. The case $i=m$ is more subtle: vector fields, whose real part is
symmetric or antisymmetric in $x_m$, and the imaginary part is, respectively,
antisymmetric or symmetric in $x_m$, constitute two invariant sets. However,
these sets are not linear subspaces (over the field of complex numbers);
in other words, this property can be used in computations, but it does not
restrict an eigenmode, since multiplying an eigenmode by the complex unity $\i$
does not give rise to a new eigenmode --- except for $\varepsilon=0$, when
only the symmetric or antisymmetric part of the eigenmode ``from which
the branch originates'' is non-zero. Consequently, for ${\bf q}={\bf e}_m$
we can use labels for branches of dominant eigenfields of $\L_{\varepsilon\bf q}$
that have the same meaning as the labels
of invariant subspaces of the domain of the small-scale magnetic induction
operator $\L$, except for the symmetry or antisymmetry in place $m$
of the label is determined only for the eigenmode for $\varepsilon=0$.
The symmetry $\gamma$, involving swapping of the horizontal Cartesian
coordinates as well as swapping of vector field components, does not distinguish
invariant subspaces of $\L_{\varepsilon\bf q}$ for $\varepsilon>0$. It maps
eigenfunctions of $\L_{\varepsilon\bf q}$ to eigenfunctions
of $\L_{\varepsilon\bf q'}$ for ${\bf q}'=(q_2,q_1,q_3)$.

\begin{figure}[t!]
\begin{picture}(100,70)(0,0)
\put(0,0){\includegraphics[width=\columnwidth]{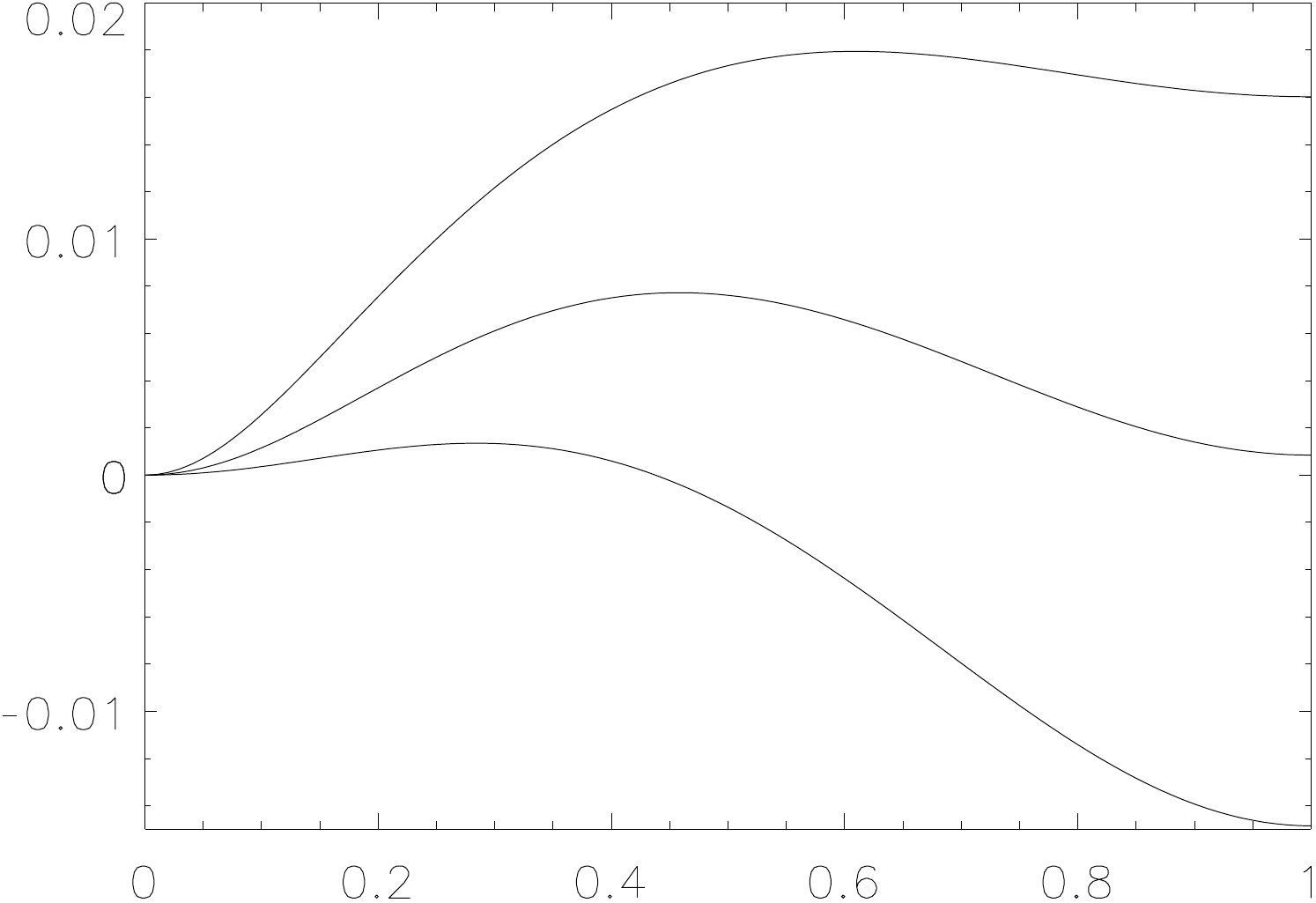}}
\put(83.5,58){$\eta=0.1$}
\put(83.5,39){$\eta=0.11$}
\put(83.5,12){$\eta=0.12$}
\put(54.5,8){$\varepsilon$}
\end{picture}
\caption{Growth rates (vertical axis) of dominant large-scale magnetic modes
({\tt SSAEE} subspace) generated by mTG~\rf{TG}, \rf{notte} for ${\bf q}={\bf e}_1$.
}\label{epsdep}

~

\begin{picture}(100,70)(0,0)
\put(0,0){\includegraphics[width=\columnwidth]{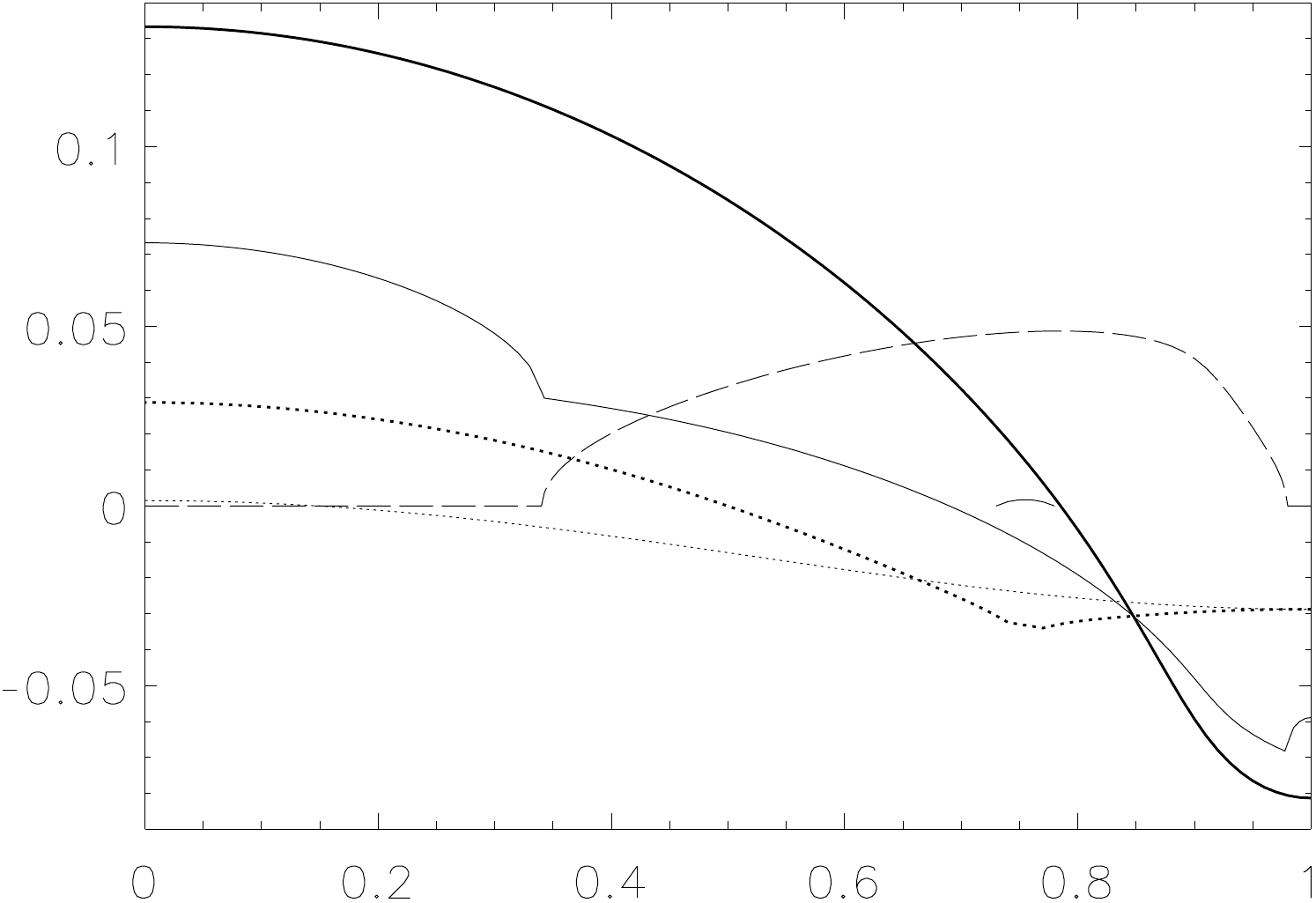}}
\put(54.5,8){$\varepsilon$}
\end{picture}
\caption{Growth rates (vertical axis) of dominant large-scale magnetic
eigenmodes generated by mTG~\rf{TG}, \rf{notte} for $\eta=0.02$,
${\bf q}={\bf e}_1$: four branches in symmetry subspaces
{\tt AAAEO} (bold line), {\tt AAAEE} (thin dotted line), {\tt SSAEO} (bold
dotted line and thin solid line for $0.73<\varepsilon<0.78$\,: the real and imaginary
parts of the associated eigenvalues; outside this interval,
the eigenvalue is real), and {\tt SSAEE} (thin solid and dashed lines:
the real and imaginary parts of the associated eigenvalues).
}\label{re50}\end{figure}

\begin{figure}[t!]
\begin{picture}(100,65)(0,0)
\put(0,0){\includegraphics[width=\columnwidth]{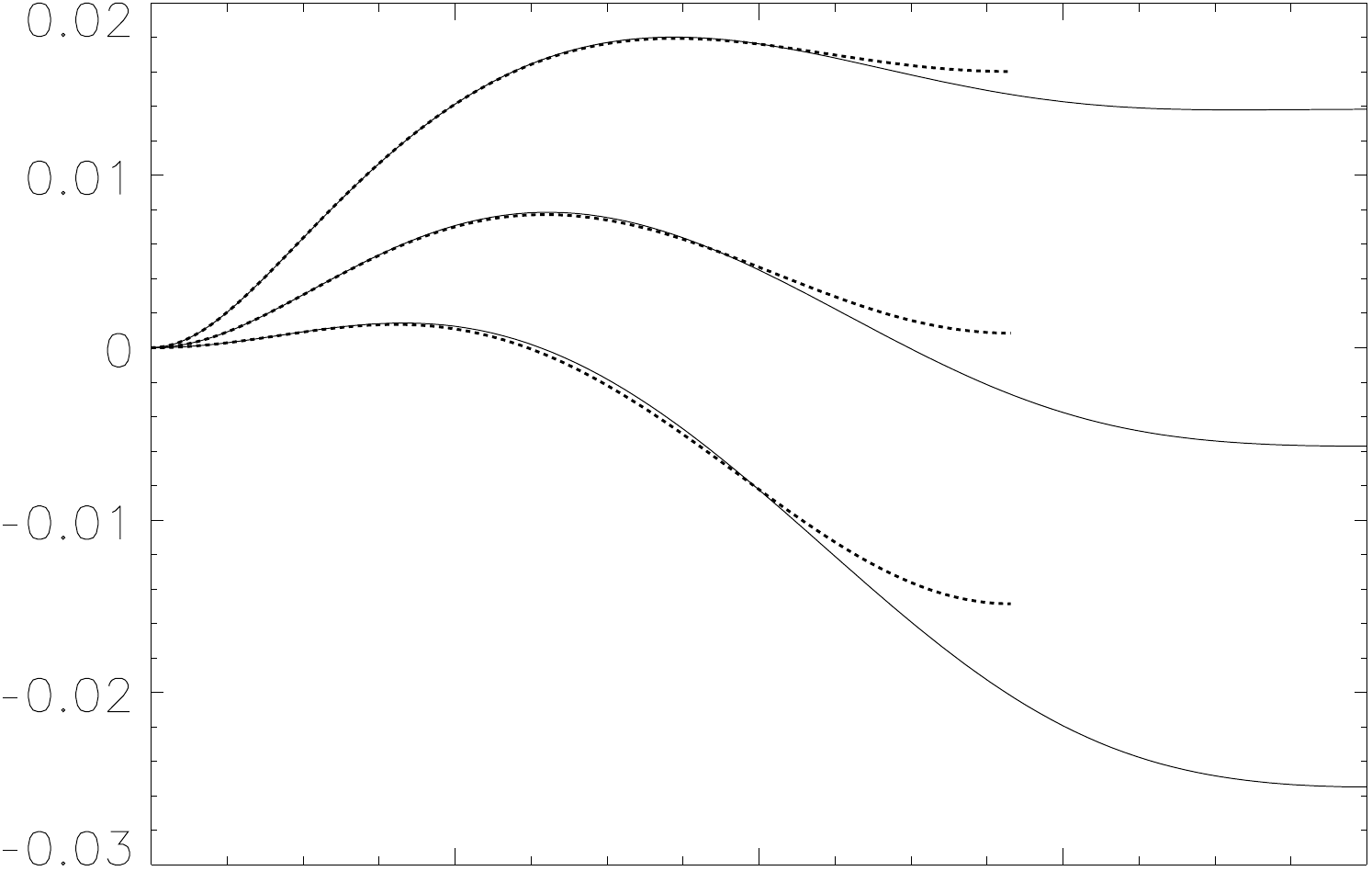}}
\put(83.5,57.5){$\eta=0.1$}
\put(83.5,34){$\eta=0.11$}
\put(83.5,10){$\eta=0.12$}
\put(54.5,3){$\varepsilon$}
\end{picture}

\vspace*{2mm}
$\displaystyle\hspace*{9mm}0
\hspace*{14mm}{\sqrt{2}\over4}
\hspace*{13mm}{\sqrt{2}\over2}
\hspace*{12mm}{3\sqrt{2}\over4}
\hspace*{9.5mm}\sqrt{2}$

\caption{Growth rates (vertical axis) of dominant large-scale magnetic modes
generated by mTG~\rf{TG}, \rf{notte} (solid lines) for
${\bf q}=(1,1,0)/\sqrt{2}$. For comparison, growth rates of dominant large-scale
magnetic modes for ${\bf q}={\bf e}_1$ and the same molecular
diffusivities~$\eta$ are shown (dotted lines; cf.\ \Fig{epsdep}).
}\label{epscomp}

~

\begin{picture}(100,65)(0,0)
\put(0,0){\includegraphics[width=\columnwidth]{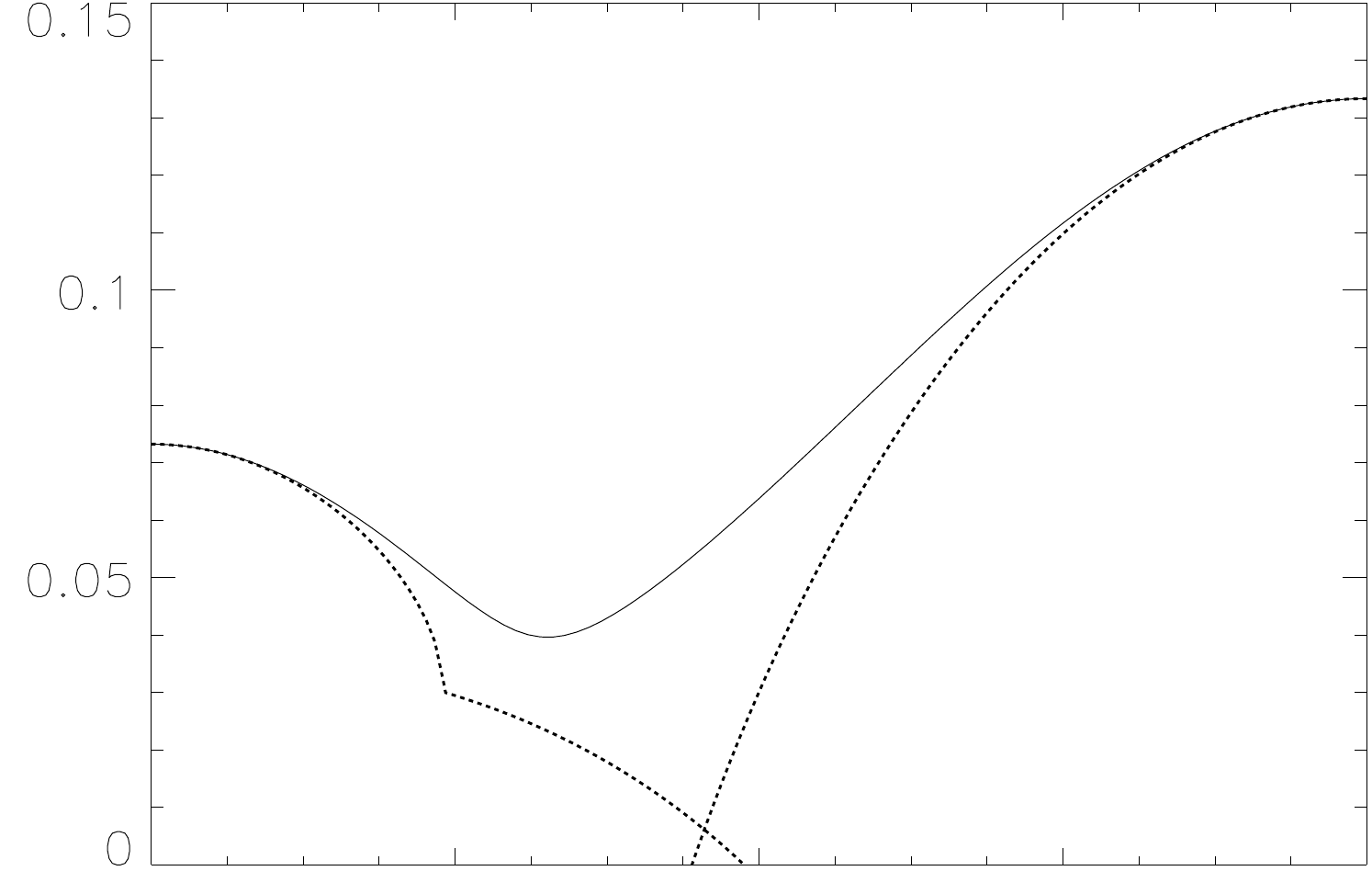}}
\put(54.5,3){$\varepsilon$}
\end{picture}

\vspace*{2mm}
$\displaystyle\hspace*{9mm}0
\hspace*{14mm}{\sqrt{2}\over4}
\hspace*{13mm}{\sqrt{2}\over2}
\hspace*{12mm}{3\sqrt{2}\over4}
\hspace*{9.5mm}\sqrt{2}$

\caption{Growth rates (vertical axis; solid line) of dominant large-scale
magnetic eigenmodes generated by mTG for
$\eta=0.02$, ${\bf q}=(1,1,0)/\sqrt{2}$.
For comparison, growth rates in the branches of dominant large-scale magnetic
modes for ${\bf q}={\bf e}_1$ in the symmetry subspaces {\tt AAAEO} (right)
and {\tt SSAEE} (left) for the same $\eta$
are shown (dotted lines; cf.\ \Fig{re50}).
}\label{epsre50}\end{figure}

We have computed the dominant eigenvalues (i.e., the ones having the maximum
real part among all eigenvalues for the given parameter values) of the magnetic
induction operator and the associated large-scale magnetic modes generated
by mTG~\rf{TG}, \rf{notte} --- the flow employed
by \cite{La} --- for the wave vectors of the large-scale
amplitude modulation ${\bf q}=(1,0,0)$ (see \Figs{epsdep}{re50})
and ${\bf q}=(1,1,0)/\sqrt{2}$ (\Figs{epscomp}{epsre50}).
Since the flow possesses the symmetries in $x_1$ and $x_2$ and
the $\gamma$-symmetry, actually the computations cover all possible choices
of~$\bf q$ from the following list:
$\pm{\bf e}_1,\,\pm{\bf e}_2,\,(\pm1,\pm1,0)/\sqrt{2}$.

Plots of growth rates of large-scale magnetic modes for ${\bf q}={\bf e}_1$
and a varying scale ratio $\varepsilon$ are shown in \Fig{epsdep}
for $\eta=0.1$ used by \cite{La}, as well as for $\eta=0.11$ and 0.12\,.
For these molecular diffusivities the dominant eigenvalues of the operator
$\L_{\varepsilon\bf q}$ are real. \cite{zpf} noticed that a graph of the dominant
growth rates is periodic in $\varepsilon$ with period 1 (because any
large-scale field $\e^{\i\varepsilon\bf q\cdot x}\,{\bf b'(x)}$, where
$\bf b'(x)$ is a small-scale field, can be also expressed as
$\e^{\i(\varepsilon-p)\bf q\cdot x}\,\left(\e^{\i p\bf q\cdot x}\,{\bf b'(x)}\right)$,
and for an arbitrary integer $p$ the field $\e^{\i p\bf q\cdot x}\,{\bf b'(x)}$
is also small-scale). Also, a graph of the dominant magnetic mode
growth rate as a function of the scale ratio $\varepsilon$ is symmetric about
the vertical axis: applying complex conjugation to equations~\rf{eee}
and~\rf{eed} shows that if, for a given scale ratio $\varepsilon$, $\bf b'(x)$
is a small-scale eigenfunction associated with an eigenvalue $\lambda$, then
$\overline{\bf b'(x)}$ and $\overline\lambda$ are, respectively, a small-scale
eigenfunction and the associated eigenvalue for the opposite ratio
$-\varepsilon$. Consequently, graphs of the dominant growth rate are symmetric
about each vertical line $\varepsilon=q/2$ for integer~$q$. By contrast,
the plots in \Figs{epsdep}{re50} show eigenvalues associated with branches
of eigenfunctions of the problem~\rf{eee}--\rf{eed}, smoothly parameterised
by $\varepsilon$. They have a period 2 in $\varepsilon$ and are
symmetric about vertical lines $\varepsilon=q$ for all integer~$q$. The
parabolic shape of the plots near $\varepsilon=0$ agrees with
expansion~\rf{bexp} for $\lambda_0=\lambda_1=0$. That $\varepsilon=0$ is a
local minimum of the plots in \Fig{epsdep} corroborates that
magnetic eddy diffusivity is negative for the molecular diffusivities
$\eta=0.1,\,0.11\,.0.12$, for which plots are presented in this figure; the
respective eigenmodes $\bf b'(x)$ constitute {\tt SSAEE} branches.

Near the origin, the plots of growth rates in \Figs{epsdep}{epscomp}
have a parabolic shape (which is a signature of magnetic
eddy diffusivity) for $\varepsilon$ below 0.1; this roughly estimates the range
of scale ratios for which the asymptotic formalism describes qualitatively
correctly the large-scale dynamo driven by an array of mTG flow cells.
A similar parabolic-shape correction of growth rates due to the action of eddy
diffusivity is observed for non-neutral magnetic modes (\Figs{re50}{epsre50})
in all the symmetry subspaces considered.

Our computations demonstrate that for $\eta=0.1$, mTG
can generate large-scale magnetic field by the mechanism
of negative eddy diffusivity in a range of parameter values.
By contrast, for $\eta=0.02$ no large-scale magnetic field generation was found
by \cite{BrMi} in DNS. We have computed four branches
of dominant eigenmodes for $\eta=0.02$ and ${\bf q}={\bf e}_1$ (see
\Fig{re50}), that belong to invariant subspaces {\tt AAAEO}, {\tt AAAEE},
{\tt SSAEO} and {\tt SSAEE} with the resolution of $96^3$ harmonics (upon
dealiasing, harmonics with wave numbers up to 45 are kept); energy spectra
of the eigenmodes decay by at least 9 orders of magnitude.

We observe two major differences with the case $\eta=0.1$\,. First,
a small-scale dynamo persists for $\eta=0.02$\,. Implementation of the TFM
procedure requires integrating equation~\rf{bprim}; the solution
converges to the dominant small-scale mode, amplitude-modulated by the
large-scale harmonic $\e^{\i\varepsilon x_m}$. Clearly, in the presence
of a small-scale dynamo, the solution is dominated by the growing small-scale
mode, and not by the neutral mode \rf{eimo}. Solutions can be expanded in the
series~\rf{bexp} in the scale ratio $\varepsilon$, the series for the
eigenvalue $\lambda$ now beginning with the respective small-scale dynamo
eigenvalue. For a parity-invariant flow this modifies the molecular diffusivity
operator, acting on the amplitude-modulating factor (called amplitude) in the
respective large-scale mode; like in the absence of a small-scale dynamo, the
correction is due to interaction of the fluctuating part of the magnetic field
and the small-scale flow, and thus again eddy diffusivity is the leading-order
eddy effect. Second, the point $\varepsilon=0$ is now a local maximum, implying
that eddy diffusivity is now positive. However, the growth rates of large-scale
magnetic modes are still positive when $|\varepsilon|$ is small, i.e.,
these modes do grow, albeit slower than the small-scale modes
for $\varepsilon=0$. In other words, the growing large-scale modes decay
relative to the faster growing small-scale mode, which explains the statement
``a dynamo is observed but it is not a large-scale dynamo'' \citep{BrMi}.

Yet another difference with the case $\eta=0.1$ is visible in the behaviour
of dominant eigenmodes constituting the {\tt SSAEE} branch. For $\eta=0.02$,
they experience two bifurcations: on increasing the scale ratio $\varepsilon$,
a pair of real eigenvalues (including the dominant one) turns into a pair
of complex-conjugate ones at $\varepsilon\approx0.34$, that are superseded again
by two real eigenvalues at $\varepsilon\approx0.98$ (only the largest of which
is shown in \Fig{re50}). We observe a characteristic feature of dependence
on the parameter near a point of such a bifurcation: the plots of real
eigenvalues and of the imaginary part of complex eigenvalues (but not of
the real part of the complex eigenvalues) have singularities of the kind of
$\sqrt x$ for $x\ge0$ near zero --- the growth rate depends on $\varepsilon$
continuously, but its derivative is infinite. This stems from the fact that
the quadratic characteristic polynomial of $\L_{\varepsilon\bf q}$, reduced onto
the invariant plane of the associated eigenfunctions, has coefficients that are
differentiable in $\varepsilon$, and hence the discriminant is approximately
a linear function of $\varepsilon$ near the point of bifurcation. Vanishing
of the discriminant at such points gives rise to the singularities
mentioned above (the real parts of complex eigenvalues are
not affected, since they are just proportional to the coefficient
of the linear term of the characteristic polynomial).
\cite{CGPZ} observed a similar behaviour in the dependence of magnetic field
generation by thermal convection on the rotation rate (see Fig.~18 {\it ibid}.).

We have also computed the short-scale parts $\bf b'(x)$ of the dominant
large-scale magnetic modes \rf{lsmm}, generated by the same instance
of mTG~\rf{TG}, \rf{notte} for the wave vector
${\bf q}=(1,1,0)/\sqrt{2}$ and molecular diffusivities
$\eta=0.1,\,0.11,\,0.12$ (using the resolution of $64^3$ Fourier harmonics)
and 0.02 ($96^3$ harmonics). As for ${\bf q}={\bf e}_1$, this has been done
by solving the eigenvalue problem \rf{eee}--\rf{eed} for the modified operator
of magnetic induction $\L_{\varepsilon\bf q}$. For all considered $\eta$ and
$\varepsilon$, the computed dominant short-scale modes of $\L_{\varepsilon\bf q}$
possess now the $\gamma$-symmetry, the antisymmetry in $x_3$ and the symmetry about
the $x_3$-axis, which is the composition of the symmetries in $x_1$ and $x_2$:
\begin{align*}f^i(-x_1,-x_2,x_3)&=-f^i({\bf x}),\quad i=1,2,\\
f^3(-x_1,-x_2,x_3)&=f^3({\bf x}).
\end{align*}
These short-scale modes are comprised of the Fourier harmonics, for which all
the three wave numbers $k_i$ in the directions $x_i$ have the same parity.
The associated eigenvalues of the operator $\L_{\varepsilon\bf q}$ are real.

For mTG, eddy diffusivity is the same for all horizontal
wave vectors (see \rf{TGe1}). Comparison of the eigenvalues computed for
${\bf q}=(1,1,0)/\sqrt{2}$ and ${\bf q}={\bf e}_1$ in \Figs{epscomp}{epsre50}
illustrates how this axisymmetry is reflected in
the eigenvalues for $\varepsilon>0$. We observe that the dependence
of the dominant eigenvalues on the direction of a horizontal wave vector
is very weak when $\varepsilon$ is as large as roughly 0.8 for $\eta=0.1$,
when $\varepsilon\le0.7$ for $\eta=0.11$ and 0.12, and only when
$\varepsilon\le0.22$ for $\eta=0.02$\,.

\section{TFM vs MST: analytic and numerical\\comparison}\label{work}

We have seen in Section \ref{flowIV} that the TFM used by \cite{BrMi}
for evaluating magnetic eddy diffusivity for R-IV yielded
the results compatible with those obtained by employing the homogenisation
techniques within the MST approach. Given that distinct types of averaging
are employed in MST and TFM, this conformity of results may seem unexpected.
In the present section we compare the two approaches.

TFM starts by computing a zero-mean
solution $\bf b'$ to equation \rf{bprim} for the test field
\BE{\bf b}_{\rm test}=\e^{\i\varepsilon x_m}{\bf e}_n\EE{tf}
(this is equivalent to employing the two real fields \rf{ritf} for
$\kk=\varepsilon{\bf e}_m$, but simplifies the algebra).
Any solenoidal small-scale zero-mean field (for instance,~0) can serve
as an initial condition for $\bf b'$. The solution will then automatically
be solenoidal at any time $t>0$. TFM assumes that \rf{bprim} does not have
growing solutions for the test fields and averaging applied. Numerical
integration of \rf{bprim} proceeds till transients decay and the solution
$\bf b'$ saturates. The eddy diffusivity correction
tensor is then deduced as the matrix
that relates the obtained mean e.f.m.~$\overline{{\bf v'}\times{\bf b'}}$
with the test fields \rf{tf}.

\subsection{TFM with volume averaging}\label{tfmv}

We now consider a variant of TFM, in which volume averaging is involved
in extracting the fluctuating part of the auxiliary fields, $\bf b'$,
and show that then the TFM values of eddy quantities
converge in the limit of large scale separation to the values yielded by MST.
The demonstration, given here for steady flows $\bf v'$, can be readily
extended to encompass time-periodic flows.

Our solutions can be obtained as the real and imaginary parts
of the fluctuating part of the field
\BE{\bf b}={\bf b}_{\rm test}+{\bf b'}=\e^{\i\varepsilon x_m}
\widetilde{\bf b},\qquad\LA\widetilde{\bf b}\ra={\bf e}_n.\EE{expsub}
Note that when extracting the fluctuating part~$\bf b'$,
we average $\bf b$ {\it after} pulling out the factor $\e^{\i\varepsilon x_m}$,
since averaging over $x_m$ any field of the form
$\e^{\i\varepsilon x_m}{\bf f}({\bf x})$,
where $\bf f$ is independent of $\varepsilon$, yields just 0.
The evolution equation \rf{bprim} for the auxiliary field $\bf b'$
is then equivalent to the equation obtained by substituting~\rf{expsub}
into~\rf{bprim} and cancelling out the exponential:
\BE{\partial\widetilde{\bf b}\over\partial t}=
(\L_{\varepsilon{\bf e}_m}\widetilde{\bf b})',\EE{Beq}
(the operator $\L_{\varepsilon{\bf e}_m}$ is defined by \rf{eee});
$\widetilde{\bf b}$ also satisfies the condition \rf{eed}, stemming
from solenoidality of $\bf b'$, and has a constant average
$\LA\widetilde{\bf b}\ra={\bf e}_n$. For small $\varepsilon>0$, the elliptic operator
$\L'_{\varepsilon{\bf e}_m}$ in the r.h.s.~of \rf{Beq} is an O($\varepsilon$)
perturbation of the operator of magnetic induction, $\L$. Consequently, this
stage of TFM can be readily understood in the framework of MST.
By the general theory of perturbation of linear operators (\citealt{kato};
see also \citealt{vi}), an eigenfunction of $\L$ and the associated eigenvalue
involved in a Jordan cell of size $M$ are altered by O($\varepsilon^{1/M}$).

TFM is applicable when no small-scale dynamo operates. In this section we
assume that the kernel of the operator of magnetic induction, defined
in the box of periodicity of the flow, is three-dimensional (for a given
$\bf v$, this holds for all $\eta>0$ except only for a countable number of
$\eta$ values).
A solution to~\rf{Beq} is a sum of a transient ${\bf b}^{\rm tr}$,
whose rate of exponential decay is O(1), and the neutral mode of the perturbed
operator, that branches from the respective neutral mode of $\L$
(for which $M=1$):
\BE\widetilde{\bf b}={\bf S}_n({\bf x})+{\bf e}_n
+{\rm O}(\varepsilon)+{\bf b}^{\rm tr}\EE{Btra}
for any permissible initial conditions for $\widetilde{\bf b}$.

{\it1. Magnetic $\alpha$-effect.}
For a generic steady flow~$\bf v$, we can now calculate the TFM estimate
of the $\alpha$-tensor using the ansatz~\rf{xemf}. By~\rf{expsub} and~\rf{Btra},
after the transient decays below O$(\varepsilon)$ at times O(ln$\varepsilon)$,
\BE{\bf b}=\e^{\i\varepsilon x_m}({\bf S}_n({\bf x})+{\bf e}_n
+{\rm O}(\varepsilon)\,).\EE{Xtra}
Large-scale computations of $\bf b'$ are usually done for a rational
$\varepsilon=i_1/i_2$ (with common factors cancelled out in integers $i_1$
and $i_2$) such that the periodicity
of $\e^{\i\varepsilon x_m}$ is compatible with that of the small-scale flow
$\bf v$. Thus, we can assume that the computational domain has the size
$2\pi i_2$ in $x_m$.
When applied to a steady field, the Fourier transform \rf{Fou} involved
in \rf{xemf} differs only by a constant factor from the inverse Fourier
transform $\F_{\kk}=(1/V){\cal F}_{\kk,0}$ that recovers coefficients
in expansion of a function in the spatial variables:
$$\F_{\kk}\left(\sum_{\bf m}\hat{f}_{\bf m}\,\e^{\i\bf m\cdot x}\right)=\hat{f}_{\kk}.$$
Here $V$ denotes the volume of the spatial periodicity domain.
Using~\rf{Xtra}, we find
\be\F_{\varepsilon{\bf e}_m}(\overline{{\bf v'}\times{\bf b'}})
=&\int_0^{2\pi i_2}\!\!\e^{-\i\varepsilon x_m}\!\int_0^{2\pi}\!\!\int_0^{2\pi}
\!\!\!\!{\bf v'}\times{\bf b'}\ {\d{\bf x}\over(2\pi)^3i_2}\nonumber\\
=&\LA{\bf v'}\times{\bf S}_n\ra+{\rm O}(\varepsilon)
\label{essal}\end{align}
(here any spatial averaging $\overline{\phantom{m}}$ is acceptable, provided
it does not involve averaging in $x_m$, or otherwise special precautions are
taken as discussed above).
Since $\F_{\varepsilon{\bf e}_m}({\bf b}_{\rm test})={\bf e}_n$,
by~\rf{xemf} the $n$-th column of the $3\times3$ matrix $\bm\alpha$
coincides in the limit $\varepsilon\to0$ with the $n$-th column of $\A$.

{\it Remark 1}.
TFM for evaluation of the $\alpha$-effect tensor in non-parity-invariant flow
in the original formulation \citep{Sch05,Sch07} prescribed the use
of constant test fields ${\bf b}_{\rm test}={\bf e}_n$, which coincides with
\rf{tf} for $\varepsilon=0$. Consequently, $\L'_{\varepsilon{\bf e}_m}=\L$
in \rf{Beq}, and thus this version of TFM with the spatial averaging
reproduces the MST $\alpha$-effect tensor precisely.

{\it2. Magnetic eddy diffusivity.}
If the flow is parity-invariant, i.e., $\bf v(-x)=-v(x)$, the three
small-scale eigenfunctions from the kernel of $\L$ are parity-antiinvariant:
${\bf S}_n(-x)={\bf S}_n(x)$. The parity-invariant part of $\widetilde{\bf b}$,
even if zero initially, is subsequently produced from the predominantly
parity-antiinvariant field~\rf{Btra} by the term
$\i\vphantom{\widetilde{\bf b^|}}\varepsilon{\bf e}_m\times({\bf v}\times\widetilde{\bf b})$
in~\rf{Beq}. Since all parity-invariant eigenmodes of $\L$ decay (by the
original assumption on the spectrum of $\L$),
the parity-invariant part of $\widetilde{\bf b}$ remains O($\varepsilon)$
at all large enough times. We can seek $\widetilde{\bf b}$ as a perturbed
truncated series for the neutral mode of $\L$,
known from MST:
$$\widetilde{\bf b}={\bf e}_n+{\bf S}_n
+\i\varepsilon{\bf G}_{mn}+{\bf b}^{\rm new}.$$
Substituting this ansatz into \rf{Beq}, we obtain an equation of the form
$${\partial{\bf b}^{\rm new}\over\partial t}=\L_{\varepsilon{\bf e}_m}{\bf b}^{\rm new}
+{\rm O}(\varepsilon^2).$$
We therefore find
\BE{\bf b}=\e^{\i\varepsilon x_m}({\bf e}_n+{\bf S}_n
+\i\varepsilon{\bf G}_{mn}+{\rm O}(\varepsilon^2)+{\bf b}^{\rm tr}),\EE{Bmk}
where ${\bf b}^{\rm tr}$ is a
transient, whose rate of exponential decay is O(1).

We now calculate the entries of the magnetic eddy diffusivity correction tensor
from the equation
\BE\F_{\varepsilon{\bf e}_m}(\overline{{\bf v'}\times{\bf b'}})=-\sum_{p,q}
{\bm\eta}_{pq}(\varepsilon)\,\F_{\varepsilon{\bf e}_m}\!\!
\left({\partial{\bf b}_{\rm test}\over\partial x_p}\right)_{\!\!\!q}\!.\EE{xemfev}
This is ansatz~\rf{xemf} for steady flow and zero $\alpha$-effect.
As in item~{\it1}, we assume
$\varepsilon=i_1/i_2$ is rational so that the periodicities of~$\bf b$ \rf{Bmk}
and the small-scale flow $\bf v$ are compatible, and the computational
domain has the size $2\pi i_2$ in $x_m$. On the one hand, we then find
\be\F_{\varepsilon{\bf e}_m}(\overline{{\bf v'}\times{\bf b'}})
&=\int_0^{2\pi i_2}\!\!\e^{-\i\varepsilon x_m}\!\!\int_0^{2\pi}\!\!\!
\int_0^{2\pi}{\bf v'}\times{\bf b'}\ {\d{\bf x}\over(2\pi)^3i_2}\nonumber\\
&=\LA{\bf v'}\times({\bf S}_n
+\i\varepsilon{\bf G}_{mn})\ra+{\rm O}(\varepsilon^2)\nonumber\\
&=\i\varepsilon\Db_{mn}+{\rm O}(\varepsilon^2).\label{esssp}\end{align}
On the other,
$$\F_{\varepsilon{\bf e}_m}\left({\partial{\bf b}_{\rm test}\over\partial x_p}\right)
=\i\varepsilon\delta^p_m{\bf e}_n.$$
By~\rf{xemfev}, $\Db_{mn}=-\lim_{\varepsilon\to0}{\bm\eta}_{mn}$
for all $m\ne n$, i.e., TFM does produce in the limit $\varepsilon\to0$
the respective entry of the tensor of magnetic eddy correction. Note, however,
that a sufficiently high spatial resolution is necessary for a satisfactory
discretisation of both the small-scale field ${\bf S}_n({\bf x})$ and at least
one period of the modulating harmonic $\e^{\i\varepsilon\bf q\cdot x}$.

Above, we have investigated the algorithms for evaluating $\Db_{nm}$
for $n\ne m$. Test fields~\rf{tf} for $m=n$ are gradients and hence incompatible
with our analysis. To evaluate $\Db_{nn}$, we can use solutions to~\rf{bprim}
for test fields, that are real and imaginary parts of
$${\bf b}_{\rm test}=(\i\varepsilon)^{-1}\Nabla\times(\e^{\i\varepsilon
(jx_{n_1}+x_n)}{\bf e}_{n_2}),$$
where $n_1\ne n$, $n_2\ne n$ and $j$ is an integer.

\subsection{TFM with other spatial averagings}\label{tfms}

We now consider briefly the canonical variants of TFM, in which the averaging
denoted by a bar is performed over one or two
Cartesian variables under the same assumptions as in Section~\ref{tfmv}.
For simplicity, we again ignore the memory effect
by assuming that the test field does not depend on time.
As before, we cancel out in \rf{bprim} the exponent $\e^{\i\varepsilon x_m}$,
involved in the unknown field \rf{expsub}, and find
\BE{\partial\widetilde{\bf b}\over\partial t}=\P\L_{\varepsilon{\bf e}_m}\widetilde{\bf b}.\EE{notmieq}
Here $\P$ denotes a projection that deletes the mean field, but preserves
the volume average:
$$\P{\bf f}\equiv{\bf f}-\overline{\bf f}+\LA{\bf f}\ra.$$

While \rf{bprim} is equivalent to \rf{notmieq}, the latter equation has
advantages: ($i$) It can be numerically integrated in the flow periodicity cell
$\T^3$ without encountering the instabilities of problem \rf{notmieq}
which may exist at larger spatial scales (note that computations must be done
in a box of size $2\pi/\varepsilon$ in $x_m$ when the exponential or
sinusoidal dependence on $\varepsilon x_m$ is preserved in~$\bf b'$). Such
instability will then manifest itself by unbounded amplification of the growing
eigenfunctions of the operator $\P\L_{\varepsilon{\bf e}_m}$ that emerge from
round-off errors, and this will progressively wipe out the contribution
from the inhomogeneity in \rf{bprim} which we are looking for.
($ii$)~It enables us to compute auxiliary fields $\bf b'$ for irrational
$\varepsilon$ without suffering from problems due to
the presence of two incommensurate spatial frequencies in the solution.
($iii$) One can apply to solutions of \rf{notmieq} spatial averaging over
any variable, including $x_m$. In turbulence computations,
which are made in the large (from the prospective of the present discussion)
computational box, averaging a field after cancelling out the exponential
is also a feasible operation that is just equivalent to computing
the appropriate Fourier transform.

For any permissible initial conditions, \rf{notmieq} admits solutions similar
to \rf{Btra}:
$$\widetilde{\bf b}={\bf e}_n+{\bf s}_n+{\rm O}(\varepsilon)+{\bf b}^{\rm tr}$$
and, for parity-invariant flows, similar to \rf{Bmk}:
$$\widetilde{\bf b}={\bf e}_n+{\bf s}_n
+\i\varepsilon{\bf g}_{mn}+{\rm O}(\varepsilon^2)+{\bf b}^{\rm tr},$$
where ${\bf b}^{\rm tr}$ are transients, whose rate of exponential decay
is O(1). The fields ${\bf s}_n$ and ${\bf g}_{mn}$ have zero averages:
$\overline{\bf s}_n=\overline{\bf g}_{mn}=0$, the fields ${\bf e}_n+{\bf s}_n$
belong to the kernel of the operator $\P\L\P$. For parity-invariant flows,
${\bf s}_n$ are parity-antiinvariant and ${\bf g}_{mn}$ are parity-invariant.
But here the similarity ends, e.g., ${\bf s}_n\ne{\bf S}'_n$ and
${\bf g}_{mn}\ne{\bf G}'_{mn}$. Consequently, in the limit $\varepsilon\to0$
we can expect a qualitative but not quantitative agreement of MST results
with those of TFM with a non-volume averaging.

{\it Remark 2}. Plane-parallel flows independent of a Cartesian coordinate
$x_m$ are a special case, for which a solution \rf{expsub}, harmonically
modulated by the factor $\e^{\i\varepsilon x_m}$, involves the small-scale
part $\widetilde{\bf b}$ that is independent of $x_m$. Consequently, for such
flows, ${\bf s}_n={\bf S}_n$ and ${\bf g}_{jn}={\bf G}_{jn}$ for $j\ne m\ne n$,
and hence TFM recovers precisely the components $\Db_{jn}$ of the eddy
correction tensor. This is the case of R-IV.

\subsection{Kinematic generation by mTG: magnetic structures}

To understand the absence of negative magnetic eddy diffusivity in the TFM
results of \cite{BrMi}, we first inspect magnetic modes obtained from
numerical solutions of the underlying eigenvalue problem
for mTG~\rf{TG}, \rf{notte} and $\eta=0.1$\,.
The modes are eigenfunctions of the magnetic induction operator and give rise
to exponential in time solutions of the magnetic induction equation
\BE{\partial{\bf b}\over\partial t}=\L{\bf b}.\EE{main}
We consider first magnetic eigenmodes with the periodicity box of size
$(2\pi)^3$. As discussed in Section~\ref{symTG}, due to the symmetries
of the flow, magnetic modes have symmetries or antisymmetries in Cartesian
coordinates $x_i$ and in each mode the sums of wave numbers $k_1+k_2$
in all harmonics have
the same parity, as well as all sums $k_1+k_3$. These 5 symmetries are
independent and split the domain of the magnetic induction operator into 32
invariant subspaces. On top of this, magnetic modes can be symmetric or
antisymmetric with respect to swapping of the horizontal coordinates
$x_1\leftrightarrow x_2$ (the symmetry $\gamma$), but this symmetry is not
independent of the 5 former ones: it splits into invariant subspaces only 8
of the 32 aforementioned invariant subspaces --- namely those, in which the sums
of wave vectors $k_1+k_2$ are even, and vector fields are either symmetric
in both $x_1$ and $x_2$, or antisymmetric in both of these Cartesian variables.
Thus, the symmetries of mTG split the domain of the magnetic induction operator
into 40 invariant subspaces. We have computed dominant (i.e., having
the largest growth rates) magnetic modes in each of them.

\begin{table}[b!]
\caption{Maximum growth rates, $\lambda$, of $2\pi$- and $4\pi$-periodic
magnetic modes with different symmetries generated by mTG~\rf{TG}, \rf{notte}
for $\eta=0.1$\,.}
\center\begin{tabular}{|c|c|c|}\hline
\!Period\!&Symmetry subspace&$\lambda$\\\hline
$2\pi$&{\tt SAAOE},\,{\tt ASAOO}& 0.01602\\
$2\pi$&{\tt AAAEO}&0.01383\\\hline
$4\pi$&{\tt AAAOOE},\,{\tt ASAOOE},\,{\tt SAAOOE},\,{\tt SSAOOE}&0.01763\\
$4\pi$&{\tt ASAOEE},\,{\tt SAAEOE},\,{\tt SSAEOE},\,{\tt SSAOEE}&0.01734\\
$4\pi$&{\tt ASAEEE},\,{\tt SAAEEE}&0.01602\\
$4\pi$&{\tt AAAOEE},\,{\tt AAAEOE},\,{\tt ASAEOE},\,{\tt SAAOEE}&0.01404\\
$4\pi$&{\tt AAAEEE}& 0.01383\\
$4\pi$&{\tt AAAOOO},\,{\tt SAAOOO},\,{\tt ASAOOO},\,{\tt SSAOOO},\!&0.00226\\
&{\tt AASOOO},\,{\tt SASOOO},\,{\tt ASSOOO},\,{\tt SSSOOO}&\\\hline
\end{tabular}\label{gr}\end{table}

\begin{figure*}[p]
\includegraphics[width=\textwidth]{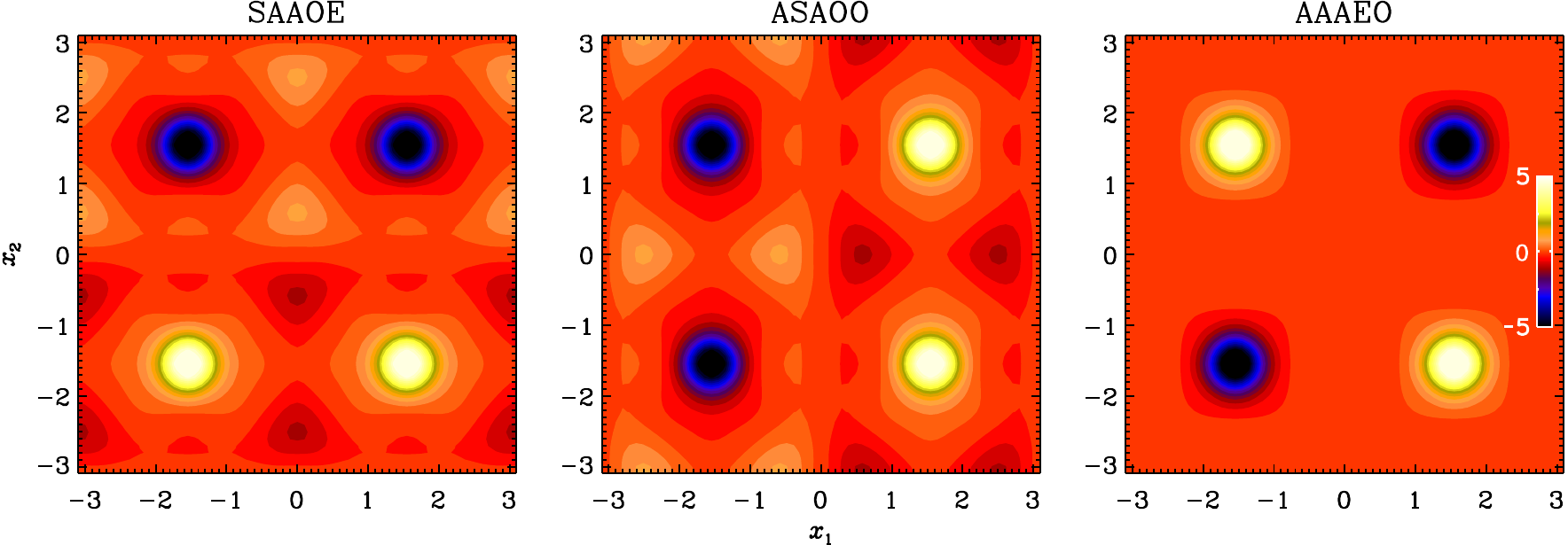}
\caption{The $x_3$-averaged normalised mean fields
$\overline{b}_3(x_1,x_2)/\langle\overline{b}_3^2\rangle^{1/2}$
for $\eta=0.1$ in the box periodicity of size~$(2\pi)^3$.
The same colour-coding scheme is used in all panels;
the data outside the interval $[-5,5]$ is clipped.
}\label{p3x1}

\medskip
\center\includegraphics*[width=.66\columnwidth]{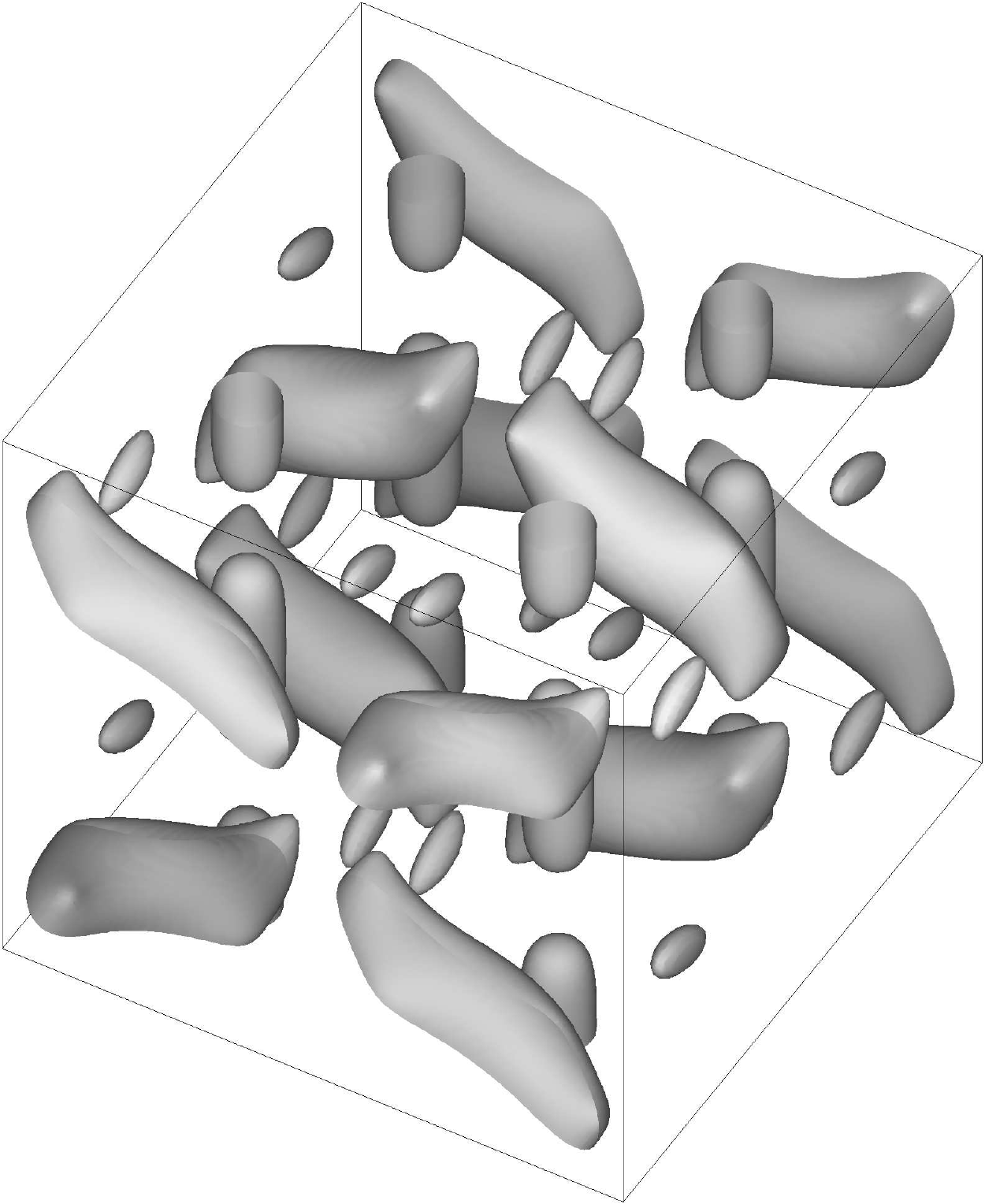}\hspace*{2ex}
\includegraphics*[width=.66\columnwidth]{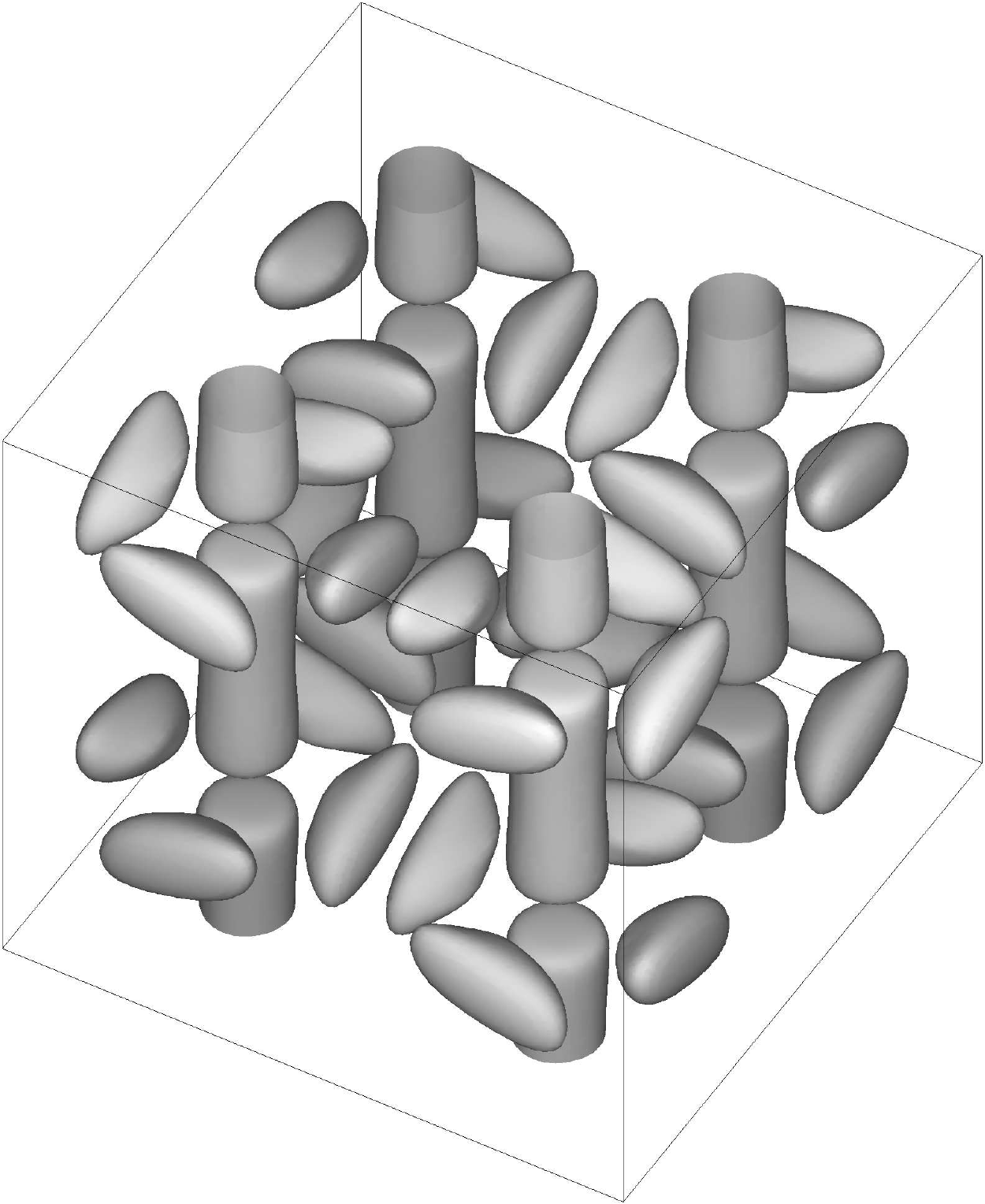}\hspace*{2ex}
\includegraphics*[width=.66\columnwidth]{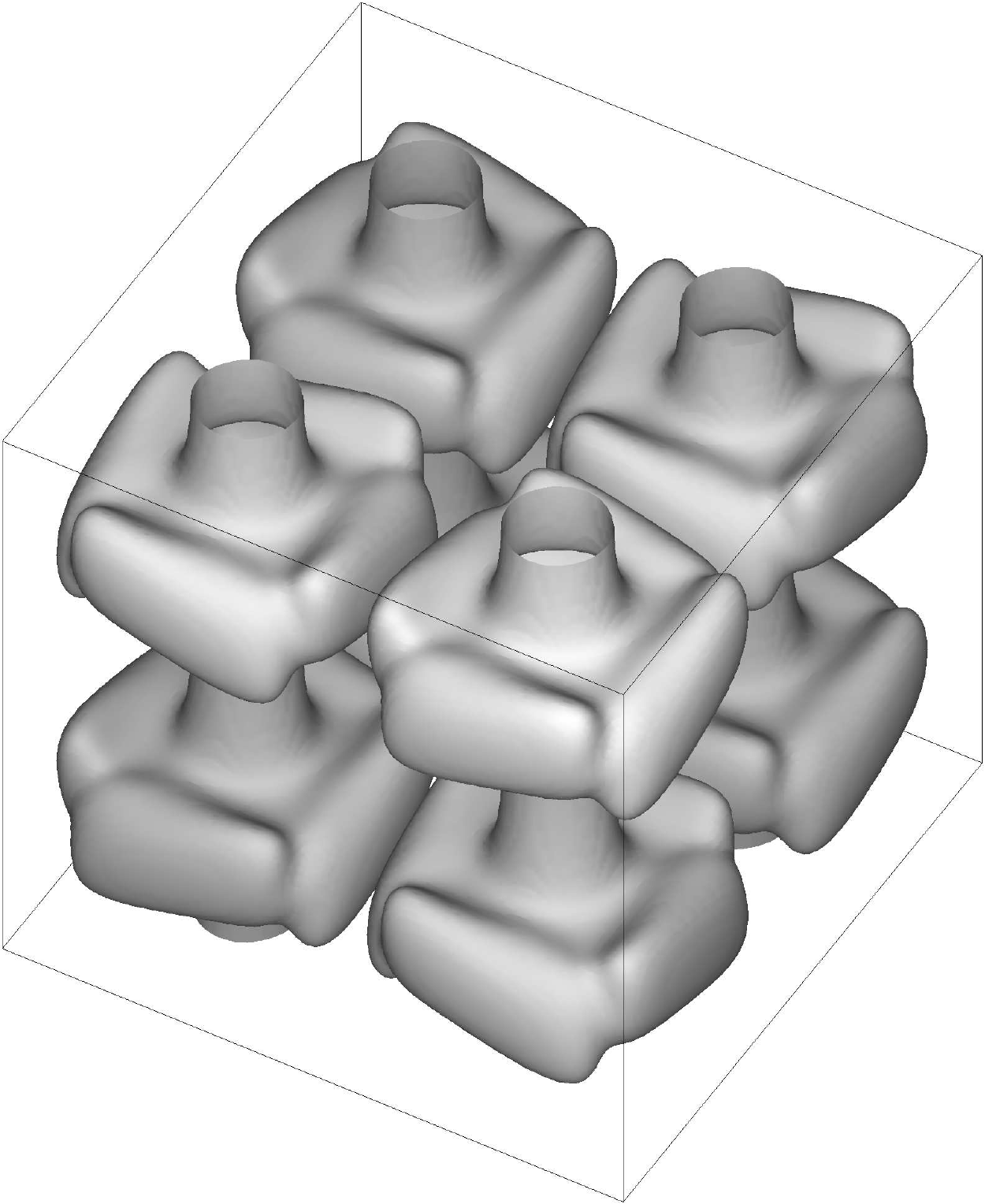}
\caption{Isosurfaces of the energy for two rms-normalised dominant
$2\pi$-periodic modes: {\tt SAAOE}, $|{\bf b}|^2=2$ (left), {\tt AAAEO},
$|{\bf b}|^2=2$ (centre) and {\tt AAAEO}, $|{\bf b}|^2=2/3$ (right).
}\label{pi2ie}

\medskip
\center\includegraphics*[width=.66\columnwidth]{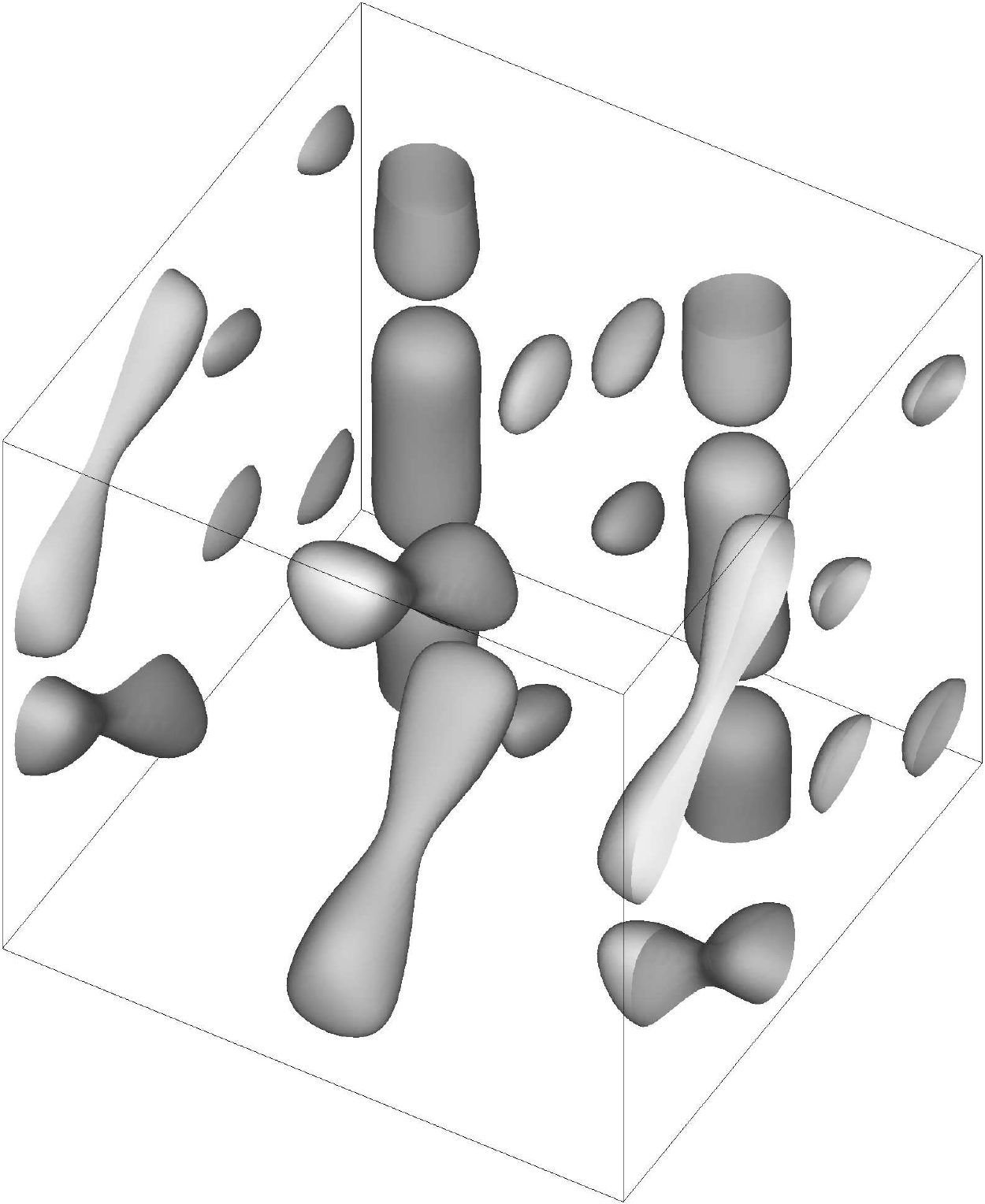}\hspace*{2ex}
\includegraphics*[width=.66\columnwidth]{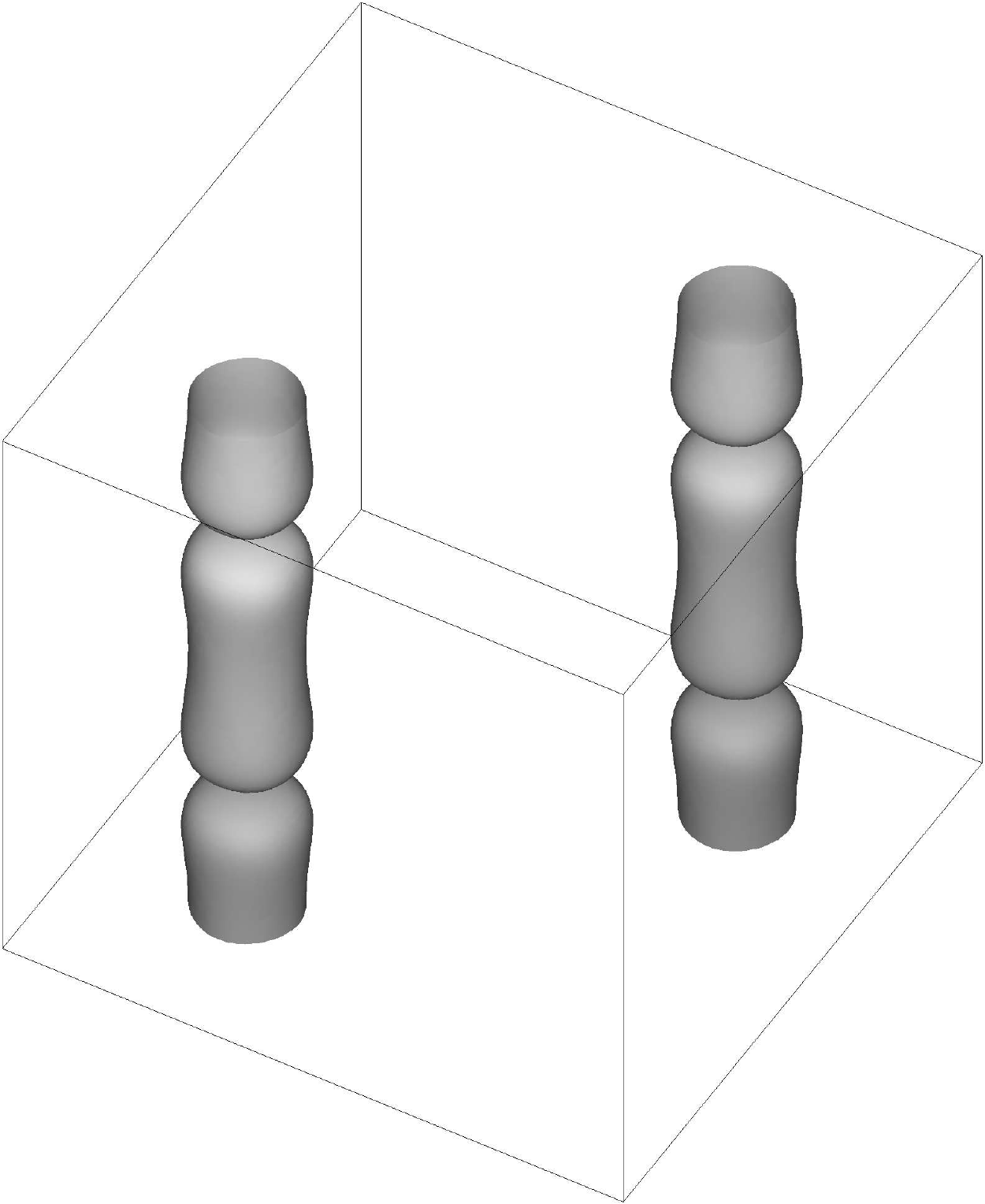}
\caption{Isosurfaces of the vertical magnetic field component, $b_3=2/3$,
for two rms-normalised
dominant $2\pi$-periodic modes, {\tt ASAOO} (left) and {\tt AAAEO} (right).
}\label{pi2i3}\end{figure*}

Only in 3 subspaces out of 40, growing $2\pi$-periodic magnetic modes
have been found (see Table~\ref{gr}). We first inspect
suitably averaged fields; as discussed in the next section
a particularly revealing average is that over the $x_3$ coordinate,
the mean field being a function of $x_1$ and~$x_2$. In \Fig{p3x1}
we show such mean fields $\overline{b}_3(x_1,x_2)$; clearly, they do not
survive horizontal averaging over the $(x_1,x_2)$ plane, because the positive
and negative contributions in \Fig{p3x1} cancel. This figure also
illustrates some of the symmetries of the invariant subspaces, to which
the 3 modes belong. Figures~\ref{pi2ie} and \ref{pi2i3} show isosurfaces
of the energy at the level $|{\bf b}|^2=2$, and of the vertical magnetic
component at the level $b_3=2/3$, for two dominant modes, that are not mutually
related by any of the symmetries. (The dominant modes in subspaces {\tt SAAOE} and
{\tt ASAOO} are mapped into each other by the symmetry $\gamma$, and the mode in
{\tt AAAEO} is $\gamma$-symmetric.)

\begin{table*}[t!]
\caption{Stagnation points of modified Taylor--Green flow \rf{TG}
for $a=b=1$, and the spectral
structure (eigenvalues, $\sigma$, and the proper subspaces) of the Jacobian
matrix, $[\partial v^m/\partial x_n]$, at these points; $j_i$ and $j$ are arbitrary integers.}
\center\begin{tabular}{|l|c|c|c|}\hline
Family&Stagnation point&Proper subspace& Eigenvalue, $\sigma$\\\hline
&&${\bf e}_1^{\vphantom{|}}$&$2+(-1)^{j_1+j_2+j_3}(2+3c)$\\
I&$\pi(j_1,j_2,j_3)$&${\bf e}_2$&$2-(-1)^{j_1+j_2+j_3}(2+3c)$\\
&&${\bf e}_3$&$-4$\\\hline
&&${\bf e}_1$&$-2$\\
II&$\pi\Big(j_1,j_2,j_3+{1\over2}\Big)$&${\bf e}_2$&$-2$\\
&&${\bf e}_3$&4\\\hline
III&$\pi\Big(j_1+{1\over2},j_2+{1\over2},j_3\Big)^{\vphantom{|}}$&
$\{{\bf e}_1,{\bf e}_2\}$&$-2\pm\i\sqrt{c+2}$\\
&&${\bf e}_3$&4\\\hline
&&${\bf e}_1$&2\\
IV&$\pi\Big(j_1+{1\over2},j_2+{1\over2},j_3+{1\over2}\Big)$&${\bf e}_2$&2\\
&&${\bf e}_3$&$-4$\\\hline
V&$\pi\Big(j_1+{1\over2},j_2,j_3+{1\over2}\Big)^{\vphantom{|}}$&
$\sigma{\bf e}_1-4(-1)^{j_1+j_2+j_3}(3c-1){\bf e}_3$&$1\pm\sqrt{1+4(3c-1)(2-c)}$\\
&&${\bf e}_2$&$-2$\\\hline
VI&$\pi\Big(j_1,j_2+{1\over2},j_3+{1\over2}\Big)^{\vphantom{|}}$&
$\sigma{\bf e}_2+4(-1)^{j_1+j_2+j_3}(3c-1){\bf e}_3$&$1\pm\sqrt{1+4(3c-1)(2-c)}$\\
&&${\bf e}_1$&$-2$\\\hline
VII&$(x_1,j\pi,x_3)$, where\vphantom{\Big)}&
${\bf e}_1(\sigma-\zeta_5(x_1))+{\bf e}_3\zeta_6(x_1,x_3)$&
$\zeta_3(x_1,x_3)\pm\sqrt{\zeta_4(x_1,x_3)}$\\
&$\sin^2x_1=\zeta_2^{-1}\!\left(\!-\zeta_1\pm\sqrt{\zeta_1^2+4\zeta_2}\right)$\vphantom{\Big)}&
${\bf e}_2$&$-2\zeta_3(x_1,x_3)$\\
&$\cos x_3=(-1)^j(3c-1)\sin x_1\tan x_1$\vphantom{\Big)}&&\\\hline
VIII&$(j\pi,x_2,x_3)$, where\vphantom{\Big)}&${\bf e}_1$&$-2\zeta_3(x_2,x_3)$\\
&$\sin^2x_2=\zeta_2^{-1}\!\left(\!-\zeta_1\pm\sqrt{\zeta_1^2+4\zeta_2}\right)$\vphantom{\Big)}&
${\bf e}_2(\sigma-\zeta_5(x_2))+{\bf e}_3\zeta_6(x_2,x_3)$&
$\zeta_3(x_2,x_3)\pm\sqrt{\zeta_4(x_2,x_3)}$\\
&$\cos x_3=(-1)^j(1-3c)\sin x_2\tan x_2$\vphantom{\Big)}&&\\\hline
\end{tabular}\label{sp}\tablecomments{
Here $\zeta_1=3c(3c+1)$, $\zeta_2=8(3c-1)(2c-1)$,
$\zeta_3(x,z)=(-1)^j\big((3c-1)((3c/2+1)\sin^2x-2\sin^4x)-\cos 2z\big)$,\break
$\zeta_5(x)=\big((-1)^j(6c-2)-4c+((-1)^j(3c-27c^3)+4-14c+54c^3)\sin^2x\big)/(2c-1)$,
~$\zeta_6(x,z)=4(3c-1)(2-\sin^2x)\sin x\sin z$,
$\zeta_4(x,z)=\big(2\sin^2z-(8c+(5-19c+3c^2+90c^3)\sin^2x)/(1-2c)\big)^2\!
+4(1-3c)(2-\sin^2x)\sin^2x\sin^2z(2+3c-(8-20c)\sin^2x)$ for odd $j$ and
$\zeta_4(x,z)=\big(2\cos^2z+(-3+3c+9c^2)\sin^2x\big)^2\!
+4(1-3c)(2-\sin^2x)\sin^2x\sin^2z(2+3c-(8-20c)\sin^2x)$ for~even~$j$.
}\end{table*}

In slow dynamos, magnetic structures can be related to stagnation points
of the flow. Eight families of stagnation points of mTG are listed
in Table~\ref{sp}; we have checked numerically that no other stagnation points
exist in mTG~\rf{TG}, \rf{notte}. Each of the first four families is
a $\gamma$-symmetric set; families V and VI are mapped by $\gamma$ into each
other, as well as families VII and VIII. Lines joining
stagnation points of family I and parallel to Cartesian axes constitute
a heteroclinic network: any such vertical line consists of heteroclinic
trajectories connecting adjacent stagnation points of families I and II, and
a horizontal line consists of heteroclinic trajectories connecting a pair of
adjacent stagnation points of family I. Each plane, parallel to a Cartesian
coordinate plane and containing stagnation points of family I, is cut
by the aforementioned heteroclinic trajectories into squares of size $\pi$,
which are invariant sets for mTG (this stems from the proportionality of $v_i$
to $\sin x_i$ for each $i$). Vertical and horizontal lines joining
stagnation points of family IV constitute another heteroclinic network:
they consist of heteroclinic trajectories connecting points of family IV with
adjacent stagnation points of families III, V and VI.

The Jacobian matrix of a solenoidal flow generically has either one positive
eigenvalue and two eigenvalues with negative real parts, or one negative
eigenvalue and two eigenvalues with positive real parts. In the vicinity
of a stagnation point of the former kind (having a one-dimensional unstable
manifold), magnetic flux ropes usually emerge \citep{Moff,GZh} that are aligned
with the unstable direction. Near a stagnation point of the latter kind
(possessing a two-dimensional unstable manifold), magnetic sheets typically emerge
\citep{ChS} spreading along the unstable manifold. (Formation of these magnetic
structures may be prohibited by symmetries.)

We observe such patterns
of asymptotic nature, foremost, vertically oriented flux ropes,
that are centred at stagnation points of family III (whose one-dimensional
unstable manifolds are segments of vertical lines),
in the plots of isosurfaces of the magnetic energy at the level
$|{\bf b}|^2=2$ (\Fig{pi2ie}, left and central panels) and
of the vertical component of magnetic field (\Fig{pi2i3}) for both modes,
shown in the figures, from the symmetry subspaces {\tt SAAOE} and {\tt AAAEO}.
These ``principal'' ropes terminate near stagnation points of family IV, whose
two-dimensional unstable manifolds are horizontal planes, and which give rise
to magnetic field sheets revealed by energy isosurfaces at the low level
$|{\bf b}|^2=2/3$ (\Fig{pi2ie}, the right panel). The sheets intermix
into vertical flux ropes centred at stagnation points of family II.
In the {\tt AAAEO} mode,
adjacent principal flux ropes are oppositely directed (see \Fig{p3x1});
consequently, the flux ropes associated with stagnation points of family II
are comprised of two pairs of oppositely oriented ``flux fibres''
(such compound flux ropes were considered by \citealt{GZh}). Since fine
structures are accompanied by enhanced dissipation, the compound ropes are weak
and not seen in the right panel of \Fig{pi2ie} --- these relatively
high-level isosurfaces only determine the region in space, where the four-fibre
flux ropes are located. Compound flux ropes consisting of two oppositely
directed fibres centred at stagnation points of families V and VI are present
in the {\tt AAAEO} mode (these individual fibres actually look more like beans
in the central panel of \Fig{pi2ie}: the width of flux ropes is
of the order of $R_m^{-1/2}$, where $R_m$ is the magnetic Reynolds number
which is clearly not high for $\eta=0.1$ considered here, and hence the magnetic
flux ropes and sheets that we observe are rather ``fat''). In the {\tt SAAOE}
mode, flux ropes centred at stagnation points of family VI (but not V)
are allowed by the symmetries defining the subspace;
these flux ropes do not have a fibre structure (in the left panel
of \Fig{pi2ie} they are cut into halves by the faces of the shown cube
of periodicity) and their energy content is even higher than that
of the principal ropes.

\begin{figure*}[p]
\center\includegraphics[height=.97\textheight]{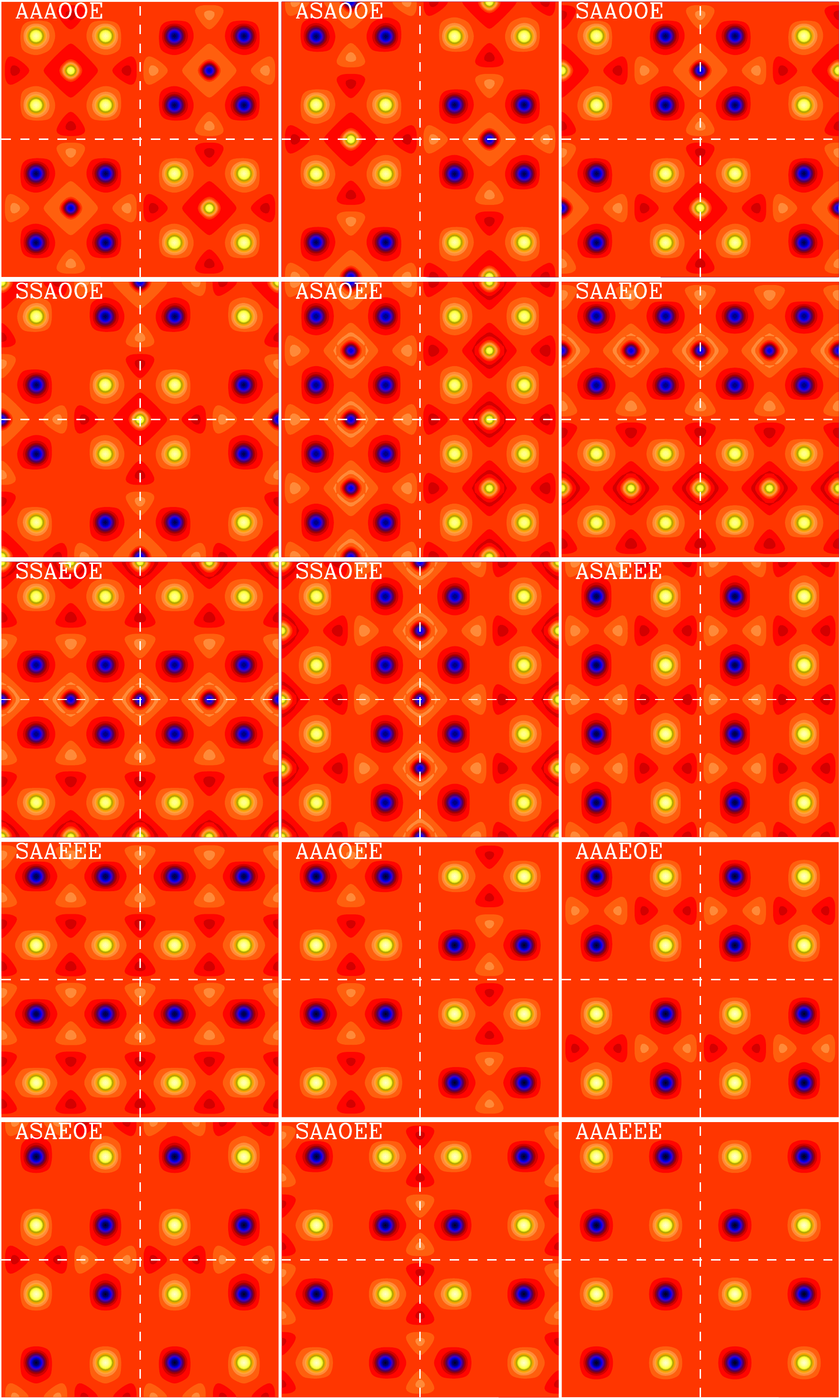}
\caption{The $x_3$-averaged normalised mean field
$\overline{b}_3(x_1,x_2)/\langle\overline{b}_3^2\rangle^{1/2}$
for $\eta=0.1$ in a domain of size~$(4\pi)^3$.
The white dashed lines mark subdomains of size $(2\pi)^2$.
Same colour-coding scheme as in \Fig{p3x1}.
}\label{pbzm}\end{figure*}

\begin{figure*}[t!]
\centerline{\includegraphics*[width=.66\columnwidth]{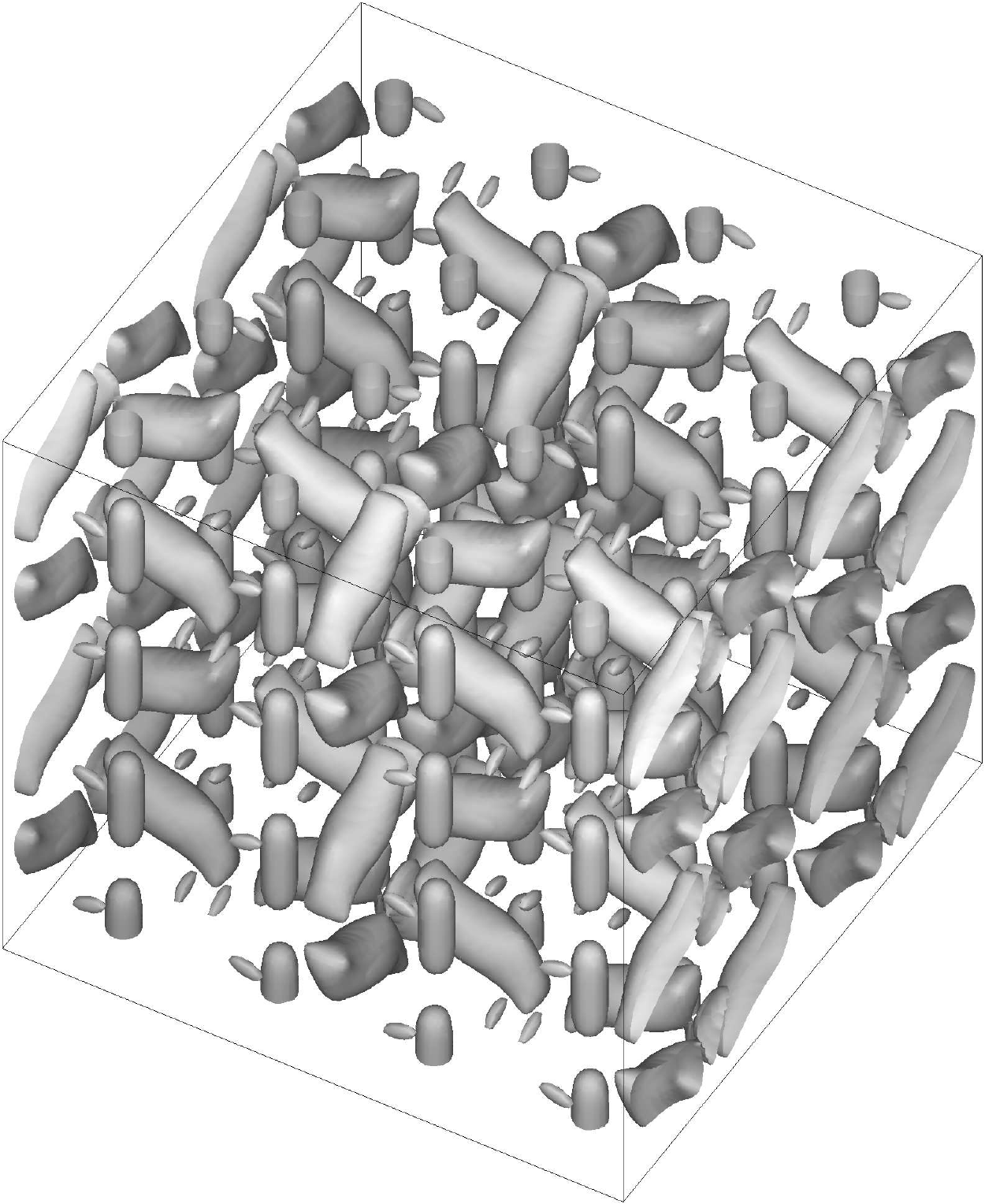}\hspace*{2ex}
\includegraphics*[width=.66\columnwidth]{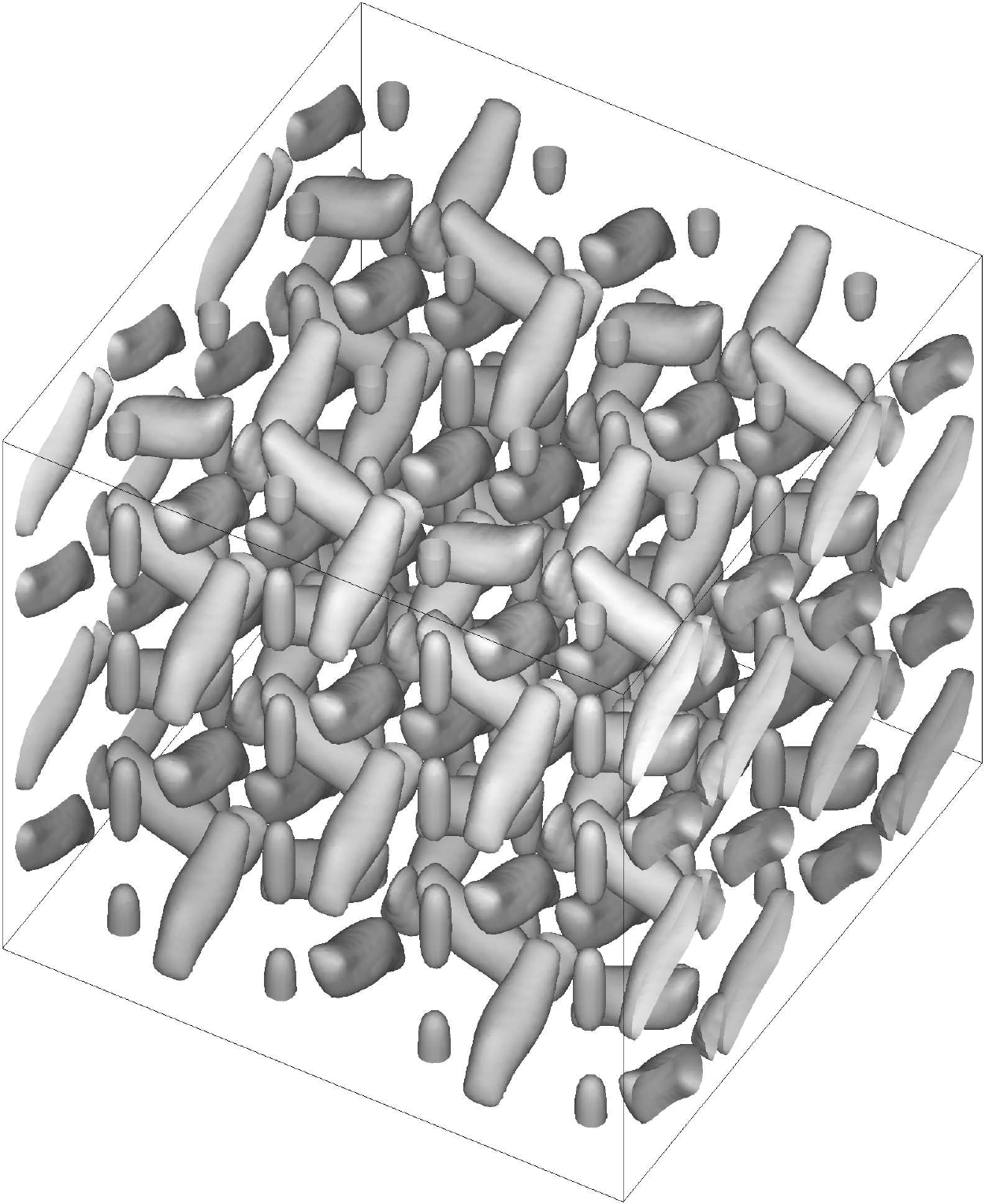}\hspace*{2ex}
\includegraphics*[width=.66\columnwidth]{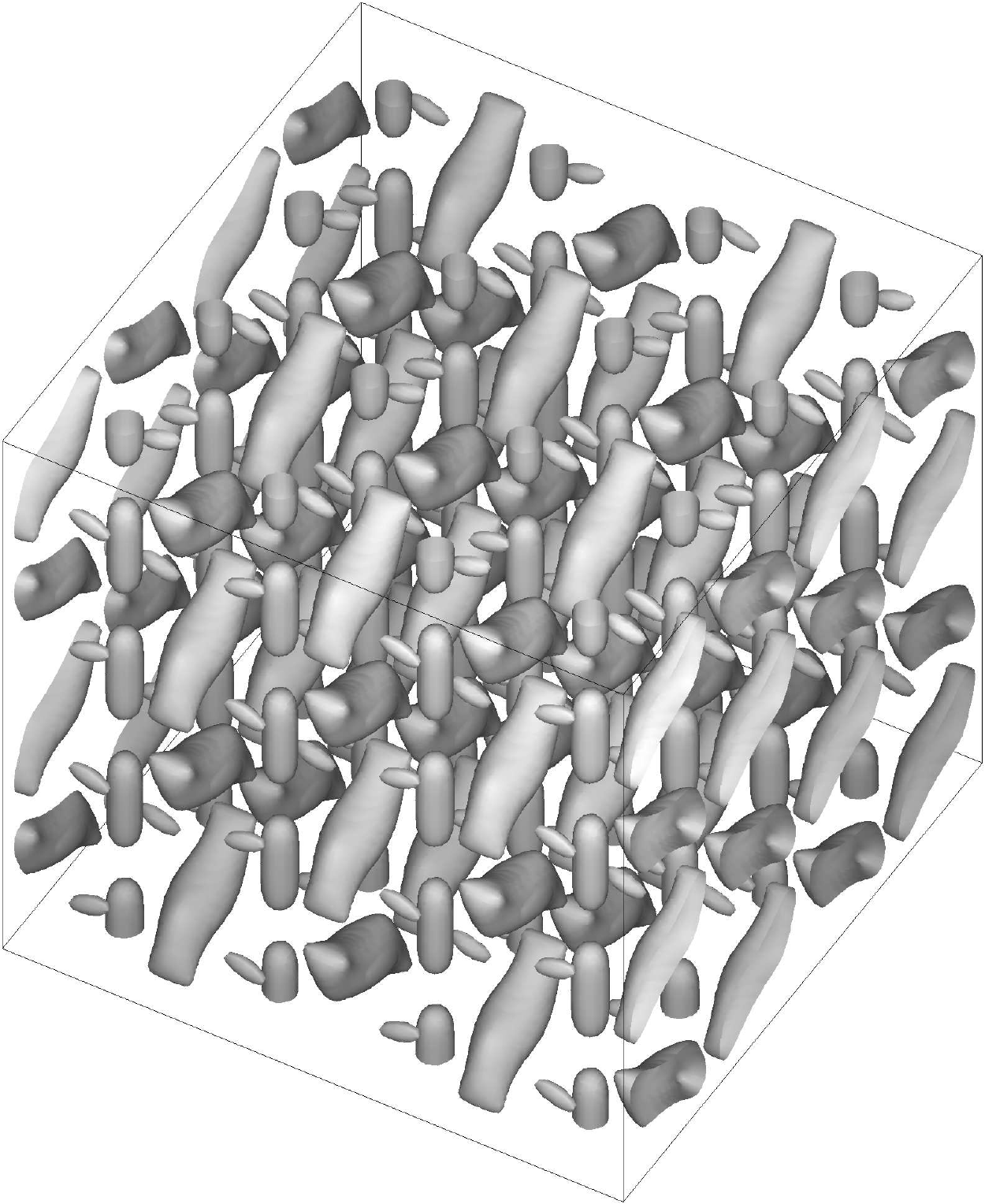}}

\vspace*{-8mm}\hspace*{2ex}{\tt ASAOOE}
\hspace*{.565\columnwidth}{\tt ASAOEE}\hspace*{.575\columnwidth}{\tt ASAEEE}

\vspace*{5mm}
\centerline{\includegraphics*[width=.66\columnwidth]{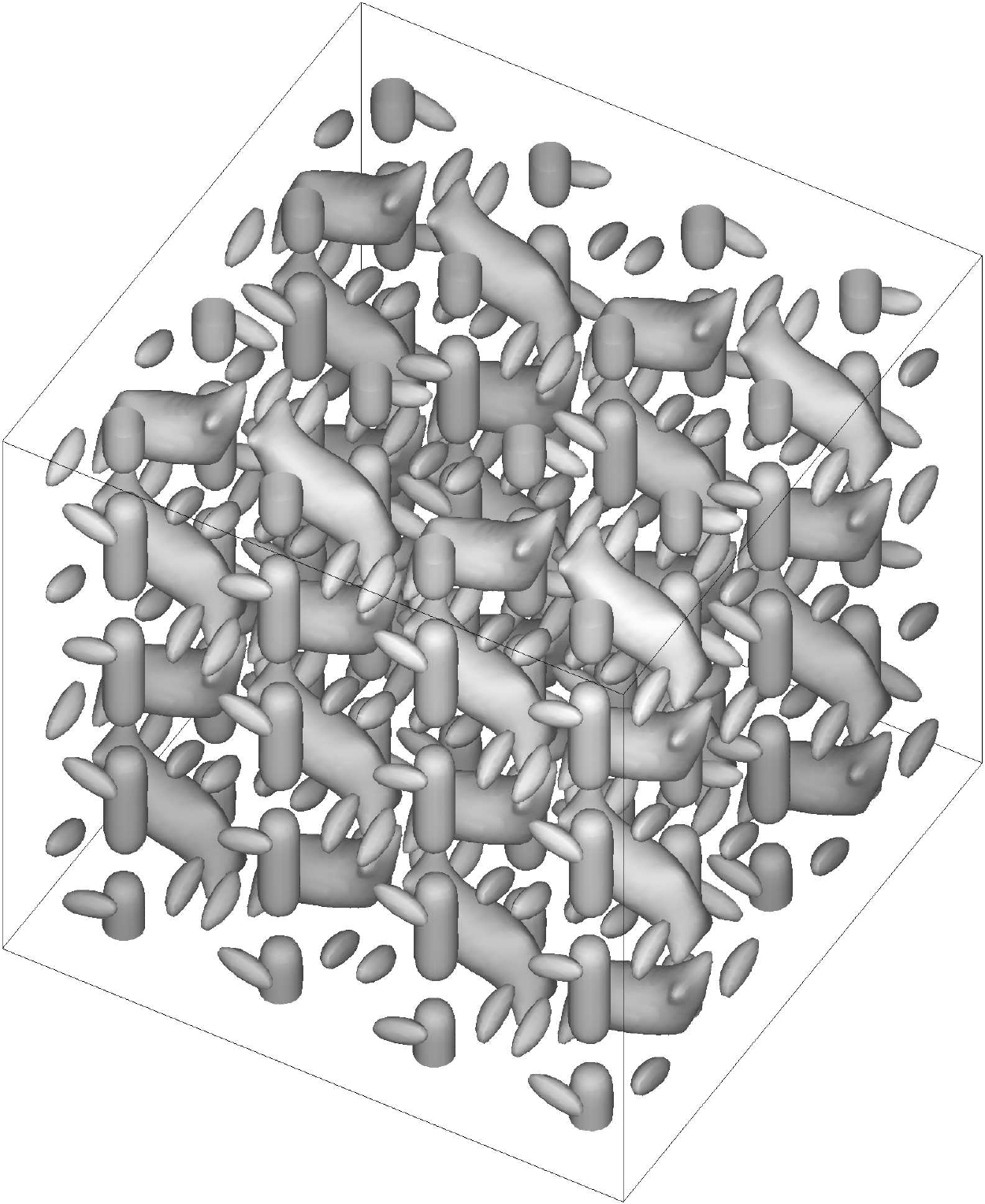}\hspace*{2ex}
\includegraphics*[width=.66\columnwidth]{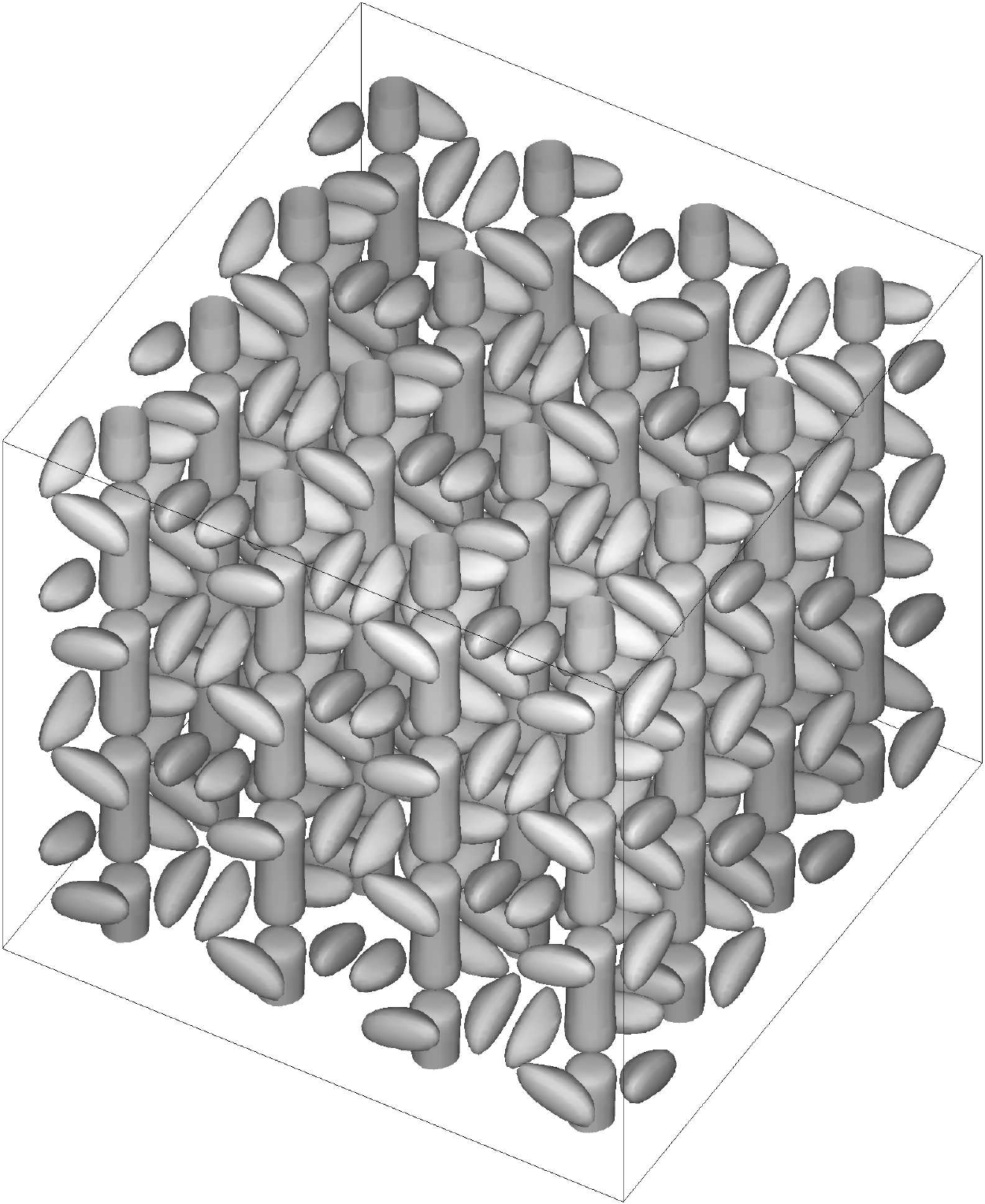}\hspace*{2ex}
\includegraphics*[width=.66\columnwidth]{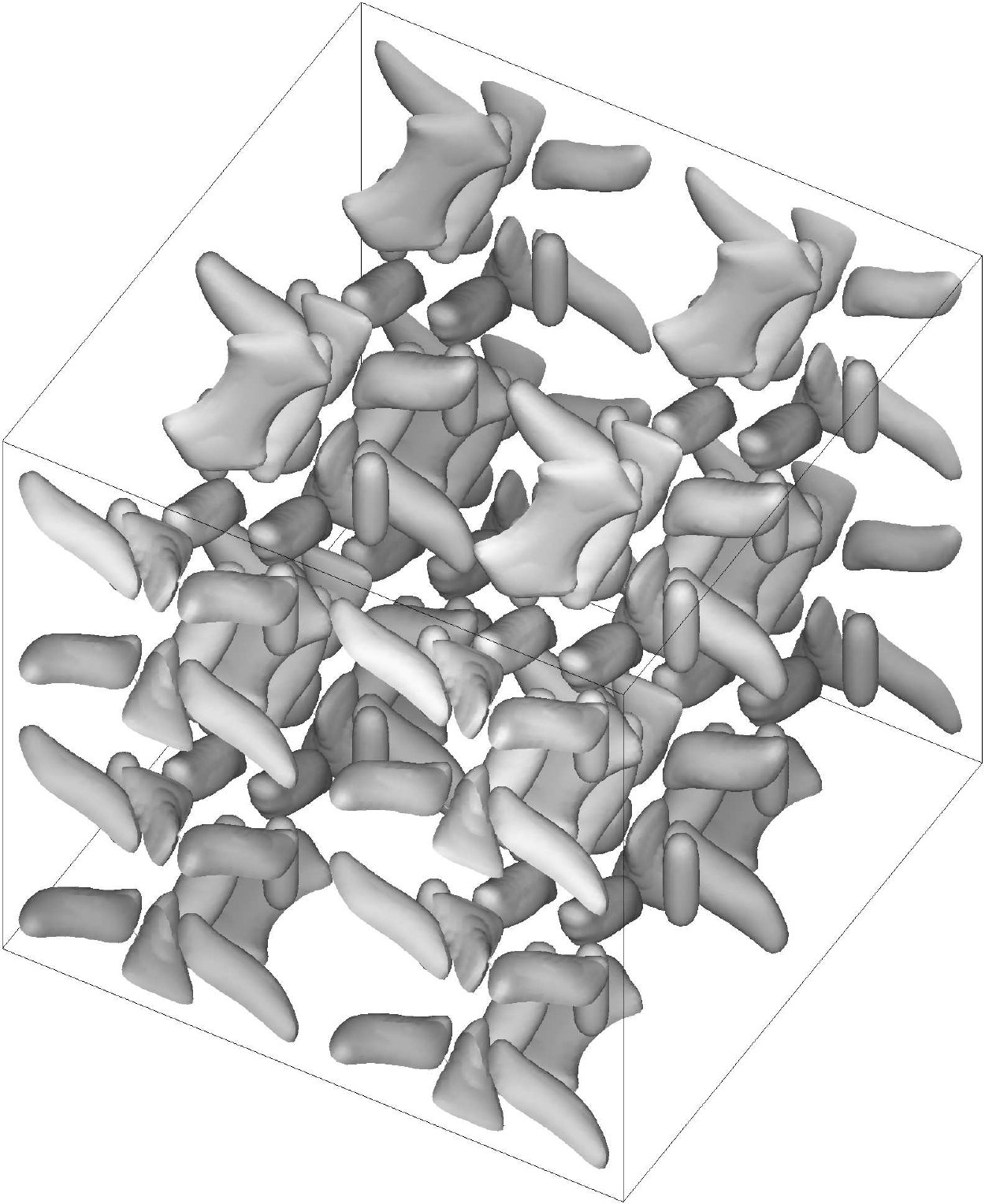}}

\vspace*{-8mm}\hspace*{2ex}{\tt AAAOEE}
\hspace*{.565\columnwidth}{\tt AAAEEE}\hspace*{.575\columnwidth}{\tt SASOOO}

\medskip
\caption{Isosurfaces of the energy for six rms-normalised isotypic dominant
$4\pi$-periodic modes at the level $|{\bf b}|^2=2$.
}\label{pi4ie}\end{figure*}

\begin{figure*}[t!]
\centerline{\includegraphics*[width=.66\columnwidth]{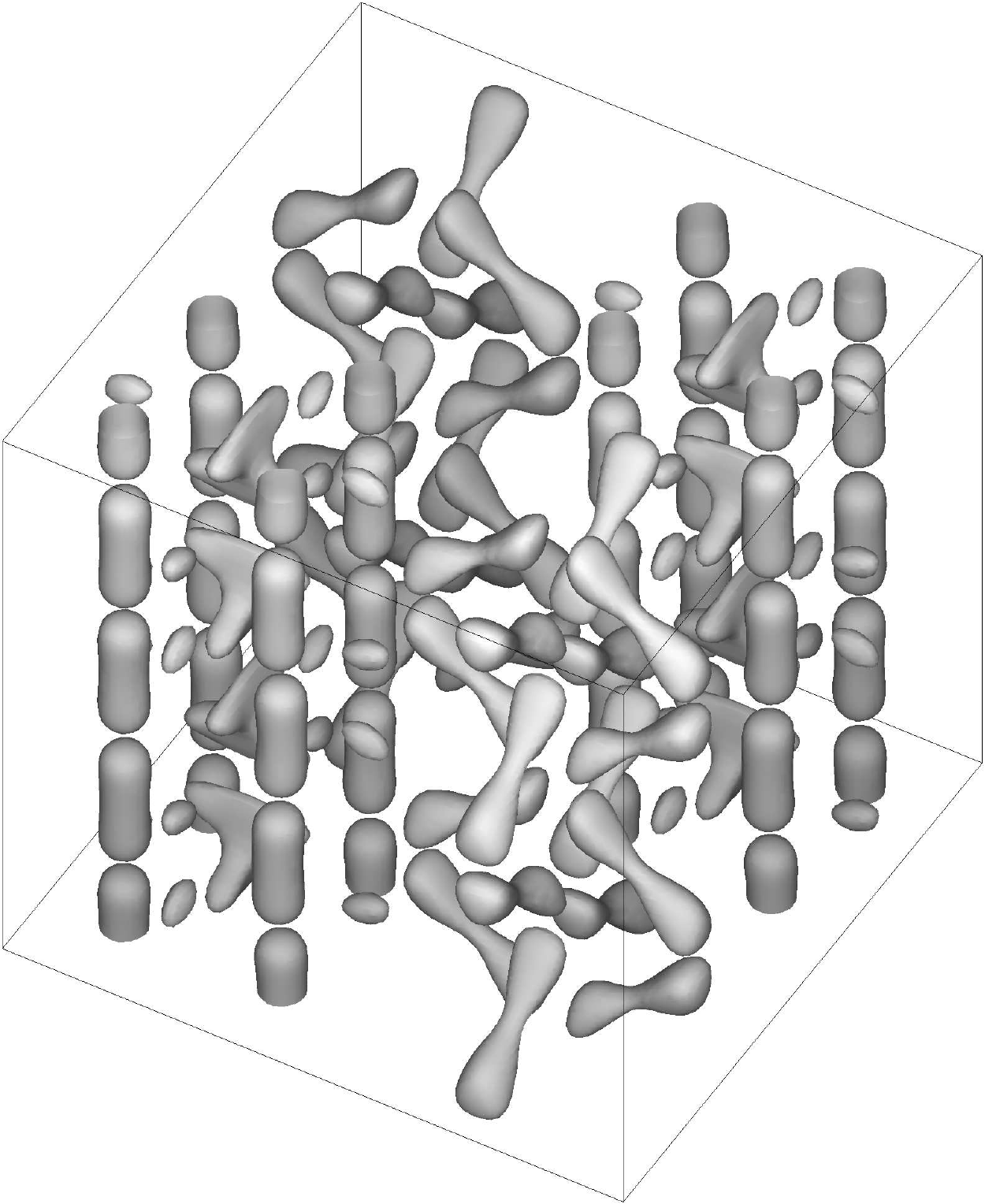}\hspace*{2ex}
\includegraphics*[width=.66\columnwidth]{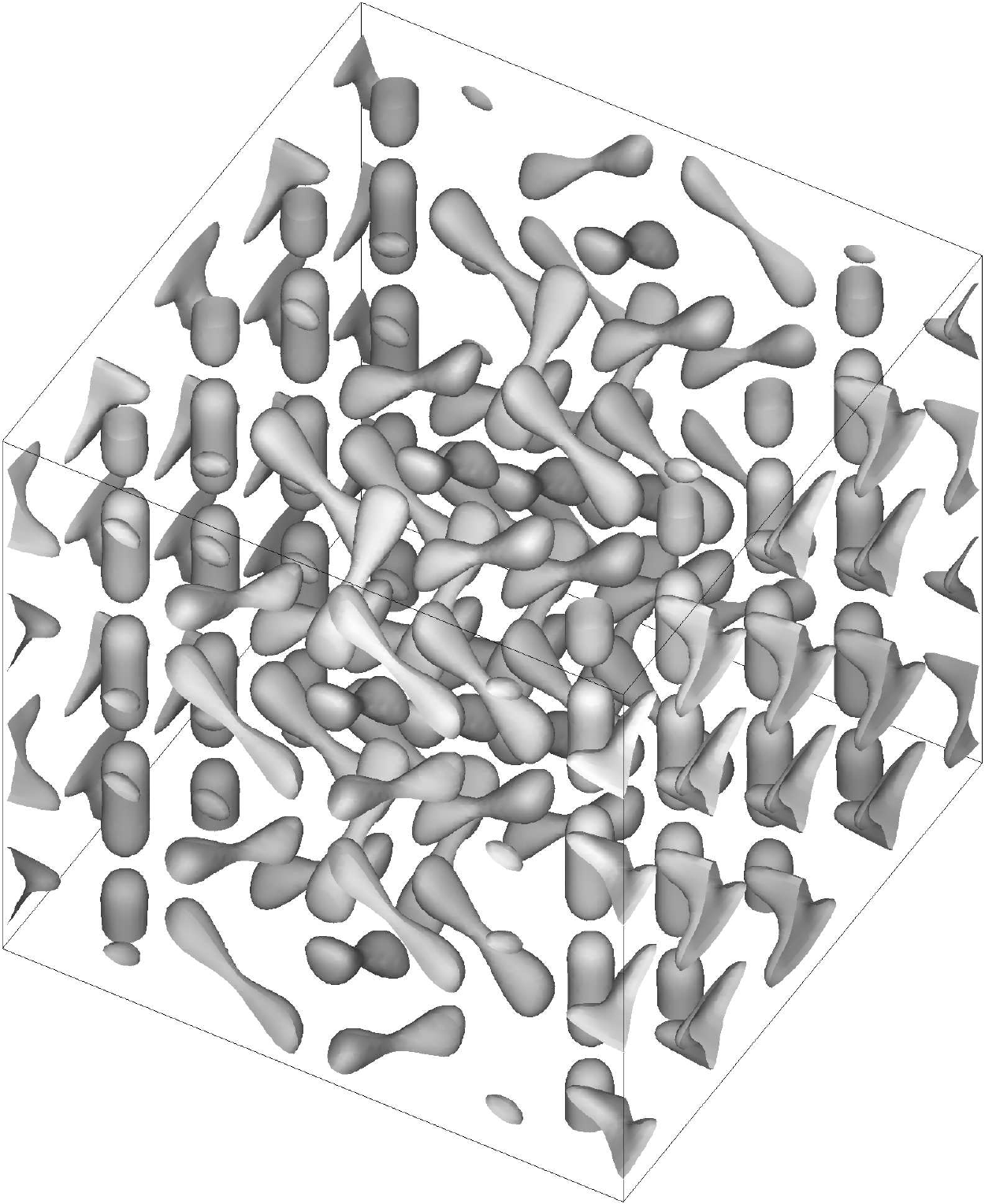}\hspace*{2ex}
\includegraphics*[width=.66\columnwidth]{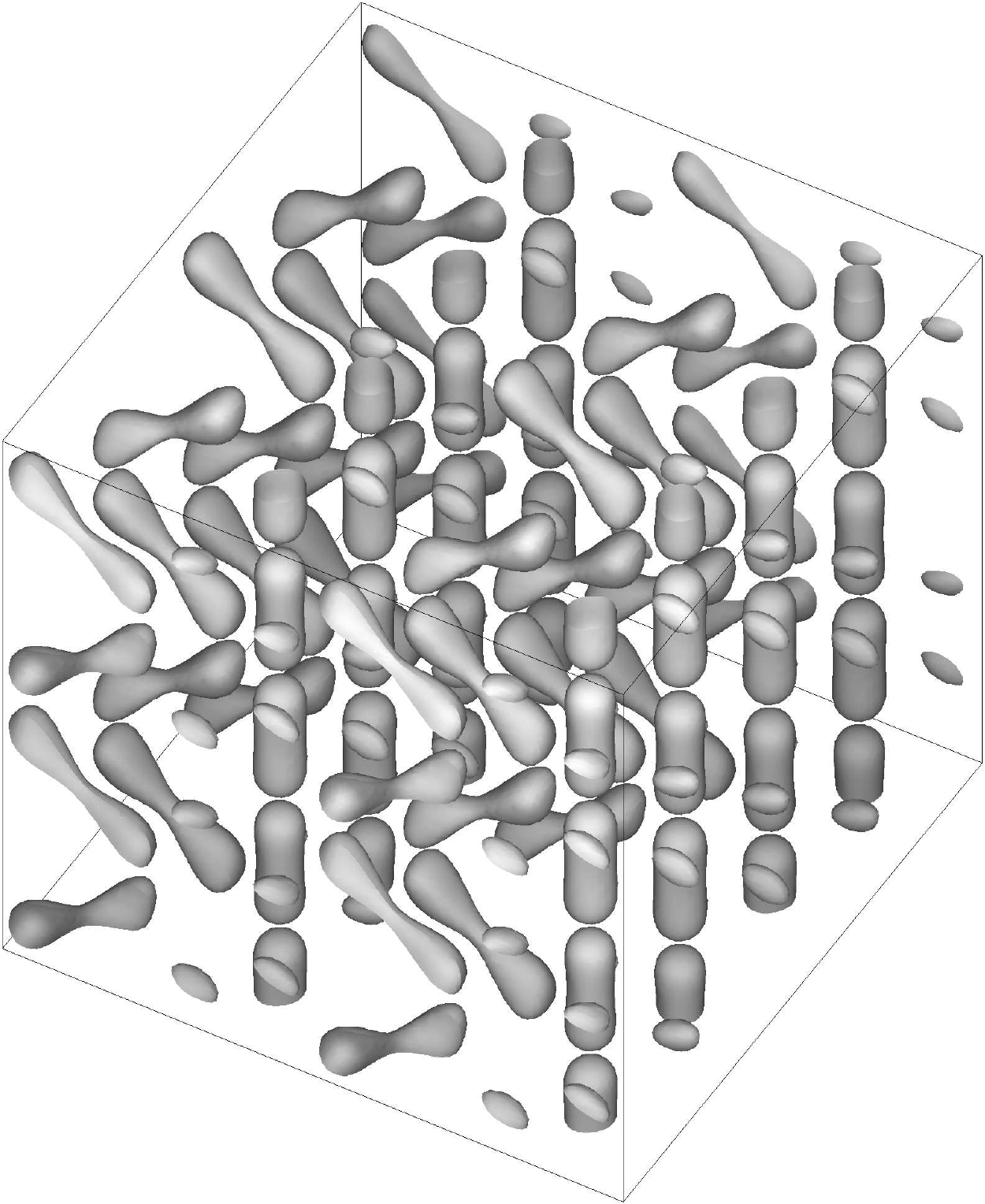}}

\vspace*{-8mm}\hspace*{2ex}{\tt AAAOOE}
\hspace*{.565\columnwidth}{\tt SSAEOE}\hspace*{.575\columnwidth}{\tt SAAEEE}

\vspace*{5mm}
\centerline{\includegraphics*[width=.66\columnwidth]{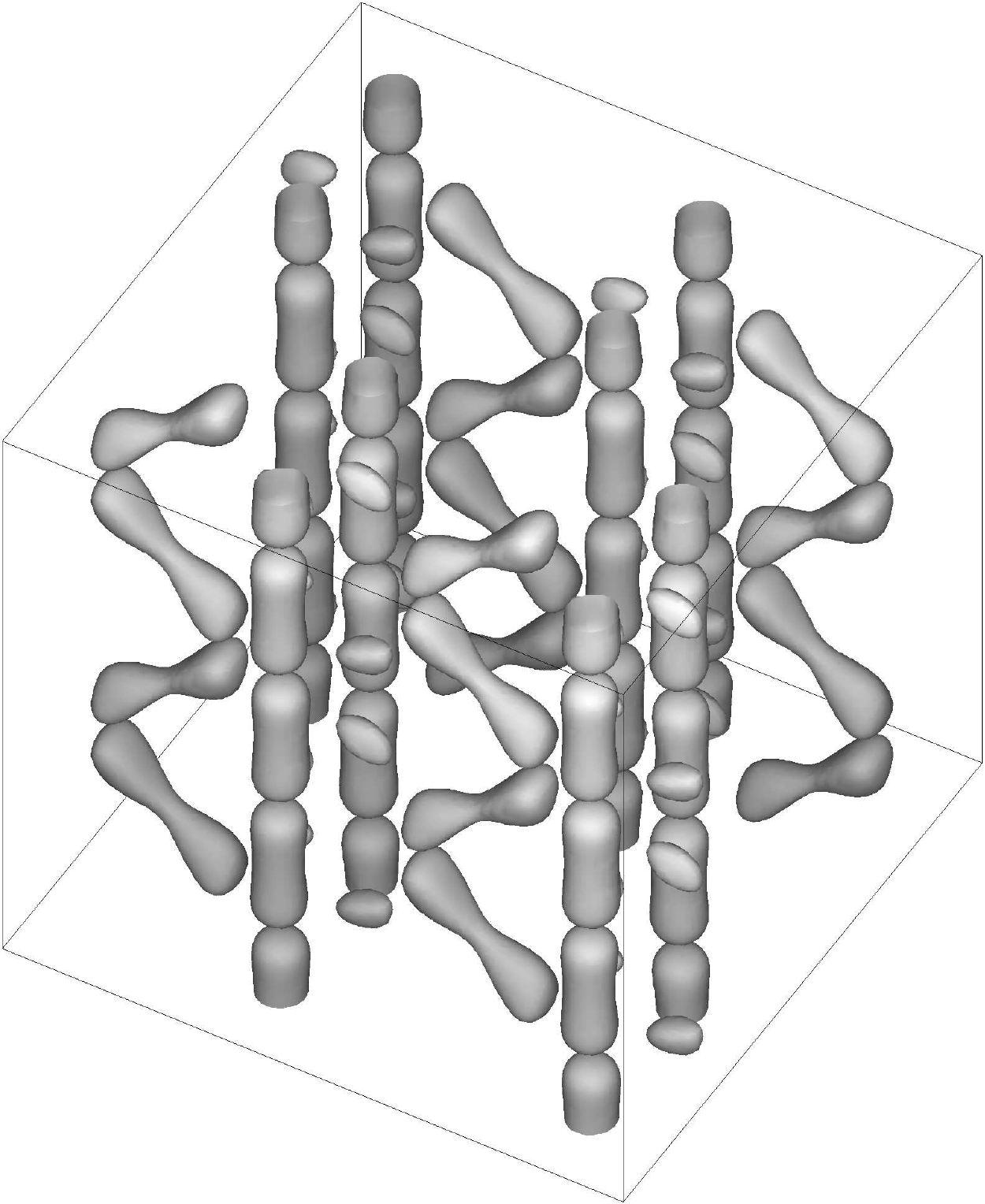}\hspace*{2ex}
\includegraphics*[width=.66\columnwidth]{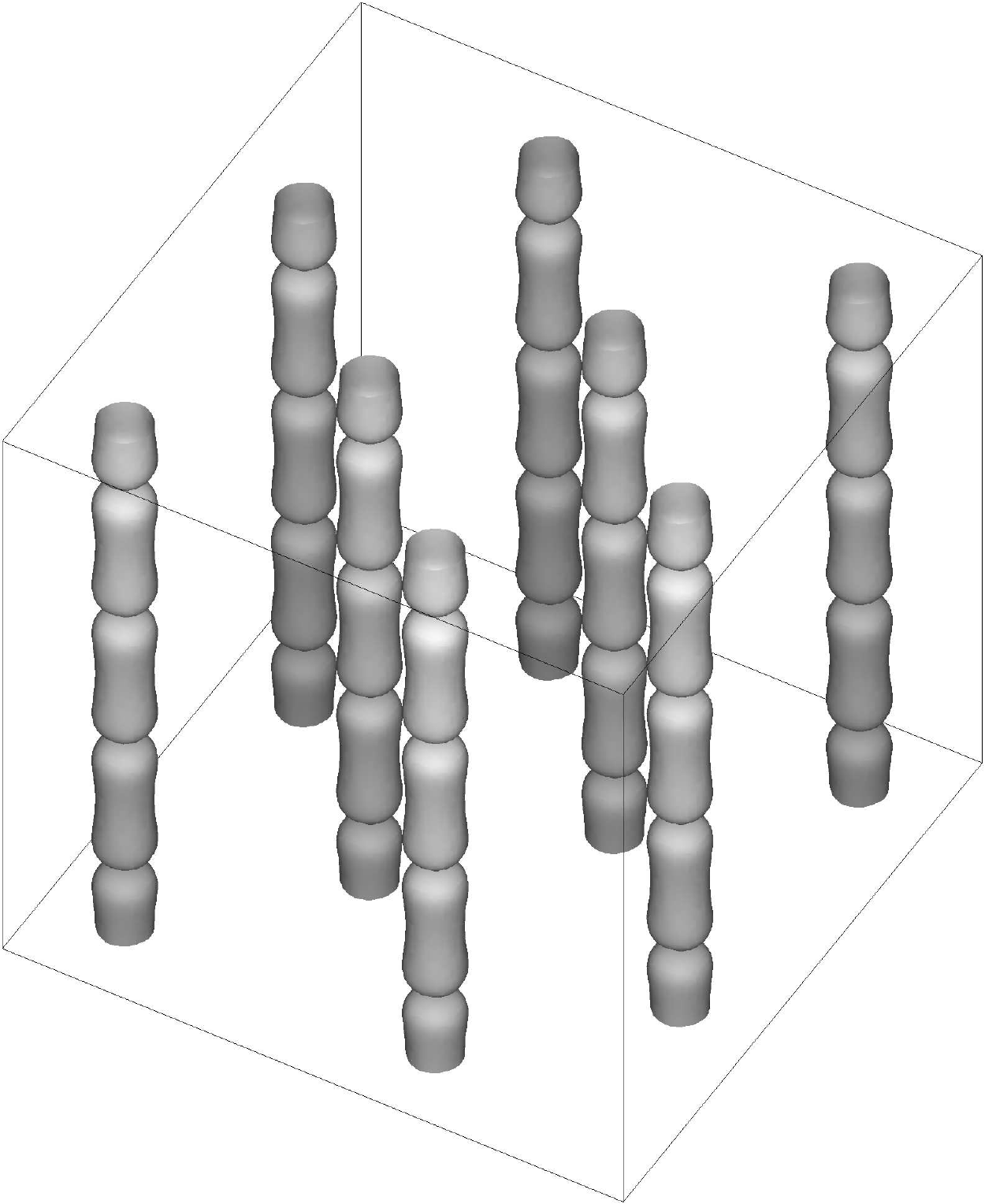}\hspace*{2ex}
\includegraphics*[width=.66\columnwidth]{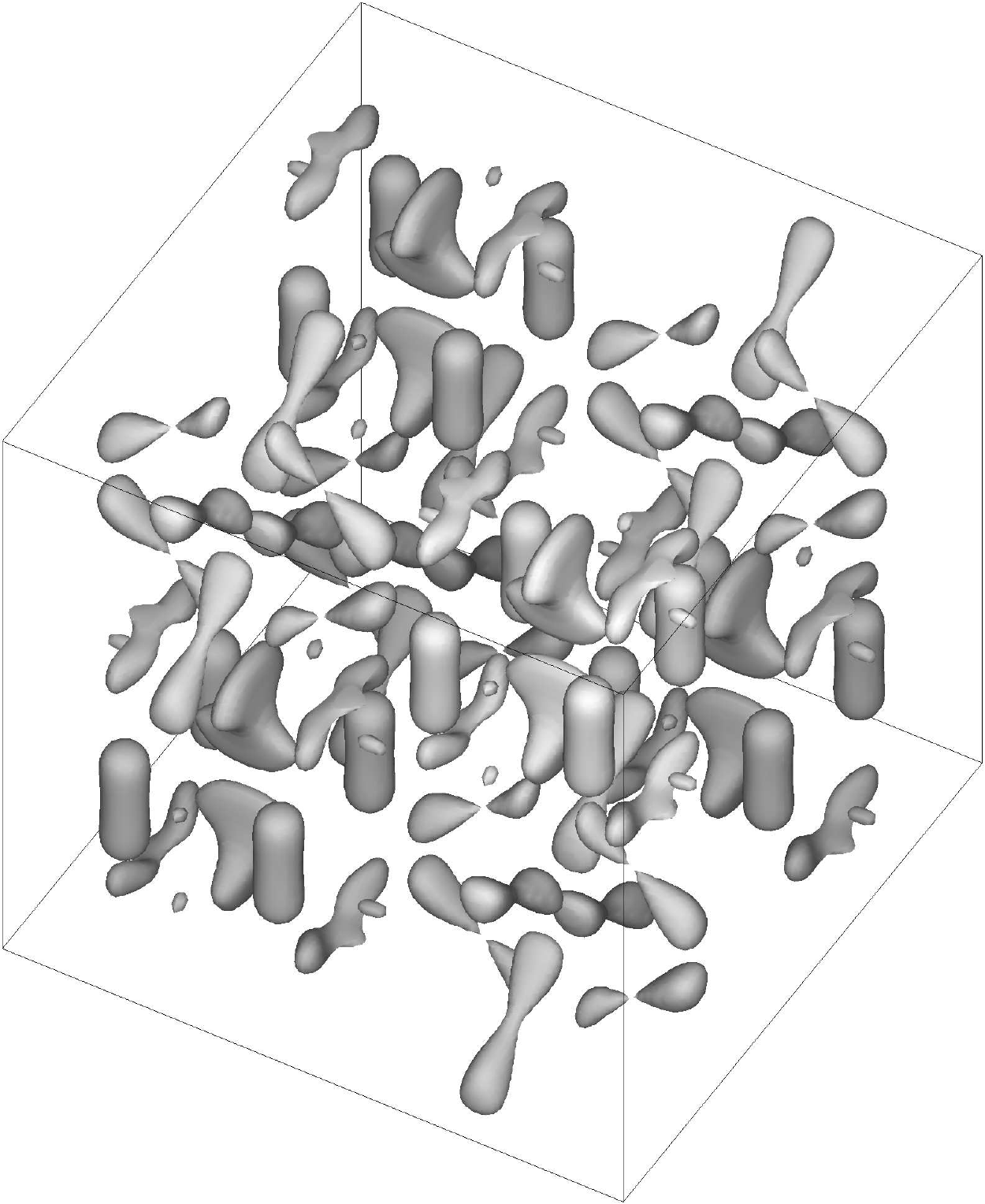}}

\vspace*{-8mm}\hspace*{2ex}{\tt AAAOEE}
\hspace*{.565\columnwidth}{\tt AAAEEE}\hspace*{.575\columnwidth}{\tt AASOOO}

\medskip
\caption{Isosurfaces of the vertical magnetic component for six rms-normalised
isotypic dominant $4\pi$-periodic modes at the level $b_3=2/3$.
}\label{pi4i3}\end{figure*}

All $4\pi$-periodic magnetic modes, growing for \hbox{$\eta=0.1$}, are also listed
in Table~\ref{gr}. Such modes can be symmetric or antisymmetric in each
Cartesian variable $x_i$; this is coded by the first 3 characters in the labels
(letters $\tt S$ and $\tt A$, respectively) of invariant subspaces, like in the
case of $2\pi$-periodic modes. The trailing 3 characters of the 6-character
labels have now a new meaning: for any fixed $i$, the wave numbers $k_i$ in all
Fourier harmonics $\e^{\i{\bf k\cdot x}/2}$ comprising a $4\pi$-periodic mode
have the same parity, which is indicated
by letters $\tt E$ or $\tt O$ (even and odd values, respectively)
in position $i+3$. We have considered neither the more subtle
parity symmetries, nor the $\gamma$-symmetry. Since the 6 aforementioned
symmetries are independent, they split the domain of the magnetic induction
operator into 64 invariant subspaces. We have computed dominant magnetic modes
in each of them using $128^3$ Fourier harmonics
(before dealiasing), which effectively provide the same spatial resolution
as $64^3$ harmonics in computations of the $2\pi$-periodic modes.

For the dominant growing $4\pi$-periodic magnetic modes, we show the same
plots as for the $2\pi$-periodic ones: the mean fields
$\overline{b}_3(x_1,x_2)$ averaged over $x_3$ for 15 dominant $4\pi$-periodic
modes (\Fig{pbzm}), and isosurfaces of the energy $|{\bf b}|^2=2$ and
of the component $b_3=2/3$ for six of them, that are not mutually related
by any symmetry (\Figs{pi4ie}{pi4i3}). Clearly, the averages over
the $(x_1,x_2)$ plane of the vertical component $b_3$ for all dominant modes
shown in \Fig{pbzm} are zero, as this was the case for the $2\pi$-periodic
modes. (For the 8 growing modes comprising the last group in Table~\ref{gr},
$\overline{b}_3=0$, because they involve only odd wave numbers~$k_3$.)
The most prominent features in \Fig{pbzm} are the averages of the vertically
oriented flux ropes centred at stagnation points of family III; all other flux
ropes cancel out upon averaging over $x_3$ either mostly or completely. It is
natural that these mean flux ropes of a similar genesis have
a similar shape in all panels in \Fig{pbzm} and, for instance, have close
extremum values, the maxima ranging from 5.72 for dominant modes from
the second group of $4\pi$-periodic modes in Table~\ref{gr} (including
subspace {\tt ASAOEE}) to 6.62 in the fifth group (subspace
{\tt AAAEEE}). (The maxima are computed for the normalised averages
$\overline{b}_3(x_1,x_2)/\LA\overline{b}_3^2\ra^{1/2}$.) It turns
out that the dominant $4\pi$-periodic modes in subspaces {\tt ASAEEE},
{\tt SAAEEE} and {\tt AAAEEE} are just the tiling of the cube of periodicity
of size $4\pi$ by 8 cubes of periodicity of size $2\pi$ with $2\pi$-periodic
modes in subspaces {\tt ASAOO}, {\tt SAAOE} and {\tt AAAEO}, respectively (note
that the growth rates of the respective $4\pi$- and $2\pi$-periodic modes
coincide). In fact, each group of $4\pi$-periodic modes that have the same
growth rate (see Table~\ref{gr}) are related by symmetries.
(For instance, the eight slowest-growing modes constituting the last group
in Table~\ref{gr} are mutually related by combinations of shifts
by $2\pi$ along the Cartesian axes.)

\subsection{DNS and TFM results for eddy diffusivity in mTG}
\label{DNSandTFM}

As noted above, horizontal averaging over the $(x_1,x_2)$ plane
cannot be applied to describe a growing mean field generated by mTG.
Indeed, averaging the solenoidality condition
for $\bf b$ we find that $\overline{b}_3$ is spatially uniform at all
times; then the spatial average of the third component of \rf{bmean} shows
that it is also time-independent. Since $\overline{b}_3$ cannot grow or decay,
such an average is unsuitable for studying the negative eddy diffusivity
dynamo for mTG (for which we are advised by MST that $\overline{b}_3\ne0$,
see \rf{eimo}). By contrast, planar averages can describe growing solutions
in the supercritical case, if one averages along $x_3$ and a diagonal direction,
or uses any of the two other planar averages, over $(x_1,x_3)$ or $(x_2,x_3)$.
Note that, for a flow with a large group of symmetries,
any planar averaging may yield, due to cancellation, identically zero averages
for modes in certain symmetry subspaces. For instance,
for mTG, no cancellation occurs for the $(x_1,x_3)$ or $(x_2,x_3)$ averagings
for the second and third groups of $4\pi$-periodic modes in Table~\ref{gr}, and
for diagonal ones for the first and fifth groups. Thus,
the average over $(x_1,x_3)$ or $(x_2,x_3)$ is adequate in 6 subspaces;
the diagonal average in 5 subspaces, including the dominant one; and none
in the remaining 12 subspaces containing growing modes.
None of the easily implementable planar averagings is universally applicable.

\begin{figure}[t]
\center\includegraphics[width=\columnwidth]{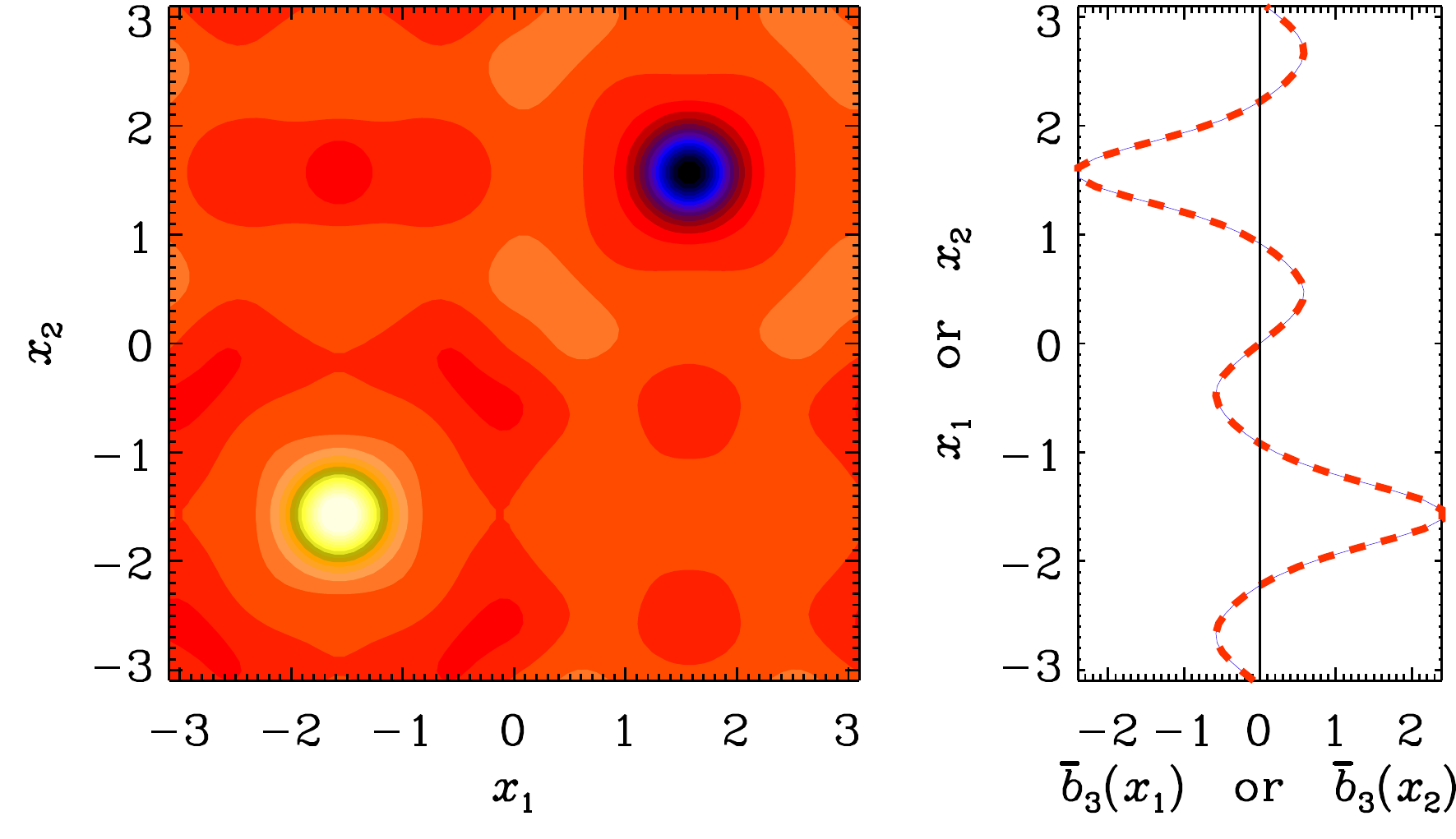}
\caption{The $x_3$-averaged rms-normalised mean field
$\overline{b}_3(x_1,x_2)/\langle\overline{b}_3^2\rangle^{1/2}$
from DNS for $\eta=0.1$ in a domain of size $(2\pi)^3$ (left panel)
and the planar average $\overline{b}_3(x_1)$ (obtained by averaging
over $x_2$ and $x_3$, black line), and $\overline{b}_3(x_2)$ (averaging
over $x_1$ and $x_3$, red dashed line overplotted).
Same colour-coding scheme as in Fig.~\ref{pbzm}.
}\label{ppzaver}\end{figure}

We now consider averaging over the $(x_2,x_3)$ plane.
The evolution of the auxiliary fluctuating field is now controlled
by the operator $\P\L\P$, where $\P$ is the projection that deletes the mean
field, but preserves the volume average, and $\L$ is the operator of magnetic
induction. The dominant modes of the new operator belong
to a symmetry subspace, different from those, where the dominant modes of $\L$
acting alone reside (see in the left panel of \Fig{ppzaver} the mean saturated
magnetic field produced by DNS with the use of the {\sc Pencil
Code}\footnote{\url{http://github.com/pencil-code}}). Despite
the additional
projections, the main visible magnetic structures are still the vertical flux
ropes centred at the family III stagnation points of mTG. The right panel
of \Fig{ppzaver} shows the mean field (which is now a function of $x_1$).
It has positive and negative extrema at $x_1=\mp\pi/2$.
Our computations also reveal that the two possible mean fields,
$\overline{b}_3(x_1)$ and $\overline{b}_3(x_2)$, have the same shape.
Again, the mean field is anharmonic and therefore the eddy diffusivity
cannot be spatially constant.

\begin{figure}[t!]
\center\includegraphics[width=\columnwidth]{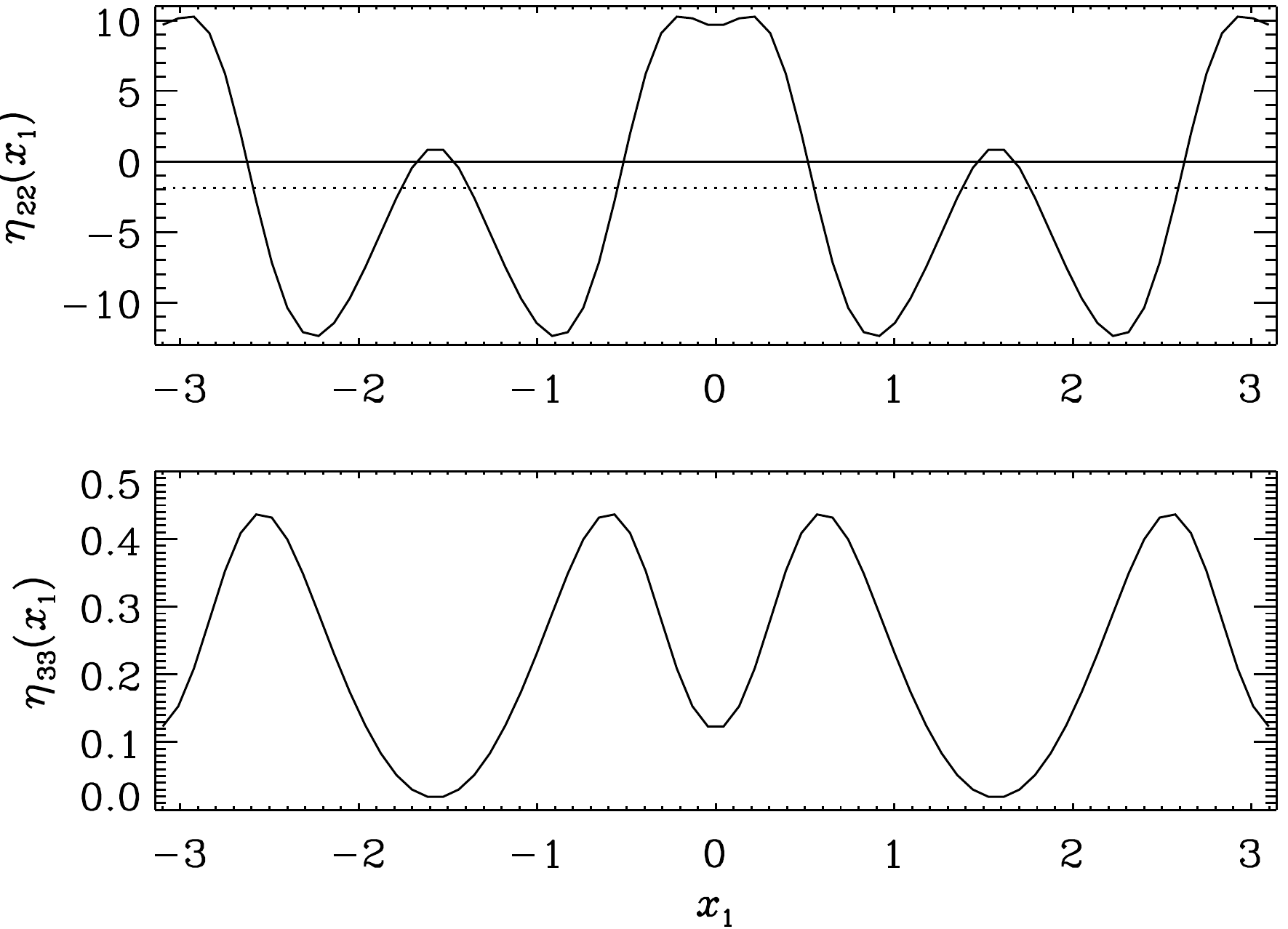}
\caption{$\eta_{22}(x_1)$ and $\eta_{33}(x_1)$ for $\eta=0.1$ and $k_1=1$.
$\eta_{22}(x_1)$ has a negative average, indicated by the dotted line.
}\label{ppxyaver}

~

\center\includegraphics[width=\columnwidth]{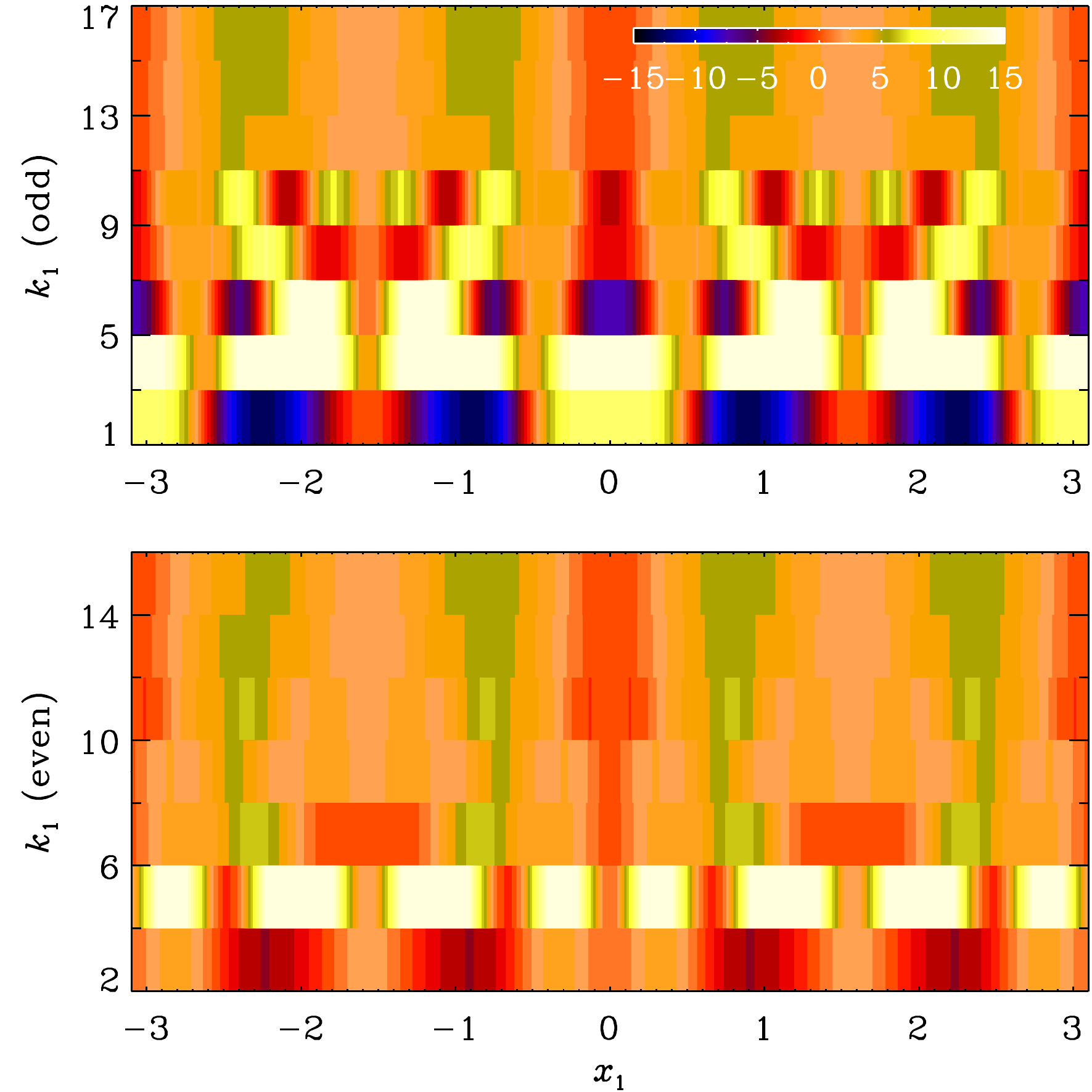}
\caption{$\eta_{22}(x_1,k_1,0)k_1^2$ for $\eta=0.1$ shown colour-coded
separately for odd (upper panel) and even (lower panel) values~of~$k_1$.
Same colour-coding scheme on both panels.
}\label{pker}\end{figure}

Owing to the anharmonic nature of the resulting mean fields,
we must consider test fields involving many Fourier harmonics.
Let us begin with the most important contribution from $k_1=1$. We use
again the {\sc Pencil Code}, where TFM is readily implemented.
In all the cases presented below we have used $72^3$ mesh points.
In \Fig{ppxyaver} we show the results for $\eta_{22}(x_1)$ and $\eta_{33}(x_1)$
for $\eta=0.1$ and $k_1=1$.
Note that both $\eta_{11}$ and $\eta_{33}$ show strong spatial variations.
However, while $\eta_{33}$ is always positive, $\eta_{11}$ has extended
regions where it is negative, giving rise to growth of $\overline{b}_3(x_1)$.

In principle, negative diffusivities can be used in a numerical
mean-field simulation. However, one would then
need to include contributions from larger wave numbers $k_1$ (or~$\epsilon$),
where $\eta_{22}$ eventually becomes positive for large wave numbers.
This was demonstrated in \cite{BrMi}, where the turbulent diffusivity
kernel was spatially constant, and so the relevant eigenvalue problem became
$$\Lambda\hat{A}_2=-(\eta+\eta_{22}(k_1,\i\Lambda))\,k_1^2\hat{A}_2$$
(cf.~\rf{eddei}). Here, $\hat{A}_2$ is the Fourier amplitude and,
for consistency (cf.~\rf{xemf}--\rf{Fou}), the eddy correction $\eta_{22}$
should be calculated for $\omega=\i\Lambda$, which is in general complex.
In the present case, we only find non-oscillatory growth, so $\Lambda$
is real and therefore the frequency $\omega$, for which $\eta_{22}$
is needed, is purely imaginary.
Since the dependence of $\eta_{22}$ on $\omega$ is in general nonlinear,
one has a nonlinear eigenvalue problem that can be solved iteratively.
Even in the simplest cases considered by \cite{HB09}, $\eta_{22}$ is
proportional to $1/(1-\i\omega\tau)$, where $\tau$ is the memory time.
To understand this proportionality for large $|\omega|$, we note that
for test fields \rf{mem} we find from~\rf{bprim}
$${\bf b'}=\i\omega^{-1}\left(\L_{\varepsilon\bf q}{\bf e}_n+{\rm O}(|\omega|^{-1})\right),$$
where $\varepsilon{\bf q}=\kk$ in the definition \rf{eee} of the operator
$\L_{\varepsilon\bf q}$.
An illustrative example of the iterative procedure was given by \cite{RDRB14}
for a more complicated case where $\omega=\i\Lambda$ was complex.
We can encounter a neutral dynamo such that Re$\Lambda=0$
(this usually occurs for a specific value of $k_1$)
by increasing $k_1$, i.e., decreasing the domain;
see Figs.~1 and 2 of \cite{RDRB14} for a related problem.

In the present case, because $\eta_{22}(x_1)$ is nonuniform, we have to allow
for all possible wave numbers of the resulting mean field and compute
the response for each wave number.
This is just opposite to the usual mean-field dynamo problem and
the MST approach where one computes the dynamo effects in the limit $\kk\to0$.
The relevant eigenvalue problem for our domain of size $2\pi$ now becomes
\be\Lambda A_2(x_1)=-\sum_{k_1=1}^\infty
&(\eta+\eta_{22}(x_1,k_1,\i\Lambda))k_1^2\nonumber\\
&\times\int_{-\pi}^\pi\e^{\i k_1(x_1-\xi_1)}A(\xi_1)\,\d\xi_1.
\label{implicit_eigenvalue_problem}\end{align}
In \Fig{ppxyaver} we have already plotted $\eta_{22}(x_1,1,0)$, but
we now need $\eta_{22}(x_1,k_1,\i\Lambda)$ for all integer values of $k_1$
and a suitable value of $\omega$.

\begin{figure}[t!]
\center\includegraphics[width=\columnwidth]{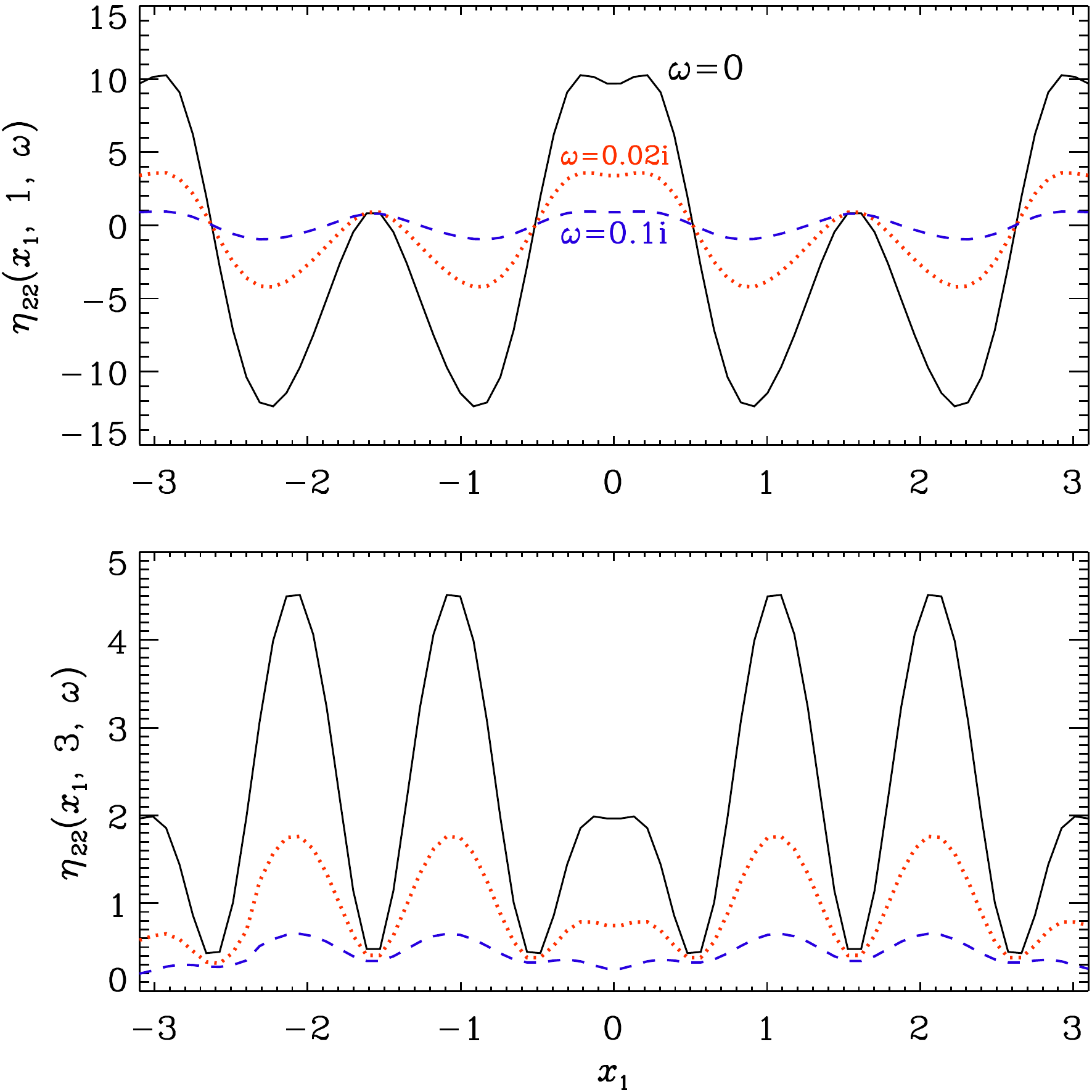}
\caption{Compensated kernel $\eta_{22}(x_1,1,\omega)$ (upper panel) and
$\eta_{22}(x_1,3,\omega)$ (lower panel)
for $\omega=0$ (black solid line), $0.02\i$ (red dotted line),
and $0.1\i$ (blue dashed line) using $\eta=0.1$.
}\label{comp_kernel_lam}

~

\center\includegraphics[width=\columnwidth]{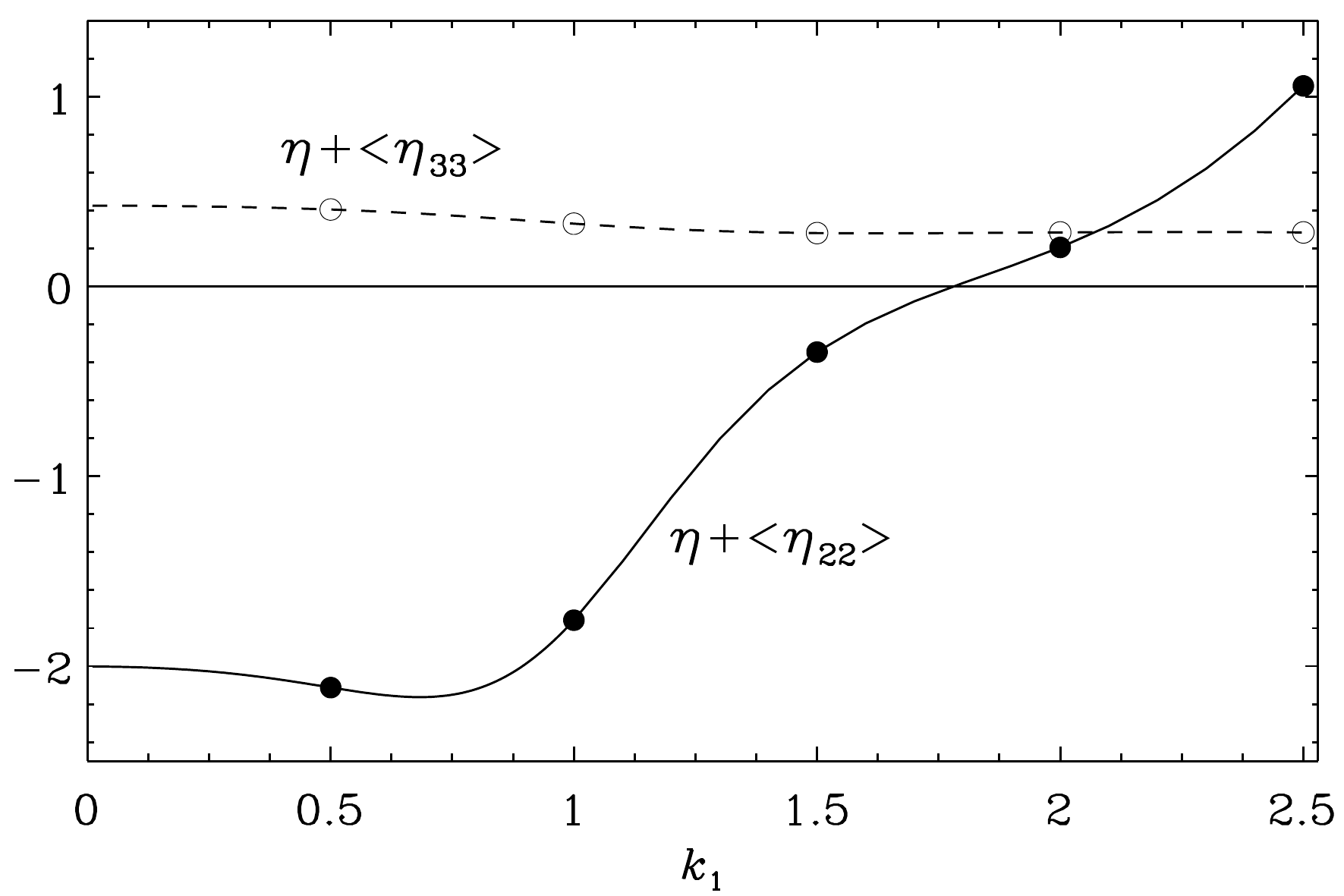}
\caption{Dependence of $\eta+\LA\eta_{22}\ra$ on $k_1$,
obtained with TFM by computing a steady solution to \rf{notmieq}
(solid line), versus the results of the {\sc Pencil Code}
(filled symbols). For comparison, $\eta+\LA\eta_{33}\ra$ is shown
(hollow symbols). Here, $\eta=0.1$, $\omega=0$.
}\label{peddydiff}\end{figure}

Note that for our
domain of size $2\pi$ the permissible wave numbers $k_1$ are integers.
Furthermore, looking at the right panel of \Fig{ppzaver}, we see
that the eigenfunction is odd about $x_1=0$.
This means that only odd values of $k_1$ contribute to the solution.
In agreement with our earlier experience, the amplitudes of the turbulent
transport coefficients fall off quadratically with increasing wave number
\citep[see, e.g.,][]{BRS08}.
We therefore expect that the compensated expression
$\eta_{22}(x_1,k_1,\i\Lambda) k_1^2$ should
be independent of $k_1$ for large values.
This is indeed the case, as can be seen from \Fig{pker}, where we plot
$\eta_{22}(x_1,k_1,0) k_1^2$ separately for odd and even values of $k_1$.

We should point out that these results are sensitive to the
values of $\eta$ and $\omega$, as will be demonstrated next.
First, in \Fig{comp_kernel_lam} we plot $\eta_{22}(x_1,k_1,\i\Lambda)$
for $\omega=0$, $0.02\i$ and $0.1\i$, and for $k_1=1$ and 3.
While the shapes of the different curves remain similar,
there is a significant reduction in the amplitude as $\omega$ increases.
Thus, it is in general impossible to omit the memory effect.
This agrees with earlier results for certain steady flows
\citep{RBDSR11,RDRB14}, although it is not a typical feature of
turbulent flows \citep{HB09}.
Second, we give in Table~\ref{TetaDep} the volume-averaged values
$\LA\eta_{22}\ra$ and $\LA\eta_{33}\ra$ for $\omega=0$ and different values
of $\eta$.
It turns out that $\eta+\LA\eta_{22}\ra=-0.017$ for $\eta=0.115$,
tentatively suggesting that this case is weakly supercritical, while for
$\eta=0.120$ we have $\eta+\LA\eta_{22}\ra=+0.053$, which would
be clearly subcritical.
These values are close to those obtained from DNS, which show that
the critical value of $\eta$ for the onset of generation of magnetic field
with the periodicities of the flow is around 0.1105 (see also Table~\ref{tabl}).
For more precise statements we would need to consider numerical solutions
to \rf{implicit_eigenvalue_problem}.

\begin{table}[t!]\caption{
$\LA\eta_{22}\ra$ and $\LA\eta_{33}\ra$ computed
for $\omega=0$ and various $\eta$.\vspace*{-1em}}
\center\begin{tabular}{|ccc|}\hline
$\eta$ & $\LA\eta_{22}\ra$ & $\LA\eta_{33}\ra$\\\hline
0.100 & $-1.858$ & 0.230 \\
0.110 & $-0.245$ & 0.239 \\
0.115 & $-0.117$ & 0.242 \\
0.120 & $-0.047$ & 0.244 \\\hline
\end{tabular}\label{TetaDep}

~

\caption{Comparison between $\eta+\LA\eta_{22}\ra$,
obtained with TFM by computing a steady solution to \rf{notmieq}
(first column, marked by an asterisk),
and $\LA\eta_{22}\ra$ as well as $\LA\eta_{33}\ra$
obtained with the {\sc Pencil Code}
for $\eta=0.1$, $\omega=0$ and various $k_1$.}
\center\begin{tabular}{|lccc|}\hline
$k_1$& $\eta+\LA\eta_{22}\ra^*$ & $\LA\eta_{22}\ra$ & $\LA\eta_{33}\ra$\\\hline
2.5 & $+1.05656$ & $+0.9559$& 0.1843\\
2.0 & $+0.20545$ & $+0.1055$& 0.1843\\
1.5 & $-0.34576$ & $-0.4460$& 0.1808\\
1.0 & $-1.75831$ & $-1.858$ & 0.2304\\
0.5 & $-2.11246$ & $-2.213$ & 0.3051\\
0.25& $-2.02750$ & $-2.127$ & 0.3210\\\hline
\end{tabular}\label{TepsDep}\end{table}

The $k_1$-dependence of eddy diffusivity $\eta+\LA\eta_{22}\ra$ is shown in
\Fig{peddydiff} and numerical values are given in Table~\ref{TepsDep}
for $\eta=0.1$ and $\omega=0$.
Here we compare the results from TFM obtained with the {\sc Pencil Code}
with those obtained by cancelling the exponential
factor and determining steady solutions to \rf{notmieq};
the latter were computed by the code by \cite{Fok}, employing
the biconjugate gradients stabilised method
BiCGstab($\ell$) for $\ell=6$ (see \citealt{SF,SV95,SV96}).
We also show $\eta+\LA\eta_{33}\ra$ which is always positive.
Note that $\eta+\LA\eta_{22}\ra$ becomes zero at $k_1\approx1.8$.
Thus in our domain of size $2\pi$, where $k_1=1$ is the smallest
wave number, the volume-averaged eddy diffusivity is clearly negative.

\section{Concluding remarks}\label{conclu}

In mean-field electrodynamics, various analytical and numerical approaches
are used to express the mean electromotive force, originally defined
in terms of small-scale fluctuations of flow velocity and magnetic
field, as functions of the large-scale mean flow and magnetic field.
Assessing the range of validity and clarifying
conflicts in application of these approaches is crucial in view of many
applications, e.g., in laboratory experiments for dynamo generation or
in astrophysics. For instance, a comparison of the traditional MFE approach
with those based on $\tau$-approximations of turbulence theory was carried out
by \cite{RR07}. Here, we have compared two different methods for estimation
of the mean e.m.f.: MST,
which explicitly considers steady or time-periodic laminar flows,
and TFM, which is not affected by such a restriction.
For instance, \cite{CSN14} recently observed in a rotating liquid sodium
experiment ``Derviche Tourneur Sodium'' a reduction of the effective magnetic
diffusivity in some regions of the flow, probably caused by
the turbulence; it is thus important to assess, under which
conditions the methods investigated here can be used to study
this kind of problems.

We have demonstrated that, in the two-scale setup, magnetic eddy diffusivities
predicted by MST
in three-dimensional small-scale steady flows are reproduced by TFM,
provided volume averaging is used. It can be similarly shown that the same
result holds true for time-periodic flows --- in this case the averaging
procedure must also involve time averaging over the temporal period.
If other types of averaging (planar or over just one Cartesian
variable) are applied, one can, in general, only expect a qualitative
agreement between the results.
One must also be aware of
the following caveat: Because of the asymptotic
character of the MST results, achieving agreement with TFM eddy diffusivities
requires small scale ratios $\varepsilon$ and thus high spatial resolution
to be used when solving the TFM test problems \rf{bprim}. This can be seen
as a drawback of TFM in comparison with MST.

Results coinciding with those of MST can also be obtained by a modified TFM
algorithm that proceeds by setting a test field as initial condition and
solving the standard magnetic induction equation \rf{main} without separating
the field into mean and fluctuating parts before the saturated regime
for the magnetic field sets in, and then computing the e.m.f.~due to
the fluctuating flow and magnetic field as in the canonical TFM. Reliance
of TFM on the integral~\rf{coint} approximation of the e.m.f., resulting
in the ansatz~\rf{xemf} (or~\rf{xemfev}) for the Fourier transforms, then
proves crucial. This feature of TFM implies that, in the multiscale limit
$\varepsilon\to0$, the
choice of the spatial variables for averaging becomes insignificant, because
the use of the Fourier transform over the remaining variables effectively
converts all the averages into averages over all the three spatial variables
(see~\rf{essal} and~\rf{esssp}). This observation does not hold for the
conventional version of MFE, where the e.m.f.~is approximated by local
differential operators. Nevertheless, the MST $\alpha$-effect tensor in small-scale
flows can be computed by TFM with the use of constant test fields and full
spatial averaging.

We have numerically confirmed the findings of \cite{La} that the modified
Taylor--Green flow possesses, in certain ranges of parameter values,
negative eddy diffusivities, and have shown that the same holds
for the G.O.~Roberts flow~IV. This is in contrast with \cite{BrMi},
who did not find negative eddy diffusivity for the former flow.
Why did \cite{La} and \cite{BrMi} arrive to different conclusions for this
flow? We have seen that the results of MST and TFM do not agree qualitatively
unless TFM applies volume averaging (R-IV depending on two spatial
variables is a special case), but a number of less important reasons
make the picture even more complicated: ($i$)
We have now obtained negative eddy diffusivities in mTG
by TFM, but we have been forced to employ a planar averaging
different from the one used by \cite{BrMi}. ($ii$) Eddy
diffusivities affecting the evolution of large-scale perturbations of distinct
short-scale magnetic modes do not coincide. While \cite{La} considered eddy
diffusivity for the {\it neutral} short-scale modes, \cite{BrMi} aimed
at evaluating it for the {\it dominant} short-scale modes, which
for $\eta=0.02$ is a distinct branch; hence, there are no reasons to expect
their results on eddy diffusivity to be interrelated. ($iii$)
Furthermore, at large scale separations (i.e., small~$\varepsilon$)
the eddy diffusivity for the dominant branch for $\eta=0.02$ is negative (see
\Fig{re50}); in this case, TFM still can be used to evaluate the
eddy diffusivity, but special precautions must be taken in its implementation:
TFM requires solving test problems \rf{bprim}, which are
likely to inherit the instability of the unperturbed magnetic induction
equation \rf{main} and have exponentially growing modes. In the course
of numerical integration, the growing modes will then set in
due to the influence of round-off errors and progressively wipe out
the contribution of the inhomogeneity in \rf{bprim}, which we intend
to determine. A feasible strategy is to compute directly a time-independent
solution to \rf{bprim} regarded as a system of linear equations
(after a suitable discretisation of the problem in space).

Investigation of eddy diffusivity is supposed to yield the effective
diffusivity which can be employed, e.g., to study nonlinear large-scale
MHD regimes. Our results demonstrate that this may be a non-realistic goal.
Magnetic field in a nonlinear MHD regime can be decomposed into a linear
combination of eigenmodes of the magnetic induction operator, where the
coefficients are time-dependent. We have presented growth rates in branches of
dominant large-scale magnetic eigenmodes generated by mTG in four symmetry
subspaces. The parabolic shape of their plots as
functions of the scale ratio $\varepsilon$ near $\varepsilon=0$ or 1 confirms
that the phenomenon of magnetic eddy diffusivity is observed for sufficiently large
scale separations in all branches. However, \Fig{re50} also shows that the
curvature of the parabola varies significantly for different branches of modes.
Hence, for a given molecular diffusivity, no universal
eddy diffusivity tensor can be assigned to a given generating flow, because the
action of eddy diffusivity significantly depends on the spectral composition of
the multiscale magnetic field itself, on which such an integral diffusivity
acts. We are thus forced to conclude that a unified description of ``average''
magnetic eddy diffusivity would only be possible in the case of a turbulent MHD regime
with a well-defined statistics of the spectral composition of magnetic field
--- for instance, when a chaotic attractor of an MHD dynamical system is
considered. (As a side remark, we note that the same holds true for the
$\alpha$-effect tensor: for large-scale permutations of different small-scale
magnetic modes generated by small-scale non-symmetric flows, generically
different tensors are obtained, and hence the integral $\alpha$-effect tensor
depends on the spectral composition of the magnetic field.)
From this prospective, the TFM approach seems advantageous, since it demands
to separately quantify the influence of each Fourier harmonics in the mean
magnetic field by nominating it as a test field and computing
the fluctuating field that it induces and the respective output e.m.f.,
and afterwards to sum up such contributions of each individual harmonics
into the integral output.

Let us note some open questions,
beginning with the most important one.

1. TFM follows MFE in relying on
approximations such as \rf{coint} (e.g., the precise kernels
in \rf{coint} are not translation-invariant). It is essential,
that nevertheless in some limits (the most important of which is the limit
of small magnetic Reynolds numbers) MFE yields results that are {\it exact},
i.e., TFM and MFE agree exactly and in all details with DNS
\citep[see the review by][]{BCDSHKR10}.
Given that ($i$) we know that MST is a precise corollary of the basic
equations, and ($ii$) we have observed that MST can quantitatively
disagree with TFM, we need to understand the mathematical reasons
for the aforementioned agreement in the respective limits. The availability
of precise mathematical demonstrations may help to determine
the conditions, under which TFM can be used reliably.

2. As we have mentioned in the Introduction,
the generated large-scale structures are described by amplitudes (depending
exclusively on slow variables) of the small-scale neutral (magnetic or MHD
stability) modes that constitute the leading term in the expansion
of perturbation in the scale ratio. In two-scale systems (such as the ones
considered here), the temporal evolution of the amplitudes is governed by equations
(mean-field or otherwise), where the $\alpha$-effect operator is never present
{\it together} with the eddy diffusivity operator: since the orders of these
differential operators are different, they emerge at different orders
of the scale ratio. (A joint action of molecular diffusivity and
the $\alpha$-effect is encountered in flows with an internal spatial scale,
see Chaps.~10 and 11 in \citealt{Zh}.)
Do the two operators appear simultaneously in amplitude
equations in a truly multiple-scale setup? In other words, can in such a setup
the mean e.m.f.~be a sum of the $\alpha$- and $\eta$-terms, as this
was assumed in the early variants of MFE?

3. Expressing entries of the magnetic eddy diffusivity correction tensor
in terms of solutions
to auxiliary problems for the adjoint operator has proved useful not only
for reducing the amount of computations, but also in analytical work,
for establishing relations~\rf{opf} between the tensors for opposite flows,
and for identifying zero entries of the tensor for translation-invariant flow
(see Section~\ref{flowIV}). In these calculations we have relied
on the similarity of the magnetic induction operator and the adjoint operator
for the reverse flow. Do solutions to the auxiliary problem for the adjoint
operator have a physical interpretation?

4. Finally, the following technical question is of certain interest:
Eq.~\rf{bprim} governing the evolution of the fluctuating magnetic
field involves the operator $\P\L\P$, where $\P$ projects out
the mean field and $\L$ is the usual operator of magnetic induction.
Suppose that there is no small-scale dynamo action, i.e., all eigenvalues
of $\L$ have non-positive real parts, and an averaging other than volume
averaging is employed. Is it then possible, for some flows and some test fields,
to have growing fluctuating solutions, i.e., can the operator $\P\L\P$ have
an eigenvalue with a positive real part?

\section*{Acknowledgments}

We are grateful to Uriel Frisch, Alessandra S.~Lanotte, Dhrubaditya Mitra and
Matthias Rheinhardt for valuable comments.
AB gratefully acknowledges financial support from the European Research
Council under the AstroDyn Research Project 227952. Part of
computations have been carried out at the National Supercomputer
Centre in Ume{\aa} and at the Center for Parallel Computers at the
Royal Institute of Technology in Sweden.
Research visits of VZ to the Observatoire de la C\^ote d'Azur (France) were
supported by the French Ministry of Higher Education and Research.

\end{document}